\documentclass[twocolumn,iop]{emulateapj}
\usepackage{color}
\usepackage{verbatim}
\usepackage{hyperref}
\usepackage{url}

\def \tlya {$\tau_{\rm eff}^{\rm Ly\alpha}$}

\def\2pr{^{\prime \prime}}
\def\deg{^{\circ}}
\def\greatsim{\mathrel{\raise.3ex\hbox{$>$\kern-.75em\lower1ex\hbox{$\sim$}}}}
\def\lesssim{\mathrel{\raise.3ex\hbox{$<$\kern-.75em\lower1ex\hbox{$\sim$}}}}
\def\gs{\mathrel{\raise0.27ex\hbox{$>$}\kern-0.70em
\lower0.71ex\hbox{{$\scriptstyle \sim$}}}}
\def\ls{\mathrel{\raise0.27ex\hbox{$<$}\kern-0.70em
\lower0.71ex\hbox{{$\scriptstyle \sim$}}}}

\slugcomment{Submitted to {\em Astronomical Journal}}

\shorttitle{Quasar Composite}
\shortauthors{Harris et al.}

\usepackage{color}
\usepackage{graphicx}
 
\newcommand{\lya}{\mbox{Ly$\alpha$}}

\newif\ifdraftmodep
\draftmodepfalse

\newif\ifapjp
\apjpfalse

\begin{document}

\title{The Composite Spectrum of BOSS Quasars Selected for Studies of the Lyman-$\alpha$ Forest}

\author{
David~W.~Harris\altaffilmark{1},
Trey~W.~Jensen\altaffilmark{1},
Nao~Suzuki\altaffilmark{2},
Julian~E.~Bautista\altaffilmark{1},
Kyle~S.~Dawson\altaffilmark{1},
M.~Vivek\altaffilmark{1},
Joel~R.~Brownstein\altaffilmark{1},
Jian~Ge\altaffilmark{3},
Fred~Hamann\altaffilmark{3},
H.~Herbst\altaffilmark{3},
Linhua~Jiang\altaffilmark{4},
Sarah~E.~Moran\altaffilmark{5},
Adam~D.~Myers\altaffilmark{6},
Matthew~D.~Olmstead\altaffilmark{1,7},
Donald~P.~Schneider\altaffilmark{8,9}
}

\altaffiltext{1}{
Department of Physics and Astronomy, 
University of Utah, Salt Lake City, UT 84112, USA.
}

\altaffiltext{2}{
Kavli Institute for the Physics and Mathematics of the Universe, 
University of Tokyo, Kashiwa, 277-8583, Japan.
}

\altaffiltext{3}{
Department of Astronomy, 
University of Florida, Gainesville, FL 32611-2055, USA.
}

\altaffiltext{4}{
Kavli Institute for Astronomy and Astrophysics, 
Peking University, Beijing 100871, China.
}

\altaffiltext{5}{
Barnard College, 3009 Broadway, New York, NY 10027, USA.
}

\altaffiltext{6}{
Department of Physics and Astronomy, 
University of Wyoming, Laramie, WY 82071, USA.
}

\altaffiltext{7}{
King's College, 
Department of Chemistry and Physics, Wilkes Barre, PA 18711, USA.
}

\altaffiltext{8}{
Department of Astronomy and Astrophysics, The Pennsylvania State University, University Park, PA 16802, USA.
}

\altaffiltext{9}{
Institute for Gravitation and the Cosmos, The Pennsylvania State University, University Park, PA 16802, USA
}

\date{\today}

\email{davidharris314@gmail.com}

\begin{abstract}
The Baryon Oscillation Spectroscopic Survey (BOSS) has collected more than 150,000 $2.1 \leq z \leq 3.5$ quasar spectra since 2009. Using this unprecedented sample, we create a composite spectrum in the rest-frame of 102,150 quasar spectra from 800 \AA\ to 3300 \AA\ at a signal-to-noise ratio close to 1000 per pixel ($\Delta v$ of 69 km~s$^{-1}$).  Included in this analysis is a correction to account for flux calibration residuals in the BOSS spectrophotometry.  We determine the spectral index as a function of redshift of the full sample, warp the composite spectrum to match the median spectral index, and compare the resulting spectrum to SDSS photometry used in target selection.  The quasar composite matches the color of the quasar population to 0.02 magnitudes in $g-r$, 0.03 magnitudes in $r-i$, and 0.01 magnitudes in $i-z$ over the redshift range $2.2<z<2.6$.   The composite spectrum deviates from the imaging photometry by 0.05 magnitudes around $z = 2.7$, likely due to differences in target selection as the quasar colors become similar to the stellar locus at this redshift.  Finally, we characterize the line features in the high signal-to-noise composite and identify nine faint lines not found in the previous composite spectrum from SDSS.
\end{abstract}

\keywords{Techniques: Spectroscopic---Quasars: General}

\section{Introduction}\label{sec:intro}

\setcounter{footnote}{0}

The material associated with quasar emission lines is ionized by the ultraviolet (UV) continuum emitted by an accretion disk surrounding a black hole. The relative flux between different emission lines can illuminate quasar structure and constrain quasar models \citep{wills85a,francis91a}.  \citet{wills85a} and \citet{francis91a} both found the ratio between Lyman-$\alpha$ flux and Fe~\textsc{II} flux to be much greater than expected given standard photoionization models.  \citet{wilhite05a} and \citet{ruan14} reported an average quasar spectral index ($\alpha_\lambda$) of -1.35 and -1.38, respectively, in the optical to Far UV (1300 \AA\ to 5800 \AA\ and 1400 \AA\ to 7000 \AA, respectively), which may be an important parameter in quasar photoionization models if the quasar spectral index at shorter wavelengths is related to the spectral index at these wavelengths.  \citet{pereyra06} and \citet{sakata11} argue that the quasar accretion disk can be modeled by a simple thin disk because the spectral index varies between repeat spectral observations of quasars.  \citet{nagao06} also constrained quasar models through observations of the broadline region in quasar spectra, constraining the gas metallicity and metallicity evolution of the broadline region.

Weaker emission lines, including iron lines, are detectable in high signal-to-noise ratio composite quasar spectra \citep[e.g.;][]{francis91a}.  Composite spectra can provide redshift estimators for individual quasar spectra when used as a cross correlation template \citep{vandenberk01b}.  Composite spectra and a record of their line profiles  allow for better quasar selection from photometric data by predicting color as a function of redshift \citep{richards02a}.  They can facilitate better continuum normalization for quasar absorption line studies, for example in studies of the Lyman-$\alpha$ forest and Broad Absorption Line (BAL) quasars \citep{grier15a}.

Quasar composite spectra also provide insight into quasar astrophysics.  \citet{francis91a} showed that quasar spectra are governed by a power-law continuum by generating a composite of 718 quasars from the Large Bright Quasar Survey \citep{foltz89}, covering a wavelength range of approximately 800 \AA\ to 5800 \AA\ in the restframe.  \citet{brotherton01a} created a quasar composite spectrum from 657 quasars in the FIRST Bright Quasar Survey.  \citet[][hereafter V01]{vandenberk01a} used 2204 spectra to generate a composite spectrum, taking data from the commissioning phase of the Sloan Digital Sky Survey 1 \citep[SDSS;][]{york00a}.  Their composite covered a wavelength range of 800 \AA\ to 8555 \AA\ in the restframe and included quasars in the redshift range $0.044 \leq z \leq 4.789$.  V01 showed systematic velocity shifts from tens to a few hundred km~s$^{-1}$  between quasar emission lines.  The optical depth blueward of Lyman-$\alpha$ increases with redshift, thus requiring a low-redshift survey from a space-based telescope to create a quasar spectrum blueward of Lyman-$\alpha$.  \citet{shull12a} created a composite spectrum from 22 Active Galactic Nuclei (AGN) using HST-COS data, with a wavelength range of 550 \AA\ to 1750 \AA\ in the restframe over $0.026 \leq z \leq 1.44$.   This high signal-to-noise ratio composite spectrum shows that the power-law continuum continues below the Lyman Limit wavelength (912 \AA) without a significant break.  \citet[][hereafter S14]{stevans14a} included 159 quasars and expanded the work of \citet{shull12a} by covering a slightly greater redshift range.  

In this work, we create a quasar composite spectrum from 102,150 quasars observed in the Baryon Oscillation Spectroscopic Survey \citep[BOSS;][]{dawson13a}, a component of SDSS-III \citep{eisenstein11a}.  These quasars cover a redshift range of $2.1 \leq z \leq 3.5$ and a restframe wavelength range of 800 \AA\ to 3300 \AA, and do not include Broad Absorption Line (BAL) quasars or Damped Lyman-$\alpha$ (DLA) quasars.  This work differs from previous efforts by increasing the number of quasars and restricting the sample to objects which lie in a smaller region of redshift and i-band absolute magnitude space.  This composite can be used in testing theoretical iron line templates \citep{bruhweiler08a} and detecting new quasar spectral lines \citep{vandenberk01b}. This spectrum will allow more detailed studies of known quasar spectral lines, including line strengths, line widths, and velocity offsets, which can be used to constrain theoretical models of quasar structure.  A key component of this work is a correction to the flux calibration, which is required in BOSS quasar observations to construct an accurate and representative continuum, as quasars were not observed in the same way as standard stars.

In Section~2, we briefly describe the surveys from SDSS-I/II and SDSS-III, define the quasar sample used in the construction of the composite spectrum, and present the properties of that sample.
In Section~3, the method for the creation of the composite spectrum is presented, along with the method for determining the uncertainty as a function of wavelength.
In Section~4, we measure the spectral index of the quasar composite and compare it to the median spectral index of the quasar sample.  We also verify the accuracy of the composite spectrum by comparing the synthetic photometry of the composite to the photometric colors of the quasar sample.
In Section~5, we present the measurement of emission line profiles and compare the results to prior work.
In Section~6, we conclude and discuss possible future work.
While the spectrophotometric correction is introduced in Section 3, details of the correction are reserved for Appendix A.
In Appendix B, the serendipitous discovery of a spectral flux offset between SDSS-I/II and BOSS is discussed and compared to the work of \citet{margala15a}.
In this work, we assume a $\Lambda$CDM cosmology with $H_{0} = 70$ km~s$^{-1}$, $\Omega_{\rm M} = 0.3$, and $\Omega_{\Lambda} = 0.7$.  These assumptions are consistent with the final WMAP results \citep{hinshaw13a} and the 2015 Planck results \citep{ade15a}.

\section{Data}\label{sec:data}

Quasar spectra used to create the composite are obtained from BOSS.  Photometry used in calibration and for assessing broad-band synthetic photometry of our quasar composite spectrum were obtained from the imaging programs in SDSS-I, -II, and -III.  Additional spectra for calibration were obtained from SDSS-I, -II, and BOSS.  Hereafter, photometry and spectra from SDSS-I and SDSS-II are abbreviated collectively as coming from SDSS.  

\begin{deluxetable}{cc}
\centering
\tablewidth{0pt}
\tabletypesize{\footnotesize}
\tablecaption{\label{tab:sdssAB} Conversion from SDSS to AB magnitudes, in the form $\delta m = m_{AB} - m_{SDSS}$.  }
\tablehead{
\colhead{Sloan Filter} & \colhead{$\delta m$}
}
\startdata
   $u$ & -0.042 \\
   $g$ & 0.036 \\
   $r$ & 0.015 \\
   $i$ & 0.013 \\
   $z$ & -0.002 \\
\enddata
\end{deluxetable}

\subsection{The SDSS Program}\label{subsec:sdssintro}

Imaging data employed in this analysis were taken in the North Galactic Cap during SDSS I/II (2000-2005), and for the South Galactic Cap in 2000-5 and 2008-9.  Data were taken in five photometric bands \citep[Sloan $u$, $g$, $r$, $i$, and $z$,][]{fukugita96a} by the Apache Point 2.5-m telescope \citep{gunn06a}.  The total imaging area was approximately 14,555 deg$^{2}$.  All images were taken on dark photometric nights with good seeing \citep{hogg01a}.  The final photometric data for the analysis were taken from SDSS Data Release 8 \citep[DR8;][]{aihara11a}.  Papers on object detection and photometric measurement include \citet{lupton01a} and \citet{stoughton02a}.  Papers discussing the photometric calibration of the SDSS Imaging Program include \citet{smith02a}, \citet{ivezic04a}, \citet{tucker06a} and \citet{padmanabhan08a}.  The astrometric calibration is discussed in \citet{pier03a}.  SDSS photometry is converted to AB magnitudes \citep{oke83a} as shown in Table~\ref{tab:sdssAB}.

In addition to photometric data, this analysis uses a small sample of spectra obtained in SDSS-I and -II.  The details of the spectra are presented in \citet{york00a}.

\subsection{The BOSS Program}\label{subsec:bossintro}

The primary science goals of BOSS are to characterize the imprint of the Baryon Acoustic Oscillation (BAO) signal on the spatial distribution of Luminous Red Galaxies and the Lyman-$\alpha$ forest detected from quasar spectra.  The BOSS BAO signal was most recently measured for Luminous Red Galaxies in \citet{anderson14a} and in the Lyman-$\alpha$ forest in \citet{delubac15a}.  The quasars utilized in the Lyman-$\alpha$ forest measurements of the BAO provide the basis for the quasar composite spectrum.

\subsubsection{Quasar Target Selection in BOSS}\label{subsec:idqso}

Objects were selected from SDSS imaging data for BOSS spectroscopy with a goal of obtaining confirmed Lyman-$\alpha$ ($z \geq 2.15$) quasars at a density of 15 deg$^{-2}$.  The selection of quasar candidates for spectroscopic observation is difficult due to large photometric errors in objects at $g=22$.  Additionally, most of the quasars observed are at $2.1 \leq z \leq 3.5$, and the quasar locus crosses the stellar locus at $z=2.7$ \citep{fan99b,richards02a}.  

To achieve the desired density of Lyman-$\alpha$ quasars, all point sources identified in the imaging data were passed through a number of different selection schemes.  The two primary methods generated the CORE and {\tt BONUS} samples \citep{ross12a}.  During the first year of operations, the CORE sample utilized the likelihood method explained in \citet{kirkpatrick11a}.  Starting in the second year of operations, a method known as Extreme Deconvolution (XDQSO) selection was employed \citep{bovy11a}.  This method attempts to identify a uniform sample by deconvolving quasars from the stellar locus in color-color space after taking into account photometric uncertainties.  The {\tt BONUS} sample is not a uniform sample, but uses ancillary imaging data from a range of sources to maximize quasar density.  Quasar candidates in the CORE and {\tt BONUS} samples contribute 20 and 18.5 targets deg$^{-2}$, respectively; a few other selections bring the total to 40 deg$^{-2}$.  These selection algorithms produce 15-18 $z>2.1$ quasars deg$^{-2}$ out to the BOSS magnitude limit of ($g \leq 22$ OR $r \leq 21.85$).

As discussed in \citet{ross12a}, the first year of BOSS observations of quasar targets was primarily used to contrast selection techniques. There were 155,019 objects targeted from the XDQSO selection after the first year and 238,553 objects targeted from the {\tt BONUS} selection over the duration of the program. Because the vast majority of quasars in BOSS were targeted according to one of these two selection schemes, and to maintain simplicity and consistency in the sample used to generate this new high signal-to-noise composite spectrum, we chose to use only objects targeted by either the {\tt BONUS} or XDQSO selections.

Because the quasar locus for high redshift ($z \approx 2.7$) quasars overlaps with the stellar locus in SDSS color-color space, many objects targeted as quasars are actually stars.  These objects were not identified as stars until after the spectra were taken.  Thus, we have a sample of stars that are selected and observed in the same manner as the observed quasars.  We call these objects ``stellar contaminants"; most are A, F, or G-type stars.  These stars tend to exhibit low-variability and can be used as a template to test fluxing errors and validate improvements to the flux calibration (see Appendix~\ref{subsec:specphot_err}).

\subsubsection{BOSS Spectroscopic Observations}\label{subsubsec:bossspec}

Spectroscopic observations with BOSS are performed using plates with a radius of 1.5$\deg$ on the sky \citep{smee13a}.  Each plate contains 1000 holes drilled for 2$''$ diameter fibers.  On each plate, approximately 160-200 fibers are allocated for quasar targets, and 20 fibers are reserved for standard stars chosen from photometric data consistent with main sequence F stars.  The holes for each plate are drilled at a position according to the estimated local sidereal time (LST) of the observation to account for hour angle and Atmospheric Differential Refraction (ADR).  

Most holes (including those for standard stars) are drilled with a position corresponding to the predicted ADR for the observation to maximize the throughput at 5400 \AA.  For convenience, we refer to fibers with ADR offsets designed to maximize 5400 \AA\ throughput as the ``red focal plane''.  Since quasar spectra are used to search for the BAO signal in the Lyman-$\alpha$ forest, the decision was made to maximize the flux at the shortest wavelengths of the quasar spectra.  Thus, holes drilled for quasar targets were drilled at slightly different positions to maximize throughput at 4000 \AA; we call this the ``blue focal plane''.  Quasar fibers in this nomenclature are simply shifted parallel to the true focal plane of the standard stars to account for ADR.  Holes drilled in the 4000 \AA\ focal plane more than 1.02$\deg$ from the plate center are also augmented with washers to change the vertical position of the fibers to match the 4000 \AA\ focus, causing a true offset from the focal plane \citep{dawson13a}.

Once observations are completed, the data reduction follows two major steps.  First, 1-D spectra are extracted and calibrated from the raw, 2-D CCD data \citep{stoughton02a}.  An integral phase of the extraction and coaddition process is flux calibration based on model spectra for the standard stars.  This step introduces a distortion in the broadband flux in spectra observed in the blue focal plane relative to the red focal plane.  In the second step of the data reduction, an automated classification and redshift is estimated for each 1-D spectrum \citep{bolton12a}.  One pixel in a BOSS spectrum has a velocity width of $\Delta~v = 69$ km~s$^{-1}$, or identically a width of 0.0001 in logarithmic wavelength.

In this work, the classification and redshift determination are not derived from the automated data reduction pipeline but instead rely on the visual inspections of all the potential quasar spectra obtained in BOSS \citep{paris12a}.  Several techniques to improve the precision of the redshift after visual inspection were used in \citet{paris12a, paris14a}; here, we use the Principal Component Analysis (PCA) redshift technique.
The PCA eigenvectors are derived from a representative sample of 8,632
visually inspected quasars with high signal-to-noise ratio, no BAL
features, and with redshift determined by the Mg~\textsc{II} line.
The redshifts used in this analysis are equivalent to those found in Table 3,
Column 12 of \citet{paris14a}.

\subsection{Sample Selection}\label{subsec:sample}

This work uses quasar data from the DR12Q catalog, presented in \citet{paris14a}.  There are 175,294 quasars in the redshift range $2.1 \leq z \leq 3.5$ included in the DR12Q catalog.  We restrict the composite to that redshift range to better understand the quasar population utilized for BOSS Lyman-$\alpha$ forest studies.  A total of 158,461 of these quasars are from the XDQSO and/or {\tt BONUS} samples.  After removing broad absorption line (BAL) quasars identified by \citet{paris14a} and Damped Lyman-$\alpha$ (DLA) quasars using the DLA catalog from \citet{noterdaeme12a}, we are left with 130,512 quasar spectra. 
The purpose of this analysis is to characterize the spectra of the most
common BOSS quasars over the
redshift range $2.1<z<3.5$.  We attempt to minimize contamination from
metals associated with
absorption systems or outflow of material within the quasar host
galaxy.  After removing broad absorption line (BAL) quasars identified
by \citet{paris14a} and Damped Lyman-$\alpha$ (DLA) quasars using the DLA catalog from
\citet{noterdaeme12a}, we are left with 130,512 quasar spectra.

We also ignore all quasars observed at an airmass above 1.2, as the spectrophotometric correction is least certain at high airmass as shown by the large scatter in the spectrophotometry in Appendix~\ref{subsubsec:photometric_comparison}.  We tested the effect of observed airmass and spectrophotometric errors on the quasar composite and found that a composite generated from quasars observed at an airmass greater than 1.2 had a 5\%\ discrepancy in the shorter wavelengths compared to a composite with the full quasar sample used.  Other airmass ranges with an equal number of quasars had discrepancies of 1\%\ or less in the same range.  The remaining sample is large enough that removing the 28,000 high airmass quasars will not significantly negatively affect the final results.  In total, 102,150 quasars were included in the composite.  These numbers are detailed in Table~\ref{tab:sampletable}.

\begin{deluxetable}{lr}
\centering
\tablewidth{0pt}
\tabletypesize{\footnotesize}
\tablecaption{\label{tab:sampletable} This table shows the number of quasars remaining in the sample after each sample selection criterion is applied.}
\tablehead{
\colhead{Criterion} & \colhead{Remaining Quasars}
}
\startdata
   $2.1 \leq z \leq 3.5$ & 175,294 \\
   XDQSO and {\tt BONUS} & 158,461 \\
   Remove BAL, DLA & 130,512 \\
   Airmass $\leq$ 1.2 & 102,150 
\enddata
\end{deluxetable}

\subsection{Properties of the Final Sample}\label{subsec:properties}

\begin{figure}[h]
\centering
\includegraphics[width=0.5\textwidth]{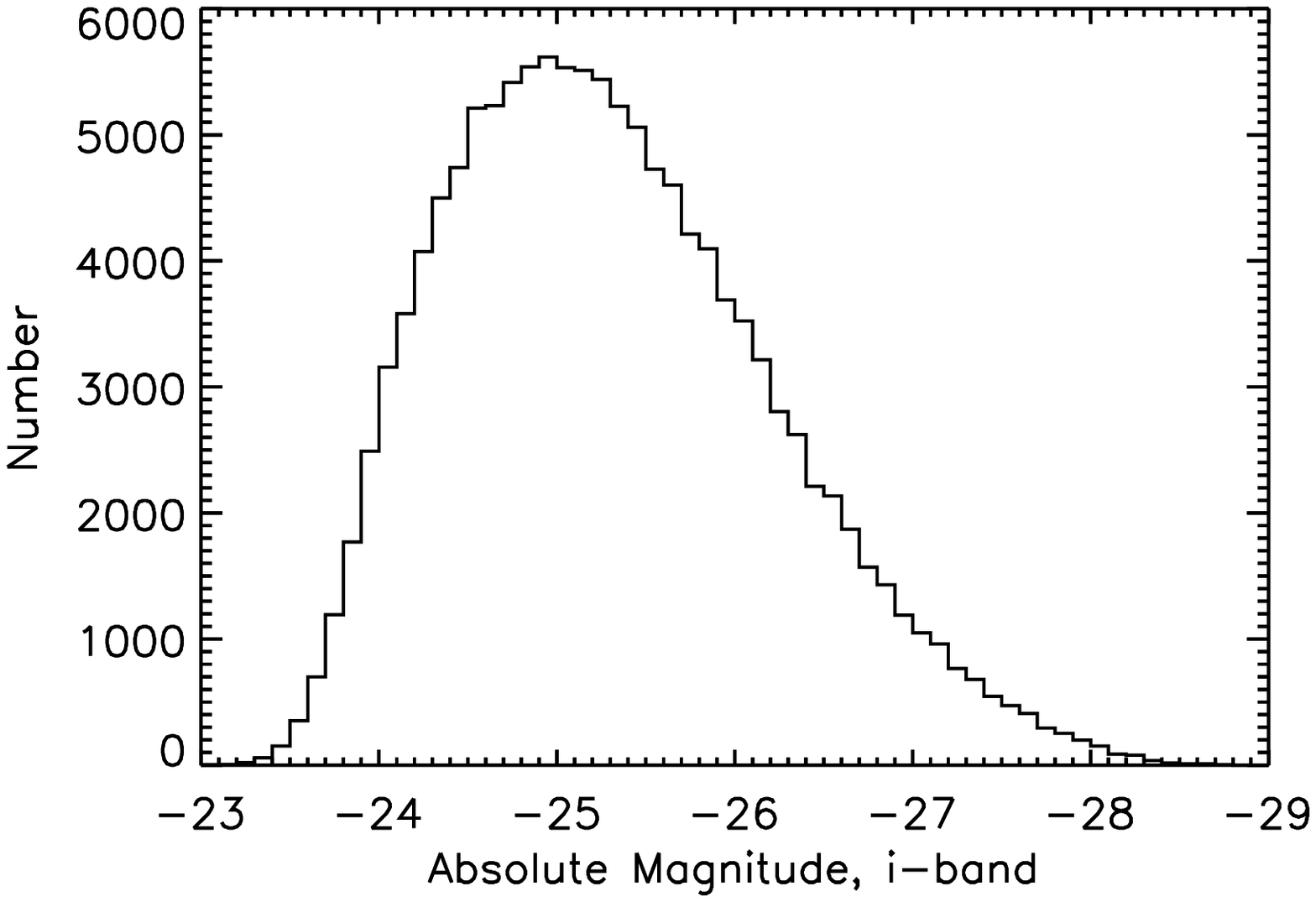}
\includegraphics[width=0.5\textwidth]{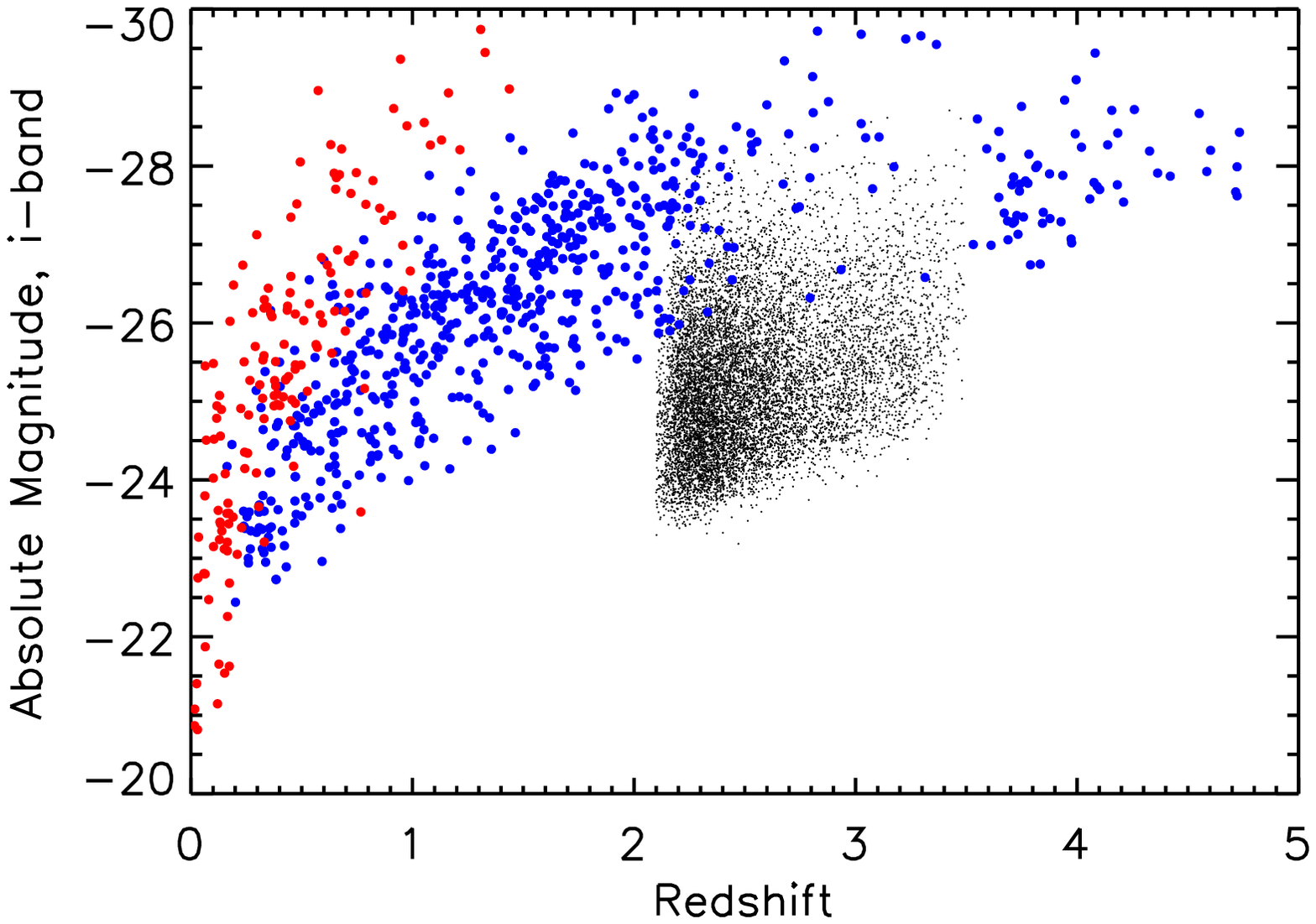}
\caption[ratiostaralpha]{{\bf First panel:}  Absolute $i$-band magnitude of the sample, as calculated by \citet{paris12a}.  {\bf Second panel:}  Absolute $i$-band magnitude vs. redshift for this work (in black), V01 (in blue), and S14 (in red).  A representative subsample (chosen by random selection) of quasars in V01 and this work are shown due to the large number of objects in each sample.}
\label{fig:fig1}
\end{figure}

The first panel of Figure~\ref{fig:fig1} displays the absolute magnitude distribution of the quasar sample used to create the composite spectrum;  68\% of quasars fall between $-24.37$ and $-26.22$ in K-corrected \citep[as in ][]{paris12a} $i$-band absolute magnitude.  The second panel of Figure~\ref{fig:fig1} shows the $i$-band absolute magnitude versus redshift distribution of the quasar sample.  
The median redshift in this sample is 2.457, and the median $i$-band absolute magnitude is $-25.20$.

\begin{figure}[h]
\centering
\includegraphics[width=0.5\textwidth]{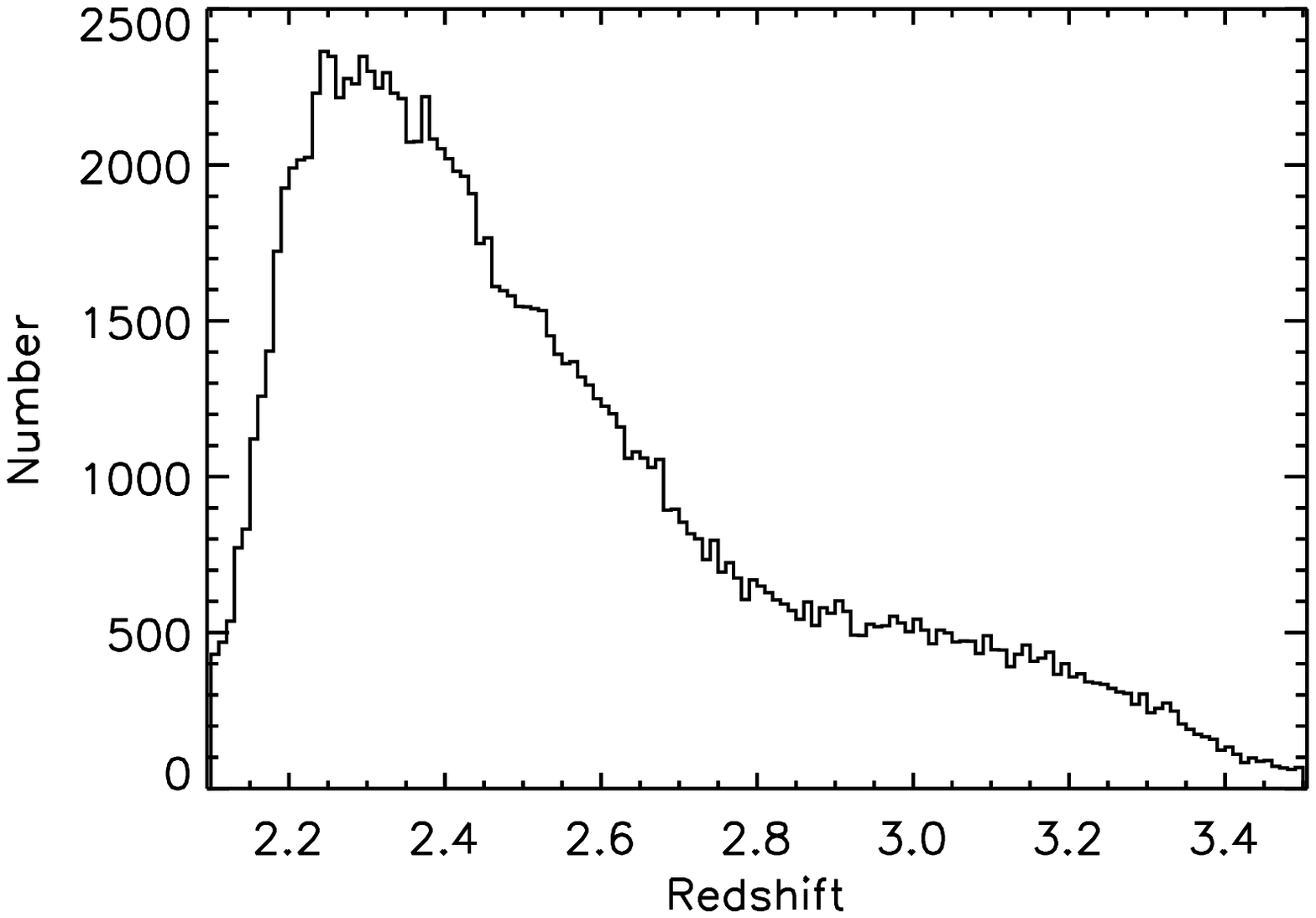}
\includegraphics[width=0.5\textwidth]{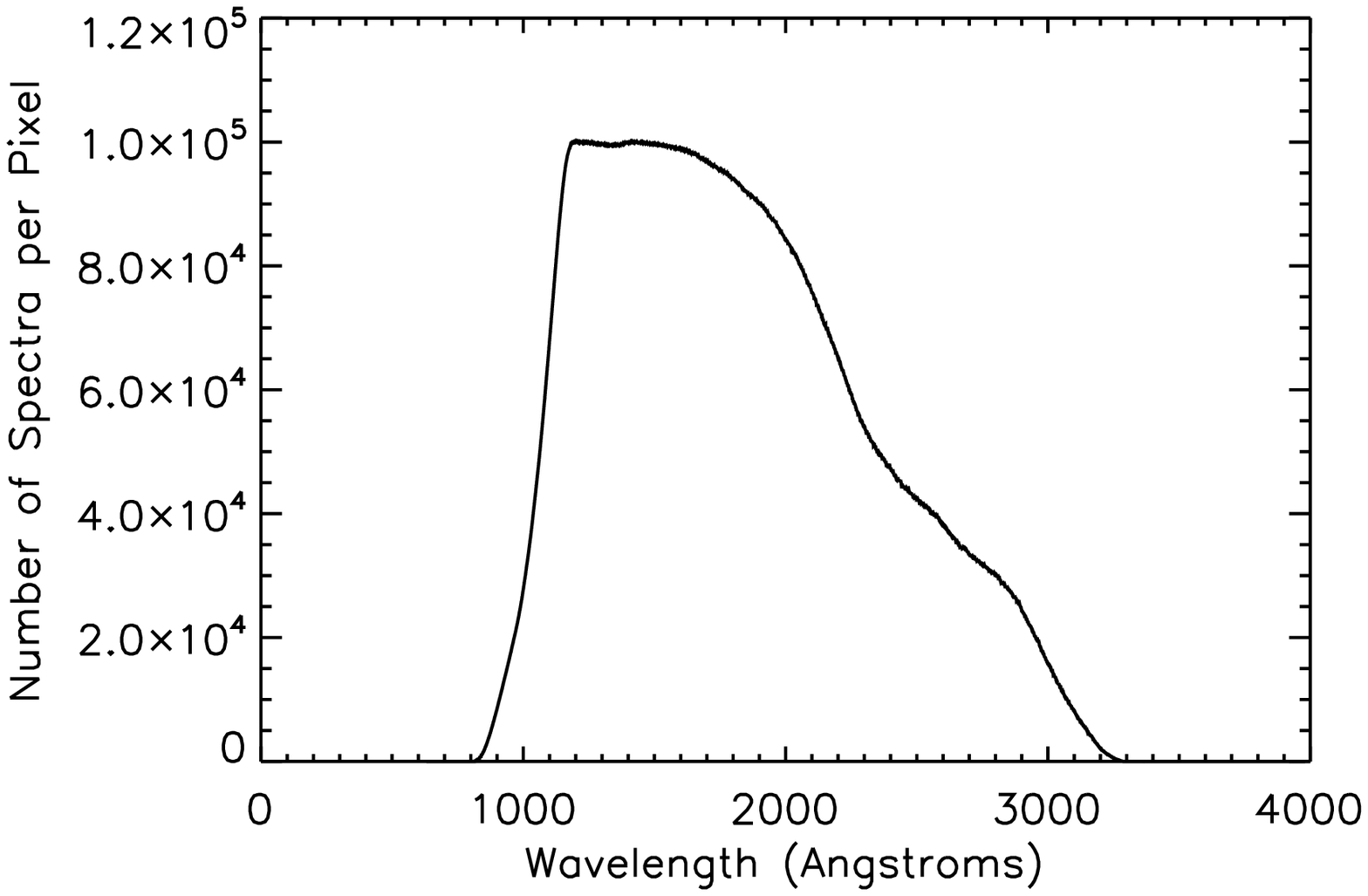}
\caption[ratiostaralpha]{{\bf First panel:}  Redshift distribution of the sample.  {\bf Second panel:}  The number of spectra at each rest wavelength of the composite.}
\label{fig:fig2}
\end{figure}

The first panel of Figure~\ref{fig:fig2} presents the redshift distribution of the quasar sample used to create the composite spectrum.  The sample peaks at $z = 2.3$ where the wavelength range includes coverage to Lyman-$\beta$.  The second panel of Figure~\ref{fig:fig2} shows the number of spectra used for the composite per wavelength pixel.  Comparatively few spectra (approximately 20,000) are used to cover the wavelength range less than 1000 \AA\ or greater than 3000 \AA.  Pixels between 1000 \AA\ and 2000 \AA\ are covered by 90,000 to 100,000 spectra, which is the greatest coverage of any work on quasar composite spectra.

\subsection{Comparison to Other Quasar Composite Spectra}\label{subsec:compare_other}

This work differs significantly from other quasar composite spectrum studies.  Primarily, this work seeks to generate a composite spectrum from a very large sample of quasars covering a relatively narrow range of redshifts and $i$-band absolute magnitudes.  In particular, we use quasars intended for BOSS Lyman-$\alpha$ forest BAO studies.  Other works have focused on different ranges in wavelength coverage, different redshift ranges, and smaller numbers of quasar spectra.  
A comparison of the coverage in luminosity and redshift between this sample
and previous composite spectra is provided in Figure~\ref{fig:fig1}.
The statistics of that comparison are provided in Table 3, along with the median equivalent width (EW) of C~\textsc{IV} of quasar spectra from this sample
taken from column 36 of the quasar catalog.
.
Here we use a narrower interval in $i$-band absolute magnitude than either V01 or S14.  The 68\% redshift range is smaller than V01 and comparable to, but slightly larger than, S14, though this work operates at a much higher redshift range than S14.

Because this composite is generated from a tighter sample of quasars than V01 in both redshift and $i$-band absolute magnitude, this composite will be more representative of quasars falling within its smaller sample selection size.  This allows for more detailed studies of quasars within this region, as there is less dispersion between fundamental quasar properties here than in V01.  S14 generates a composite from a comparably sized redshift range but at much lower redshift.  The $i$-band absolute magnitude range in S14 is also substantially larger. 

\begin{deluxetable*}{lccccccc}
\centering
\tablewidth{0pt}
\tabletypesize{\footnotesize}
\tablecaption{\label{tab:compositetable_all} The respective ranges of redshifts and K-corrected $i$-band absolute magnitudes in recent composites, including this work.  Also included is the median equivalent width of the C~\textsc{IV} emission line of quasar spectra in each sample.}
\tablehead{
\multicolumn{1}{c}{} & \multicolumn{3}{c}{Redshift Distribution} & \multicolumn{3}{c}{$i$-band Abs. Mag Distribution} & \multicolumn{1}{c}{} \\
\hline \\
\colhead{Composite} & \colhead{16\%} & \colhead{Median} & \colhead{84\%} & \colhead{16\%} & \colhead{Median} & \colhead{84\%} & {C~\textsc{IV} EW (\AA)}
}
\startdata
   V01 & 0.679 & 1.472 & 2.298 & -24.47 & -26.38 & -27.75 & 23.8 \\
   S14 & 0.104 & 0.366 & 0.717 & -23.12 & -25.32 & -27.51 & 36.3 \\
   This Work & 2.249 & 2.457 & 2.895 & -24.37 & -25.20 & -26.22 & 37.4 \\
\enddata
\end{deluxetable*}

As V01 is the most direct predecessor of this work, it is instructive to examine the differences in greater detail.  This work differs from V01 in the following ways:

\begin{enumerate}

\item This work is more restrictive in classifying an object as a quasar, requiring the object to be a point source.  In V01, an object was declared to be a quasar if it was found to be extra-galactic (through redshift determination by visual inspection) with at least one broad emission line and had a non-stellar dominated continuum.  This selection includes Seyfert galaxies.  V01 also restricted their quasar sample to objects with C~\textsc{IV}, Mg~\textsc{II}, and/or Balmer line FWHM greater than about 500 km~s$^{-1}$.  In this work, quasar identification is made by visual inspection described in \citet{paris12a} and uses only objects at $z > 2.1$.

\item This work uses quasars covering a narrower range in redshift and $i$-band absolute magnitude.  The V01 sample drew quasars from the SDSS survey which used quasars selected from regions in color-color-color-color space ($u-g$, $g-r$, $r-i$, and $i-z$) \citep{richards02a}.  BOSS employs a rigorous quasar target selection scheme which has been shown to be $\sim 40-50\% $ complete to $g < 22$ \citep{ross13a}.

\item More quasars are included in this work than in V01.  V01 used 2204 spectra, whereas this work uses 102,150 spectra, allowing this work to generate a composite at a much higher S/N.

\item The redshifts used in this work are more accurate.  V01 redshifts came from two sources.  For objects with a visible [O~\textsc{III}] line, that line was used for redshift.  The [O~\textsc{III}] line is not visible in higher redshift quasars, above $z=0.7$ in SDSS spectra.  A composite of 373 quasars with [O~\textsc{III}] redshifts was generated and used as a template to determine the redshifts of the higher redshift quasars.  In BOSS a variety of a methods largely tested against visual inspection were used \citep{paris12a}.  For this work, we use PCA redshifts found from the catalog described in \citet{paris12a} which is calibrated against Mg~\textsc{II}, an emission line not known to display velocity shifts with respect to the systematic redshift.

\end{enumerate}

\section{The Composite}\label{sec:composite}

\subsection{Spectrophotometric Correction}\label{subsec:specphot_corr_intro}

\begin{figure*}[h]
\centering
\includegraphics[width=1.0\textwidth]{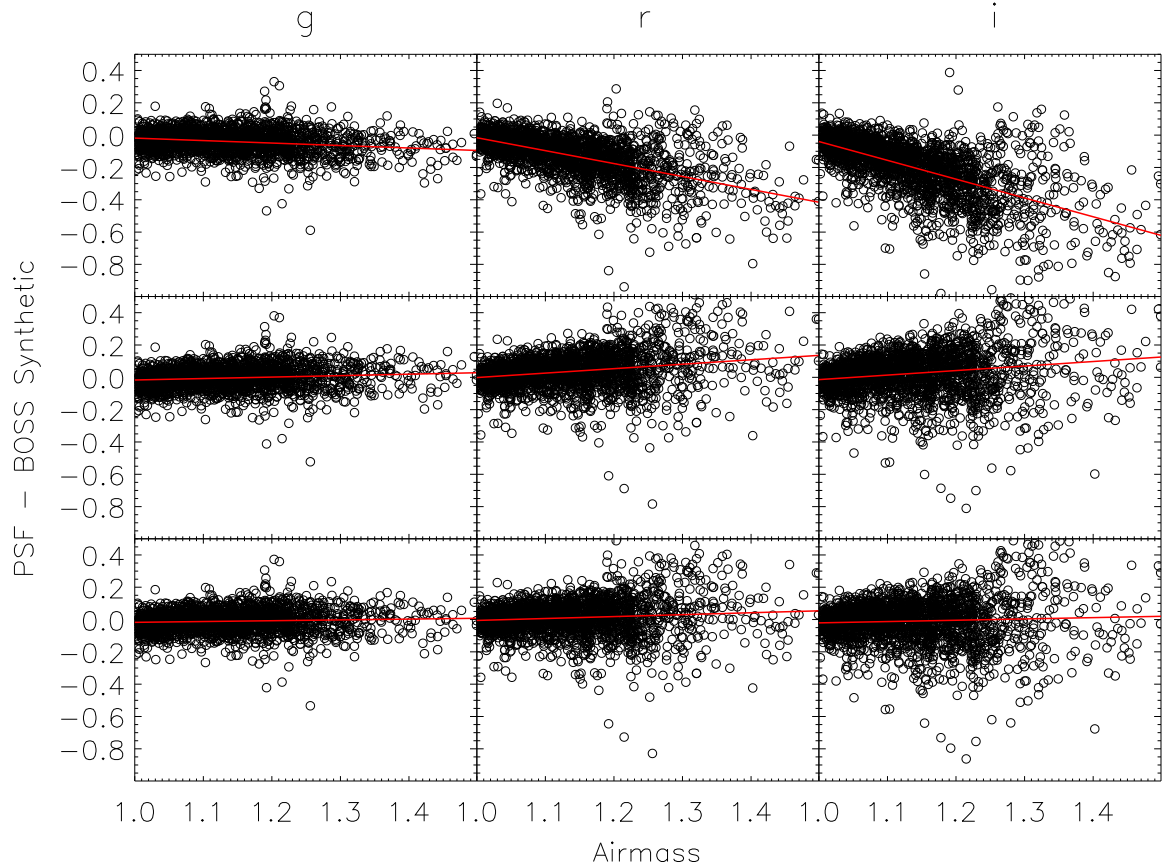}
\caption[ic]{The top row of panels shows the difference in magnitudes between SDSS imaging data and BOSS synthetic photometry in filters $g$, $r$, and $i$.  The middle row shows the same after applying the initial correction found in Appendix~\ref{subsec:correct_specphot_error} and steps 2, 3, and 4 of Section~\ref{subsec:specphot_corr_intro}.  The bottom row shows the magnitude difference after the final correction as discussed in Appendix~\ref{subsubsec:finalcorrection}.  Fits to these data are presented in Table~\ref{tab:correction_table} and shown in the panels as red lines.}
\label{fig:fig3}
\end{figure*}

The method used in the BOSS data processing pipeline to generate standard-star calibrations does not take into account the differences in ADR offsets between objects in the red focal plane and objects in the blue focal plane.  The difference in the placement of holes for fibers assigned to quasars with relation to standard star fibers discussed in Section~\ref{subsubsec:bossspec} necessitated a spectrophotometric correction that increases with airmass.  This correction is discussed in detail in Appendix~\ref{subsec:specphot_err}.  Briefly, the procedure used to generate the spectrophotometric correction is:
\begin{enumerate}
\item The spectrophotometric errors are quantified through an examination of stellar contaminants.  Stellar contaminants observed in both BOSS and SDSS were identified to allow a direct comparison between the two data reductions.  Two independent analyses of stellar contaminants are made:  
\begin{itemize}
\item First, we find the ratio of the spectrum of each stellar contaminant in BOSS relative to the spectrum obtained in SDSS.  Because the calibration was performed uniformly across all targets in SDSS, these spectral ratios constrain the broadband fluxing errors introduced by the BOSS fiber offsets for the blue focal plane.  This ratio is close to or slightly greater than one from 3800~\AA\ to 4200 \AA\ and declines to approximately 0.82 at 8500 \AA, indicating a significant loss of red photons.  See Appendix~\ref{subsubsec:spectral_comparison} and Figure~\ref{fig:fig13} for more information.
\item Second, synthetic magnitudes of all stellar contaminants in BOSS are compared to photometric imaging data in $g$, $r$, and $i$ from SDSS.  The difference between these is averaged by plate and analyzed with respect to airmass.  At low airmass, the offsets are close to zero.  As the airmass increases, the SDSS imaging flux becomes brighter than the synthetic flux from BOSS spectra.  At an airmass greater than 1.4, offsets in $g$ are approximately -0.05 magnitudes, offsets in $r$ are larger than -0.3 magnitudes, and in $i$ the offsets are approximately -0.5 magnitudes.  Thus, the offset is most significant in the longer wavelengths of the spectra and at higher airmass.  See Appendix~\ref{subsubsec:photometric_comparison} and the first row of panels in Figure~\ref{fig:fig3} for more information.
\end{itemize}
\item Some BOSS plates were specially designed with a second set of standard stars:  One set drilled in the usual manner (``red standards'') and another set drilled in the blue focal plane (``blue standards'', ANCILLARY\_TARGET2 = 20).  A new flux calibration is performed using the blue focal plane standard stars (``blue reduction'') instead of the red focal plane standard stars (``red reduction'').  This analysis is described in more detail in Appendix~\ref{subsec:correct_specphot_error}.
\item To perform the flux calibration, the blue standards' spectra reduced using the blue reduction are compared to the blue standards' spectra reduced using the red reduction.  This comparison is analyzed against airmass of the observation and wavelength.  This comparison allows us to create an initial spectrophotometric correction with respect to wavelength and airmass.  See Appendix~\ref{subsubsec:param_corr} for more details.
\item This spectrophotometric correction is applied to the BOSS blue focal plane spectra and reanalyzed as in step one.  This analysis shows the initial correction overcorrects the fluxing errors, as can be seen in the second row of panels in Figure~\ref{fig:fig3}.  This effect is explained further in Appendix~\ref{subsubsec:initial_test}.
\item The wavelength and airmass dependence of the correction is assumed to be correct, but the amplitude of the correction is treated as a free parameter.  When an arbitrary scale factor of 0.83 is applied, the color bias (the difference in the synthetic photometric color of objects in BOSS and the photometric color of objects in SDSS) reaches a minimum.  The reduced spectrophotometric correction at that point is used as the final spectrophotometric correction.  This correction is described in Appendix~\ref{subsubsec:color_bias}.  The photometric and spectroscopic tests are run again and the results of these tests to this correction are shown in Appendix~\ref{subsubsec:finalcorrection} and Figures~\ref{fig:fig3} and \ref{fig:fig16}, along with the final correction.
\item The spectrophotometric correction (a function of wavelength and airmass) is applied to all quasar spectra used in the composite.  
\end{enumerate}

The results of the tests to the photometric offsets between BOSS and SDSS data are presented in Figure~\ref{fig:fig3}.  The goal of the correction is to produce flux-calibrated spectra that on average have synthetic photometry in $g$, $r$, and $i$ consistent with imaging photometry.  The first row of figures shows the difference between the SDSS photometric magnitudes and the synthetic photometry from the unmodified spectra in BOSS.  There is a significant trend in this difference with airmass.  The second row lists the same data after the BOSS spectra have been corrected using the initial correction, as discussed in Appendix~\ref{subsubsec:param_corr}.  The difference between SDSS photometry and BOSS synthetic photometry is still dependent on airmass, but not as much as before the initial correction.  The slope of the difference with airmass has also changed sign, indicating the initial correction overcorrected the problem.  The third row shows the difference between SDSS photometry and BOSS synthetic photometry as a function of airmass after the final correction. 

\begin{figure}[h]
\centering
\includegraphics[width=0.5\textwidth]{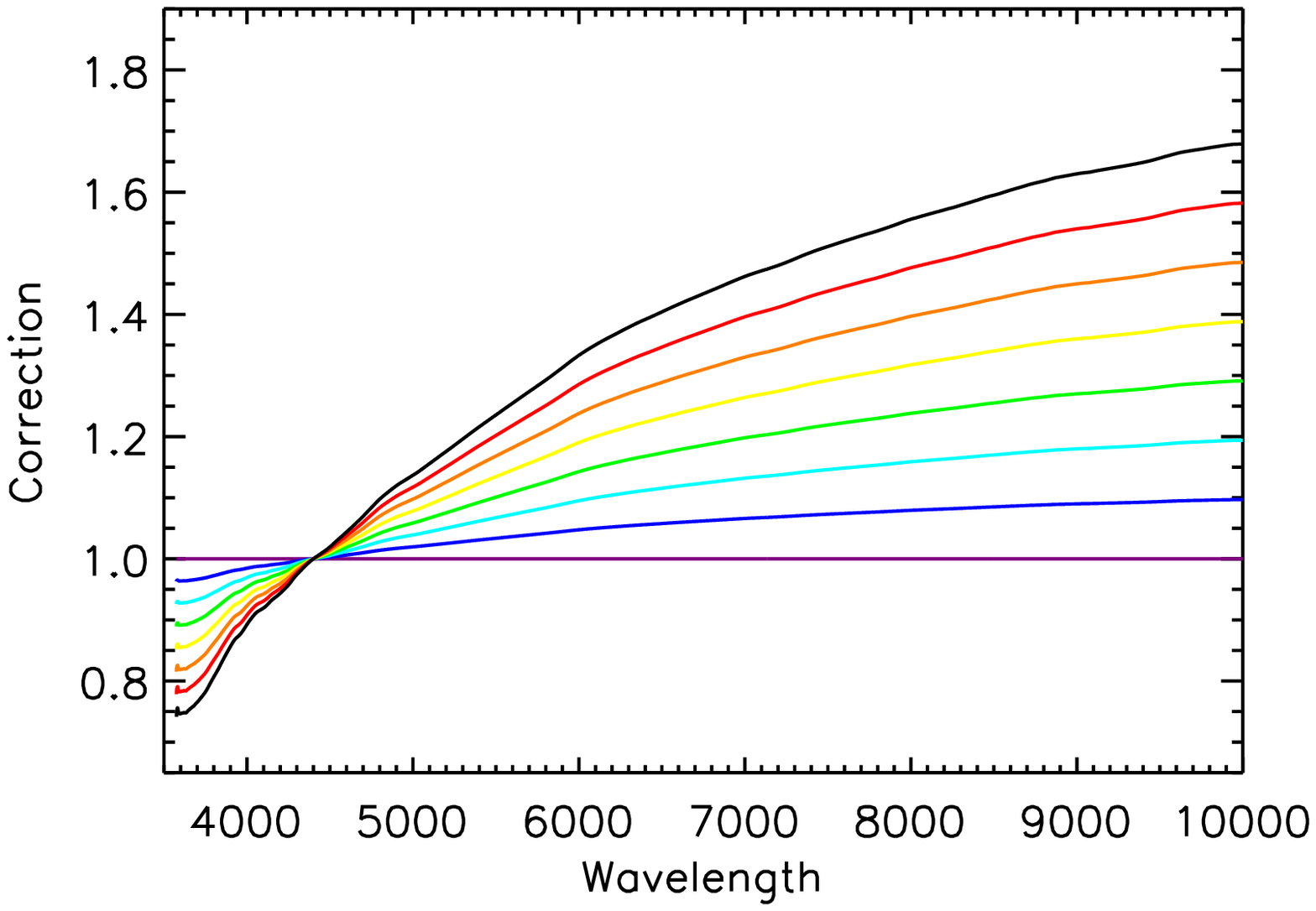}
\caption[fanplot2]{Example corrections for spectra at different airmasses using the final spectrophotometric correction described in detail in Appendix~\ref{subsubsec:finalcorrection}.  Airmasses of bottom (purple) to top (black) lines: 1.00, 1.05, 1.10, 1.15, 1.20, 1.25, 1.30, 1.35.}
\label{fig:fig4}
\end{figure}

The resulting spectrophotometric correction is displayed in Figure~\ref{fig:fig4}.  This plot presents example corrections as a function of airmass. For a given airmass, the flux of a spectrum at a particular wavelength is multiplied by the correction at that airmass and wavelength.  Despite correcting the flux of BOSS objects in the blue focal plane such that the SDSS imaging data and the BOSS synthetic photometry agree, the flux calibration is inconsistent between BOSS spectra and SDSS spectra.  See Appendix~\ref{sec:boss_sdss_ratio} for more information.  We maintain all spectra on the BOSS spectral system and do not correct them to the SDSS spectral system.

\subsection{Constructing the Composite}\label{subsec:construct_composite}

In this section, we construct the quasar composite spectrum.  Quasar spectra are dereddened using the extinction model of \citet{fitzpatrick99} and the extinction map of \citet{schlegel98a}.  A mask from \citet{delubac15a} is applied to each quasar to mask skylines.  Each quasar spectrum is shifted to the restframe and rebinned by log-linear interpolation to a logarithmic binning scheme with the same pixel width as all BOSS spectra.  The quasars in this sample cover a wide redshift range where there is significant evolution in optical depth (\tlya) due to Lyman-$\alpha$ absorption; however the exact form of this evolution is unknown and remains a focus of current research \citep{becker13a,faucher08a}.  We therefore prepare the sample of quasar spectra both with and without a correction for optical depth.  

For the composite spectrum that includes a correction for \tlya, we use
the parameterization
\tlya = $\tau_0 * \sum_{i=2}^{\infty} i^{-1.6} (1+z)^{\beta}$
for each order of the Lyman transition at each pixel in the spectrum.
In this formalism, $i$ represents the final
energy state of the transition and $z$ is the redshift of that transition at the observed wavelength
of the pixel.  The sum is performed over all transitions where the wavelength of the transition
is longer than the wavelength at that pixel in the rest frame of the quasar.
We assume values $\tau_0 = 0.0067$ and $\beta = 3.0$.  We assume a
correction beyond the Lyman limit
that scales as $0.394(1+z)^{2} * 0.937 log_{10}[(1+z)\lambda_{\infty}]$,
where $\lambda_{\infty}$ is the wavelength of the Lyman limit.
Each quasar spectrum is then rescaled at restframe wavelengths $\lambda< 1216$ \AA\
according to the \tlya\ estimates computed for the quasar redshift.

\begin{figure*}[h]
\centering
\includegraphics[width=0.9\textwidth]{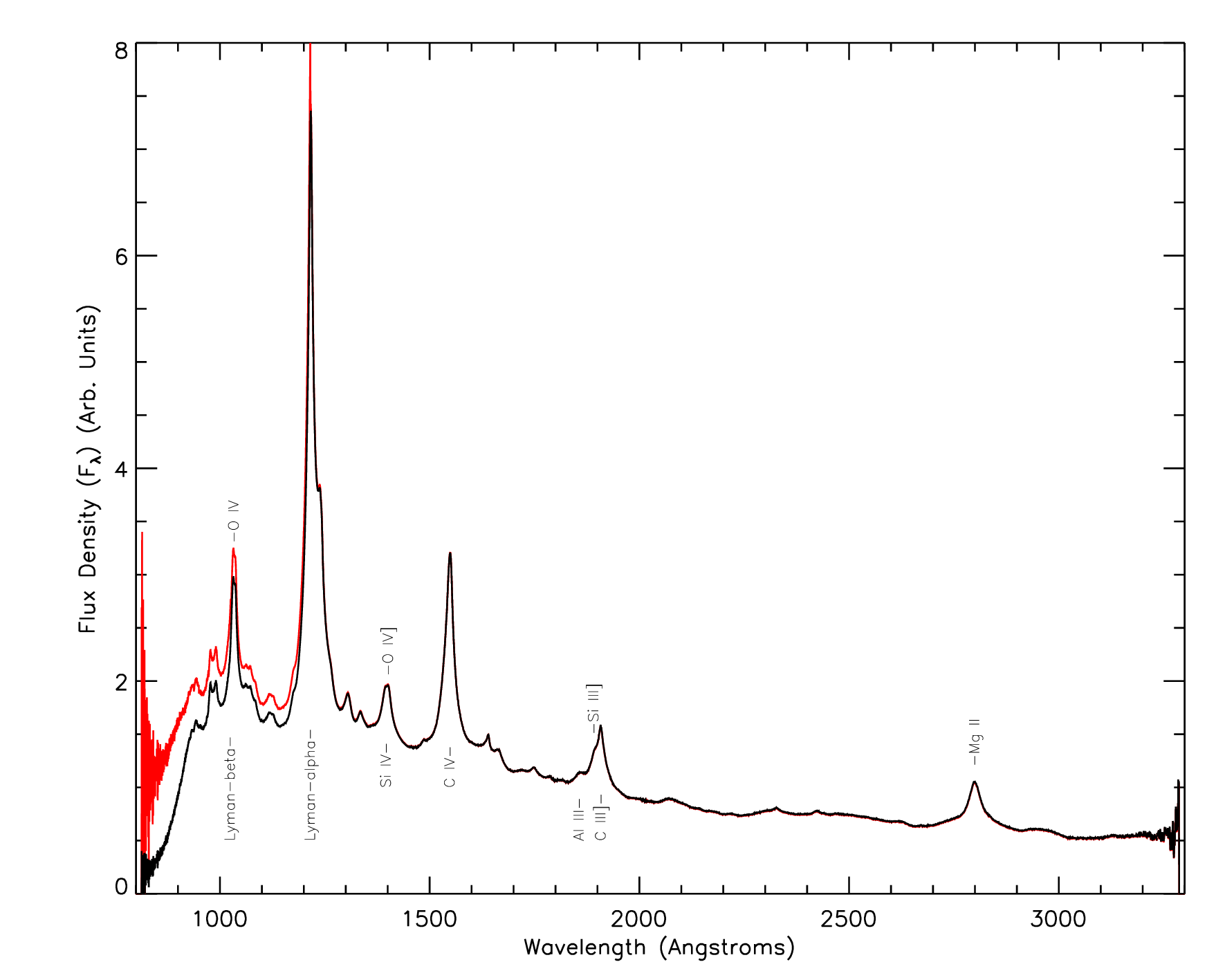}
\includegraphics[width=0.9\textwidth]{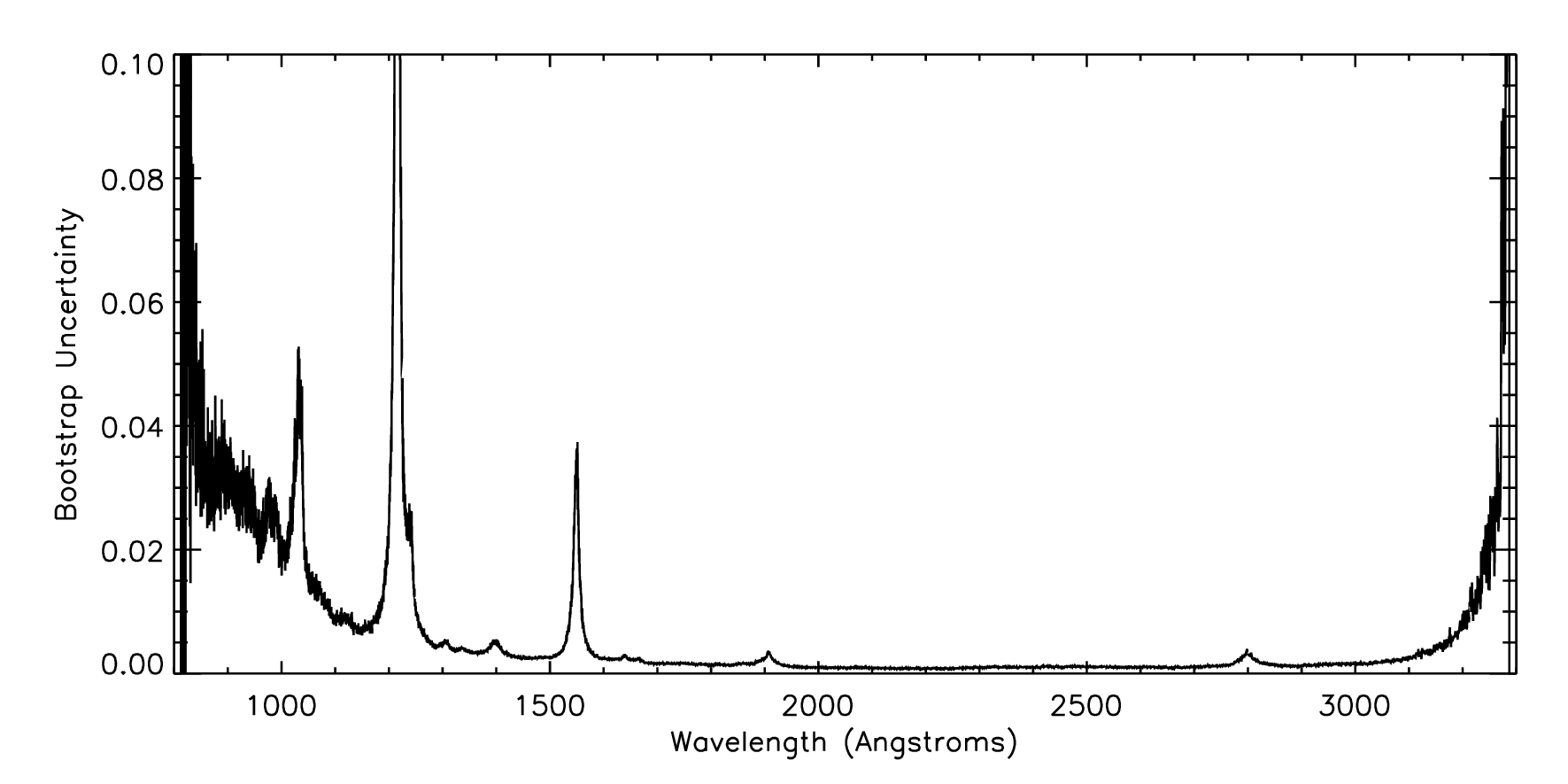}
\caption[comp1]{{\bf First panel:} The initial composite generated by this work.  The composite without optical depth correction is shown in black, and the composite with optical depth correction is indicated in red.  This composite spectrum has not yet been warped to match the median specral index of the quasar sample, as performed in Section~\ref{subsec:composite_spectral_index} and shown in Figure~\ref{fig:fig8}.  {\bf Second panel:} The fractional error of the median flux measurement in each pixel, generated by the bootstrap resampling method.  The creation of the composite and the error array are discussed in Section~\ref{subsec:construct_composite}.}
\label{fig:fig5}
\end{figure*}

The method employed to create the quasar composite spectrum in this work is based on the method used by V01 and \citet{shull12a}.  In both cases, spectra were first ordered by redshift and used to create an intermediate composite each time a new quasar was added.  The overlapping region between the composite and the subsequent spectrum was found, and the subsequent spectrum was scaled such that it had the same mean flux as the composite in the overlapping region.  This work uses approximately 50 times as many quasars as V01, thus following the same procedure becomes computationally expensive.  Instead, we bin quasars in groups of 500 to reduce the computation time.  The quasar composite spectrum is constructed using the following steps:
\begin{enumerate}
\item Spectra are ordered by increasing redshift, from lowest to highest.
\item An ``overlap region'' is chosen for the first 500 spectra.  This overlap region covers $\lambda_{\rm blue}$ to $\lambda_{\rm red}$, where $\lambda_{\rm blue}$ is the bluest pixel with measured flux in the 1st spectrum and $\lambda_{\rm red}$ is the reddest pixel with measured flux in the 500th spectrum.
\item The first 500 spectra are individually multiplied by a scaling factor chosen such that each spectrum has a weighted mean of one over the wavelength range $\lambda_{\rm blue}$ to $\lambda_{\rm red}$.  
\item For the next 500 spectra a new overlap region (again defined as $\lambda_{\rm blue}$ to $\lambda_{\rm red}$) is chosen.  In this case, $\lambda_{\rm blue}$ is the bluest pixel with data in the 1st spectrum, and $\lambda_{\rm red}$ is the reddest pixel with data in the 1000th spectrum.  The calculated weighted mean of the first 500 spectra in the overlap region is computed and used as a scale factor by which to normalize each of the following 500 spectra.
\item Each spectrum from the 501st to the 1000th is multiplied by a scaling factor such that the weighted mean of the flux in the overlap region of that spectrum is made equal to the overlap target from the previous step.
\item Every subsequent spectrum is normalized in batches of 500 following the above procedure.
\item After all the spectra are ordered and scaled, the clipped median value at each wavelength is recorded as the value of the composite at that wavelength.  
\end{enumerate}
The resulting composite spectrum for the 102,150 quasars is shown in Figure~\ref{fig:fig5}.

A composite created with the median flux value at each wavelength preserves the line ratios, while a composite created with the geometric mean flux value at each wavelength preserves the mean spectral index of the quasar continuum (see V01).  These spectra individually have a low signal-to-noise ratio, and sometimes the recorded flux is less than zero.  Thus, a standard geometric mean cannot be used in this case.  Instead, we use the median composite spectrum to preserve the line ratios, then distort the composite to match the appropriate spectral index as described in Section~\ref{sec:spectral_index}.  

\begin{figure}[h]
\centering
\includegraphics[width=.5\textwidth]{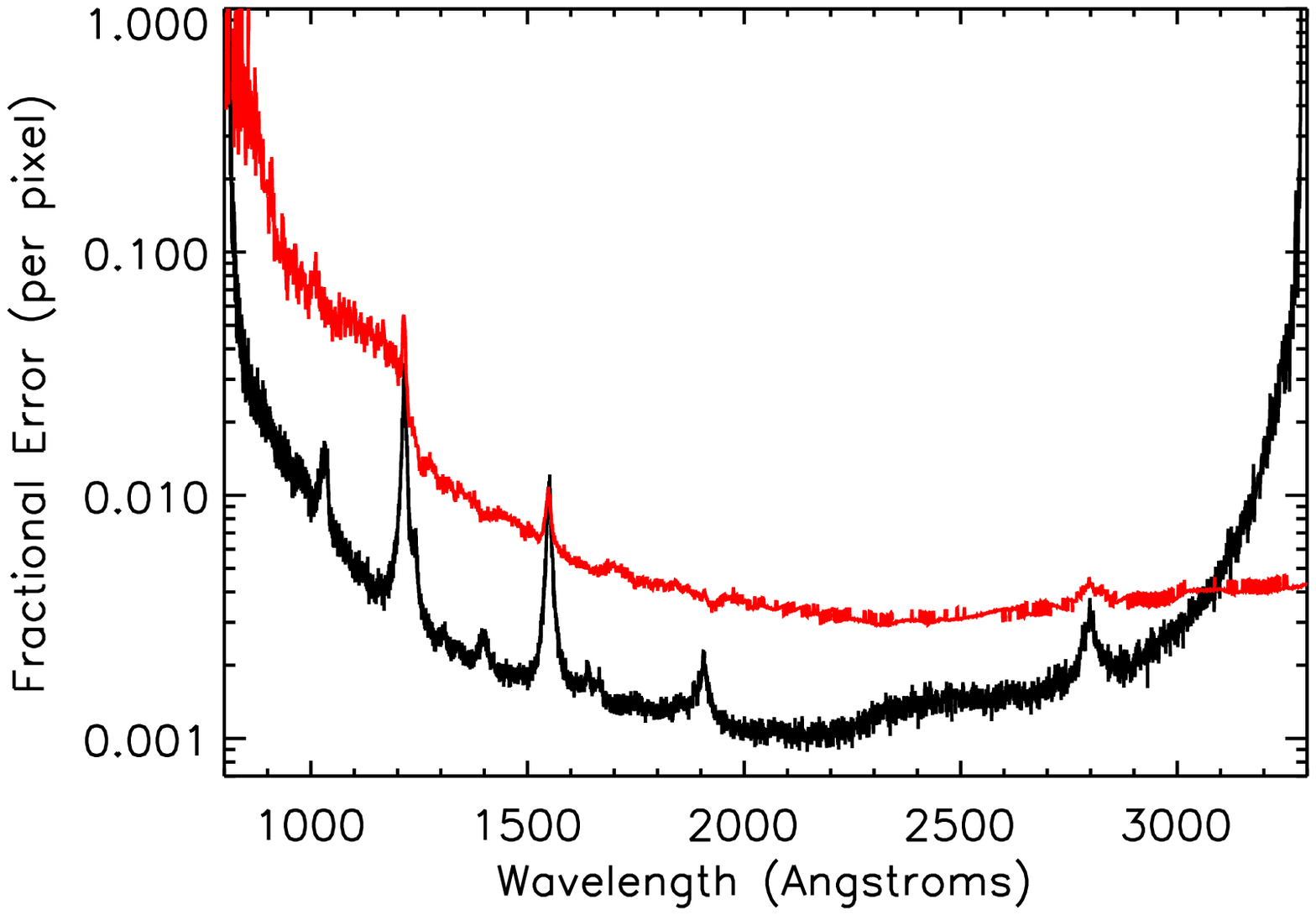}
\includegraphics[width=.5\textwidth]{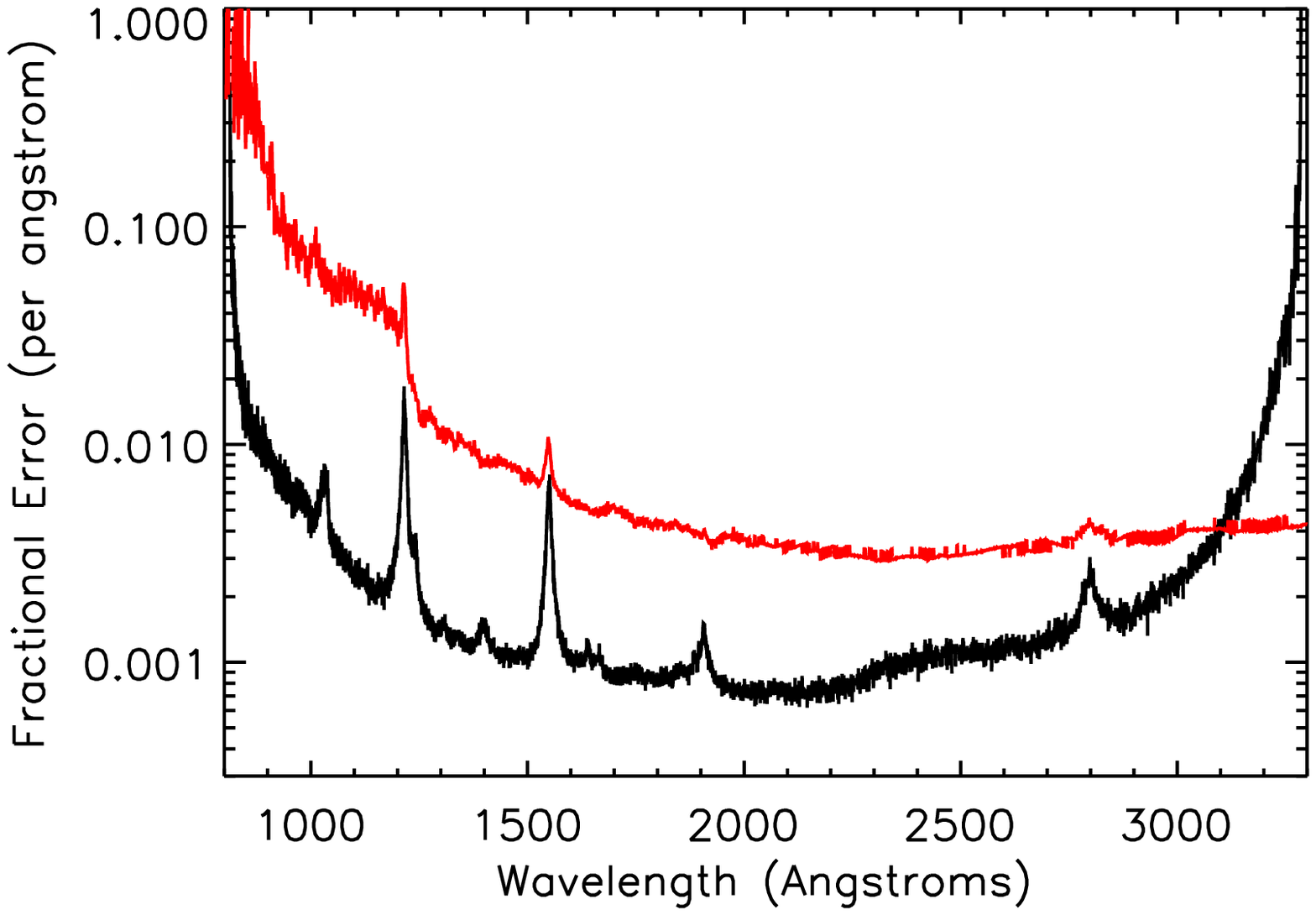}
\caption{{\bf First Panel:} Comparison between fractional errors in V01 (red) and this work (black).  The errors in this work are as small as 0.001 in the continuum, and are smaller than V01 at almost all wavelengths, the exception being at greater than 3160 \AA, which is at the red edge of the wavelength range of the BOSS Lyman-$\alpha$ forest sample.  {\bf Second Panel:} Fractional error scaled by the square root of the relative pixel scales.}
\label{fig:fig6}
\end{figure}

Errors are computed through a bootstrap resampling examination of the median.  A total of 200 new median composites are made with the same number of quasar spectra as the main quasar composite, but the quasar sample is taken randomly from the main sample, one at a time, with the possibility of repeats.  This is the most computationally expensive aspect of this analysis.  The standard deviation of the 200 composite spectra at each wavelength is recorded as the error on the normalized flux estimate at that wavelength.  Errors are presented in Figure~\ref{fig:fig5}, and compared to the errors from V01 in Figure~\ref{fig:fig6}.  This work has errors approximately half the fractional error compared to V01 in the region from 2000 \AA\ to 3000 \AA.  In addition, this work samples the quasar composite at finer wavelength intervals, especially at short wavelengths.  Thus, the signal-to-noise ratio per angstrom (rather than the signal-to-noise ratio per pixel) is much higher in this work.  At wavelengths blueward of Lyman-$\alpha$, this work has fractional errors approximately ten times smaller than V01.  The mean fractional error between 2000 \AA\ and 2300 \AA\ is 0.11\%.

\begin{figure}[h]
\centering
\includegraphics[width=.5\textwidth]{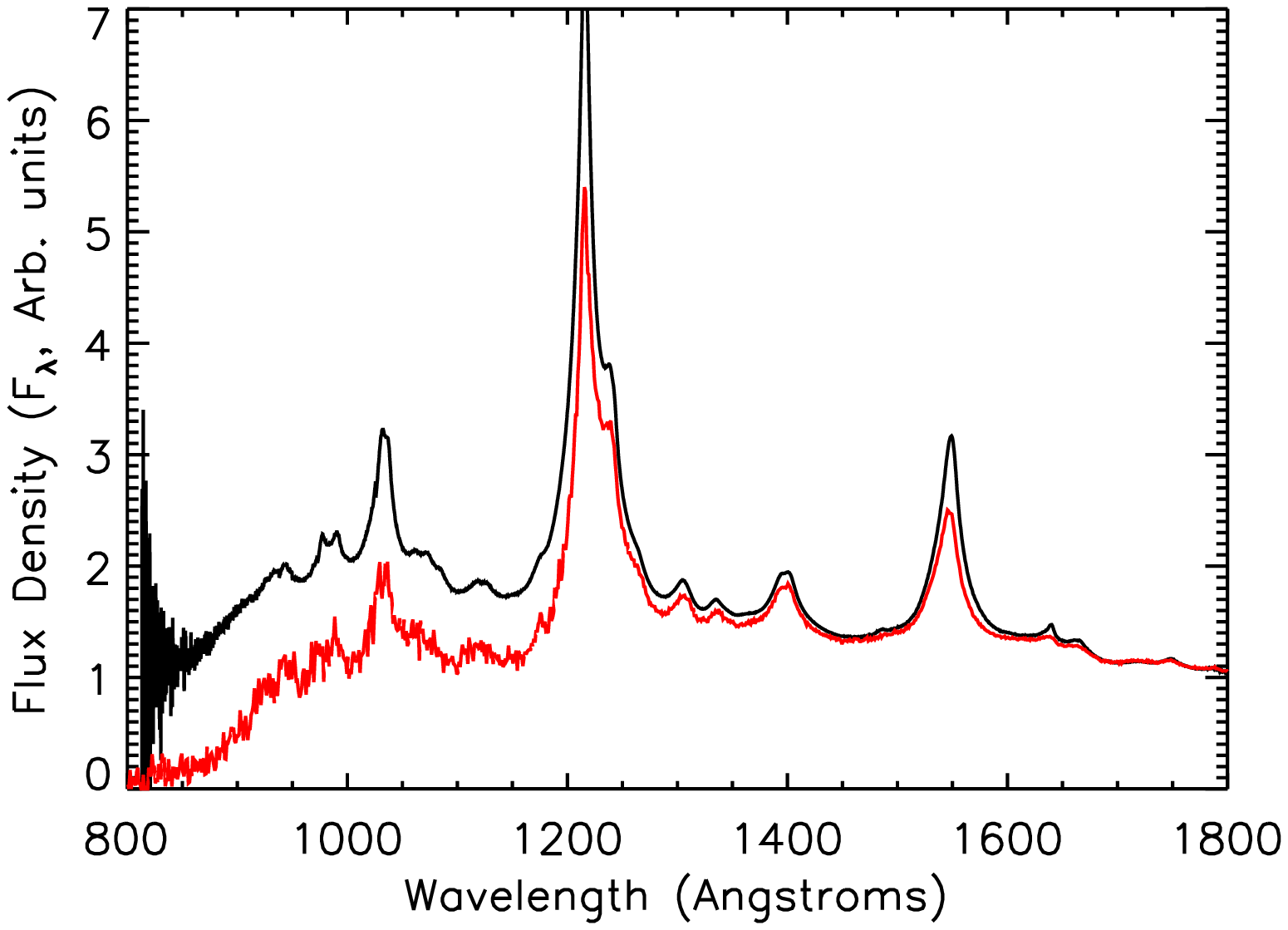}
\includegraphics[width=.5\textwidth]{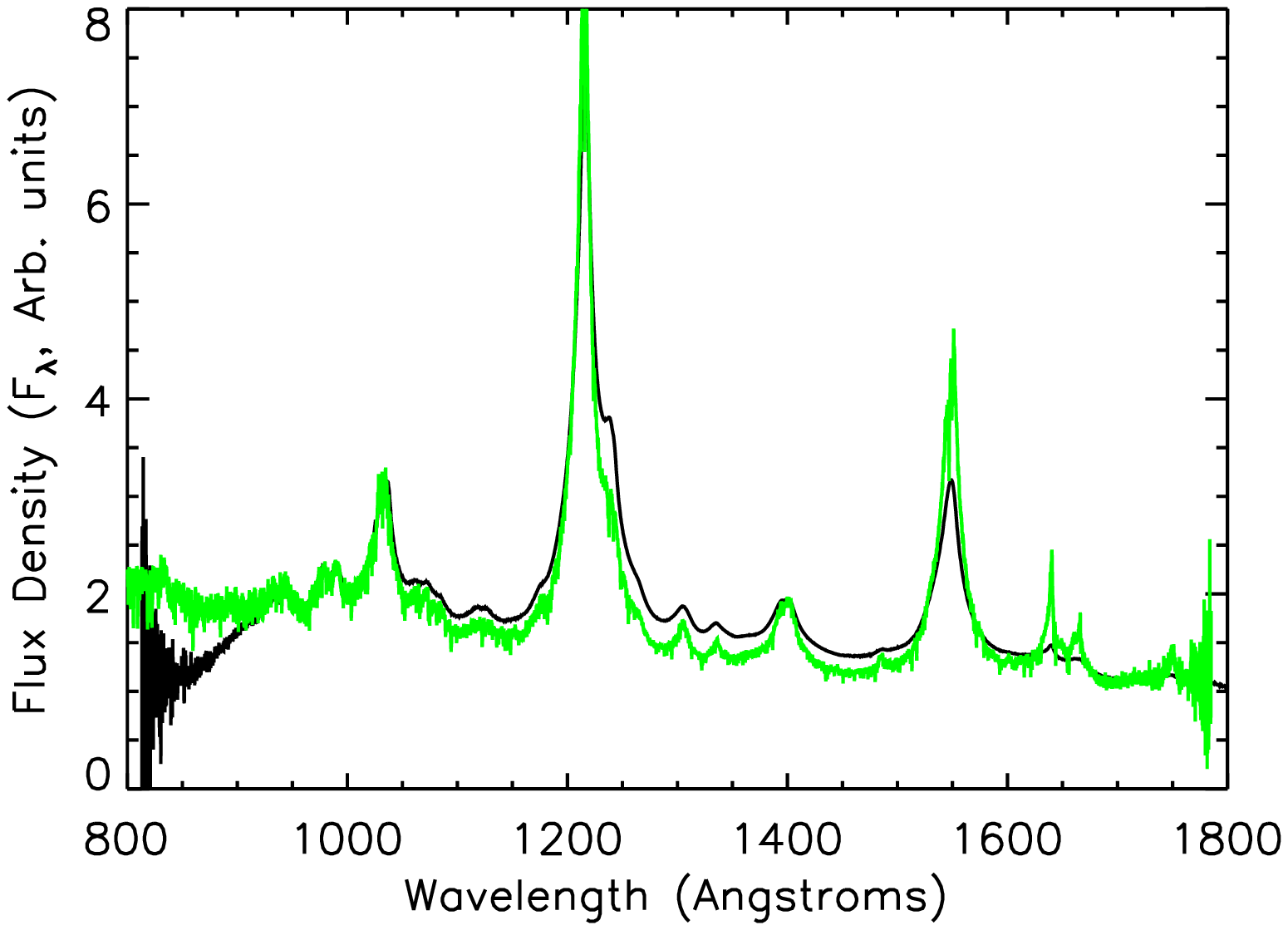}
\caption{{\bf First Panel:} Comparison between the quasar composite spectrum generated in this work without optical depth correction (in black) and V01 (in red).  {\bf Second Panel:} Comparison between the quasar composite spectrum generated in this work without optical depth correction (in black) and S14 (in green).  The quasar composite spectrum generated in this work has been warped such that the spectral index of this composite spectrum is the same as the median spectral index of the quasar sample, as explained in Section~\ref{subsec:composite_spectral_index}.  Arbitrary scaling factors have been applied to V01 and S14 for illustrative purposes.}
\label{fig:fig7}
\end{figure}

Figure~\ref{fig:fig7} shows the composite spectrum generated in this work along with the composites generated by V01 and S14.  Most notably, S14 is not constrained by atmospheric absorption of UV light, as data taken for S14 was from the Cosmic Origins Spectrograph \citep{green12a} on the Hubble Space Telescope.  Thus, S14 can explore the Far and Extreme UV, but their work excludes longer wavelengths as the spectrometer is specifically designed for the UV.  S14 recovers flux lost due to \tlya\ by modeling absorption features in individual quasars and restoring the flux based on the estimated H~\textsc{I} column density.

These three composites target different ranges of absolute $i$-band magnitude and redshift, as shown in Table~\ref{tab:compositetable_all}.  This is the primary contributor to the difference between these composites.  Some emission lines (C~\textsc{IV}, Mg~\textsc{II}, C~\textsc{III}], and others) have 20-60\% greater equivalent widths in this work than in V01.  In particular, the reconstructed C~\textsc{IV}] and Mg~\textsc{II} emission lines have equivalent widths about 60\% larger in this work than in V01; C~\textsc{III} is about 20\% larger compared to V01.  These differences originate from the Baldwin Effect \citep{baldwin77a}, as the ranges of i-band absolute magnitude in each composite spectrum differ.  The continuum shape is consistent to the several percent level.

\section{Spectral Index}\label{sec:spectral_index}
For this composite to be useful for studies involving the quasar broad band continuum, we further modify the composite spectrum to have a spectral index consistent with the median spectral index of the full sample of quasars.  The spectral index is characterized by fitting a powerlaw to the composite spectrum and to each individual quasar.  We confirm the choice of spectral index for the final quasar composite by comparing the spectrum to the photometric colors of the quasar sample as a function of redshift.

\subsection{Composite and Sample Spectral Index}\label{subsec:composite_spectral_index}

We use the same rest frame wavelength range for the composite spectrum and for each quasar to determine spectral indices.  The wavelength range was chosen after a visual inspection of the composite and of a subsample of individual spectra in the sample.  The range of wavelengths we chose to model the quasar continuum is 1440 \AA\ to 1480 \AA\ and 2160 \AA\ to 2230 \AA.  These two intervals are sufficiently separated from O~\textsc{IV}] to avoid contamination from emission lines even in the case of quasars with very broad lines.  To determine the spectral index, a simple power law is fit to the flux density and errors in the above range.  The spectral index of the composite itself is $\alpha_{\lambda} = -1.4854 \pm 0.0004$.  

\begin{figure*}[h]
\centering
\includegraphics[width=1.0\textwidth]{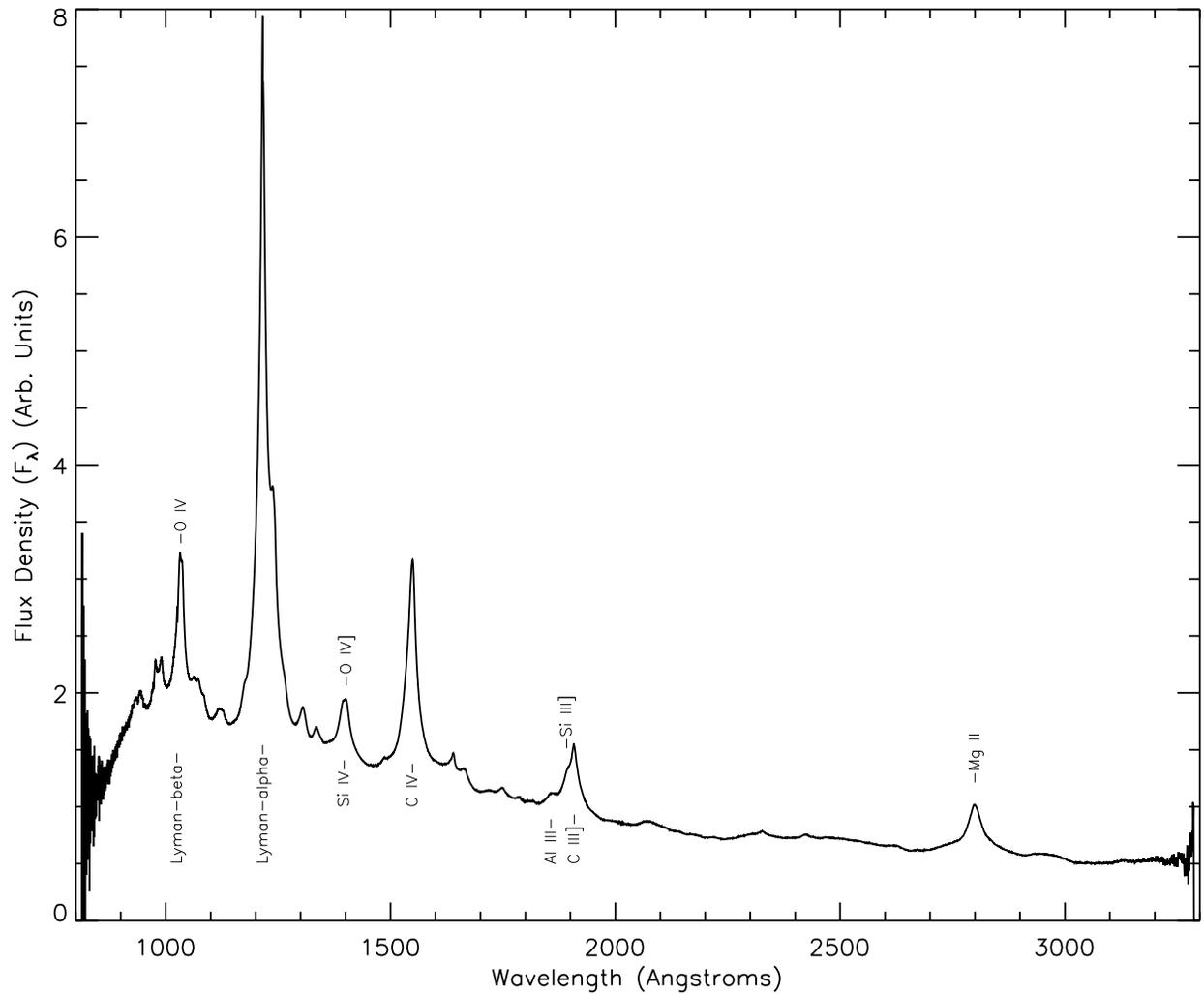}
\caption{Optical depth corrected quasar composite spectrum after having been warped to match the median spectral index of the quasar sample as described in Section~\ref{subsec:composite_spectral_index}.  This composite spectrum represents the final product of this work and is available for download at a publicly accessible website\footnote{http://data.sdss3.org/sas/dr12/boss/qso/composite/}.}
\label{fig:fig8}
\end{figure*}

The spectral index of each quasar is measured using the same wavelength interval as above.  The median spectral index of the 102,150 quasars in this sample is $\alpha_{\lambda} = -1.4613 \pm 0.0017$.  The error on the median spectral index was found by resampling the quasar sample 200 times and finding the median spectral index each time.  The error on the median spectral index is the standard deviation of the 200 measurements of the median spectral index.  We then warp the composite spectrum using the simple relation:
\begin{equation}
f(\lambda)_{final} = f(\lambda)_{\rm initial}\lambda^{\delta\alpha}
\end{equation}
where $\delta\alpha$ is the difference between the spectral index of the quasar sample and the spectral index of the composite.  This results in a new composite which is displayed in Figure~\ref{fig:fig8}.  
Because the composite spectrum had a spectral index so close to that of the median of the parent sample,
the magnitude of this spectral distortion is very small.  The ratio of flux at 800 \AA\ relative to 3300 \AA\
decreases by only 3.47\% after warping the composite spectrum.

\subsection{Quasar Sample Median Spectral Index Binned by Redshift}\label{subsec:mediansamplespectralindex}

We next characterize the distribution of intrinsic quasar spectral indices as a function of redshift.  The spectral index for individual quasar spectra were calculated as above and binned by quasar redshift in increments of $\delta z = 0.01$.  The true variation in quasar spectral indices contributes only a fraction of the observed RMS dispersion in the population.  We assume the three contributions add in quadrature as follows:
\begin{equation}
\sigma_{\rm total}^2 = \sigma_{\rm stat}^2+\sigma_{\rm sys}^2+\sigma_{\rm intrinsic}^2
\end{equation}
where $\sigma_{total}$ is the observed standard deviation in the spectral indices in a given bin, $\sigma_{\rm stat}$ is the median statistical error of the fit, $\sigma_{\rm sys}$ is the systematic error remaining after the spectrophotometric correction, and $\sigma_{\rm intrinsic}$ is the actual RMS dispersion of the spectral indices in the quasar population for that redshift bin.  

The systematic error ($\sigma_{\rm sys}$) of the spectrophotometric correction is found in a study of the ratio between BOSS and SDSS spectra of the stellar contaminants.  The selection algorithm to choose appropriate stars for the comparison is described in Appendix~\ref{subsubsec:spectral_comparison}.  These ratios between BOSS and SDSS stars are called ``star ratios'' and should be identically equal to one at all wavelengths.  The spectrophotometric correction used on each quasar spectrum in the composite is applied to the BOSS spectra before computing the ratio. 

A spectral index is fit to each star ratio spectrum over the wavelength ranges that would correspond to a quasar at $2.1 \leq z \leq 2.64$ in bins of 0.01.  The fit continues only to $z = 2.64$ because the star ratio spectra are truncated at the maximum SDSS wavelength of approximately 8500 \AA.  The weighted median spectral index at each redshift increment is also determined.  The systematic error ($\sigma_{\rm sys}$) is found with the following formula:
\begin{equation}
\chi_{\rm red}^2 = \frac{1}{N-1} \sum\limits_{i} \frac{(\alpha_{i}-\alpha_{\rm median})^2}{\sigma_{\rm fit,i}^2+\sigma_{\rm sys}^2}
\end{equation}
where $\chi_{\rm red}^2$ is the reduced $\chi^2$, $N$ is the number of quasars in each bin, $\alpha_{i}$ is each spectral index of the $i^{th}$ quasar, $\alpha_{\rm median}$ is the median spectral index in the given redshift bin, $\sigma_{\rm fit,i}$ is the uncertainty in the fit to the spectral index of the $i^{th}$ quasar, and $\sigma_{\rm sys}$ is the systematic error.  In each redshift bin, the value of $\sigma_{\rm sys}$ is computed such that $\chi_{\rm red}^2 = 1$.  We repeat the process until successive iterations converge on the same values of $\alpha_{\rm median}$ and $\sigma_{\rm sys}$ to $< 1\% $ tolerance.  The final value of $\sigma_{\rm sys}$ is taken to be the systematic error of the spectrophotometric correction in a given redshift bin.  The parameter $\sigma_{\rm intrinsic}$ is found to range between 0.2 to 0.5, generally declining at higher redshifts.

\begin{figure}[h]
\centering
\includegraphics[width=.5\textwidth]{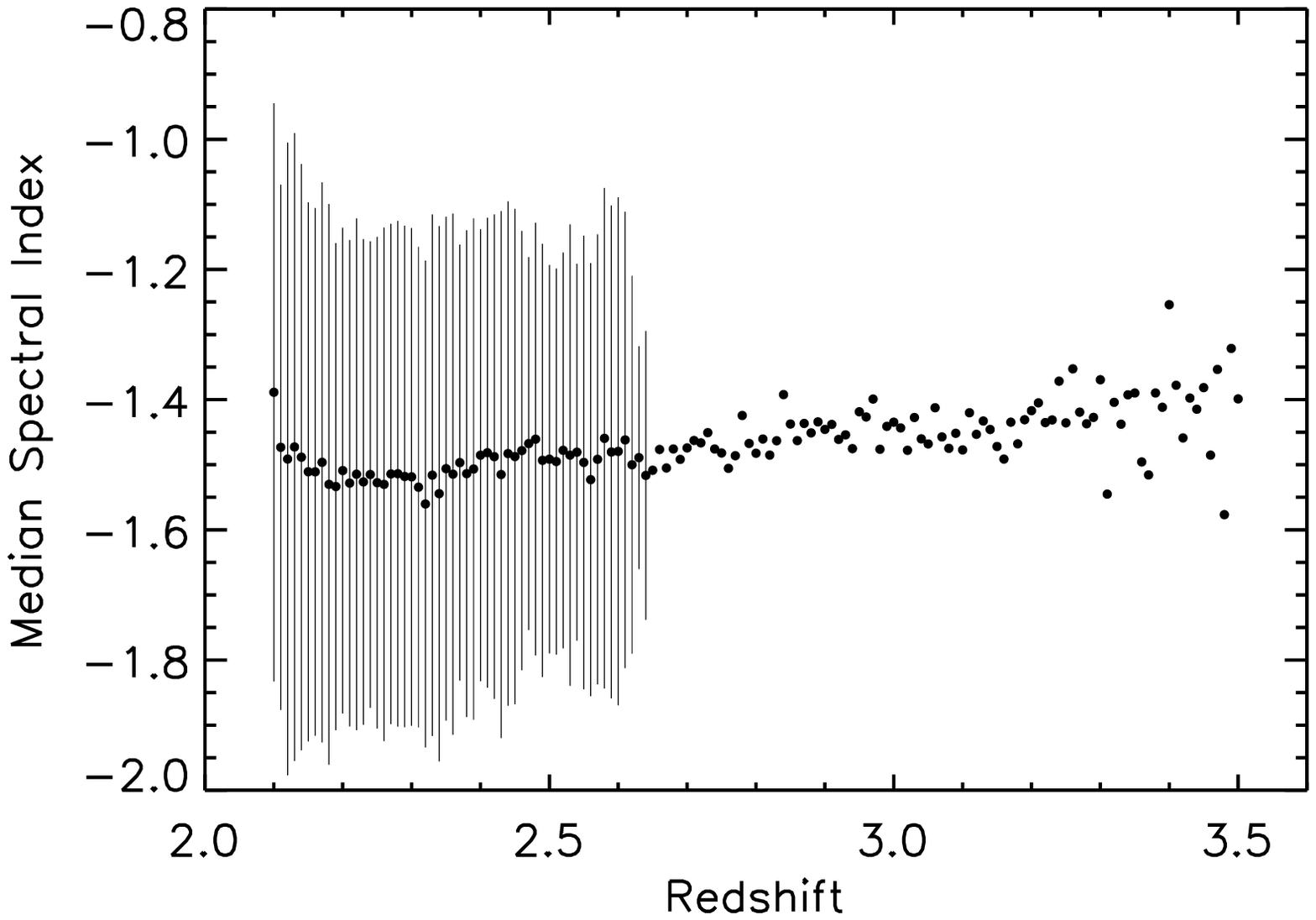}
\caption{Median spectral index of quasars, binned by redshift.}
\label{fig:fig9}
\end{figure}

The median spectral indices of the quasars used in this work binned by redshift are shown in Figure~\ref{fig:fig9}.  The median spectral indices vary from approximately $\alpha_{\lambda} = -1.45$ ($2.1 \leq z \leq 2.6$) to $\alpha_{\lambda} = -1.40$ ($2.7 \leq z \leq 3.5$).  There is an apparent trend with redshift, but we hesitate to attribute that trend to redshift evolution in the spectral index.  Because flux calibration uncertainty can introduce broadband errors, it is possible that the trend is simply due to the residual errors in synthetic photometry on $g-r$ and $r-i$ presented in Table~\ref{tab:correction_table}.  In fact, the blue residual in $g-r$ would introduce a bias toward more pronounced spectral indices for the lower redshift quasars and the red residual in $r-i$ would introduce a bias toward less pronounced spectral indices at higher redshift, exactly the trend that is shown in Figure~\ref{fig:fig9}.  The biases in these two color measurements are of order 0.02 magnitudes, which in fact could cause larger trends in spectral index than seen here.  However, given the uncertainty in the exact form of the residual fluxing errors, we are unable to improve the flux calibration any further in this analysis.  We reserve that effort for future pipeline development by the BOSS team.

\subsection{Comparing Spectral Index to Quasar Color}\label{subsec:color_spectralindex}

In each redshift bin, a broadband correction described by a powerlaw is applied to the composite spectrum such that the composite has the same spectral index as the median of the quasar sample at that redshift.  We do the same to warp the composite spectrum to match the spectral index at the $+1\sigma$ and $-1\sigma$ bounds of the distribution of spectral indices in each redshift bin.  We only have accurate estimates of the distribution of the spectral indices over $2.1 \leq z \leq 2.64$ so for higher redshifts we assume the same range of $\alpha_{\lambda}$ as that in the $z = 2.64$ redshift bin.  As described in Section~\ref{subsec:construct_composite}, the quasar composite spectrum has been corrected for the mean optical depth of the Lyman-$\alpha$ forest region.  No correction has been made to the photometric data to account for mean optical depth, so we model the suppression due to mean optical depth of neutral hydrogen by applying the same model as that presented in Section~\ref{subsec:construct_composite} to the \tlya-corrected composite.

\begin{figure}[h]
\centering
\includegraphics[width=.5\textwidth]{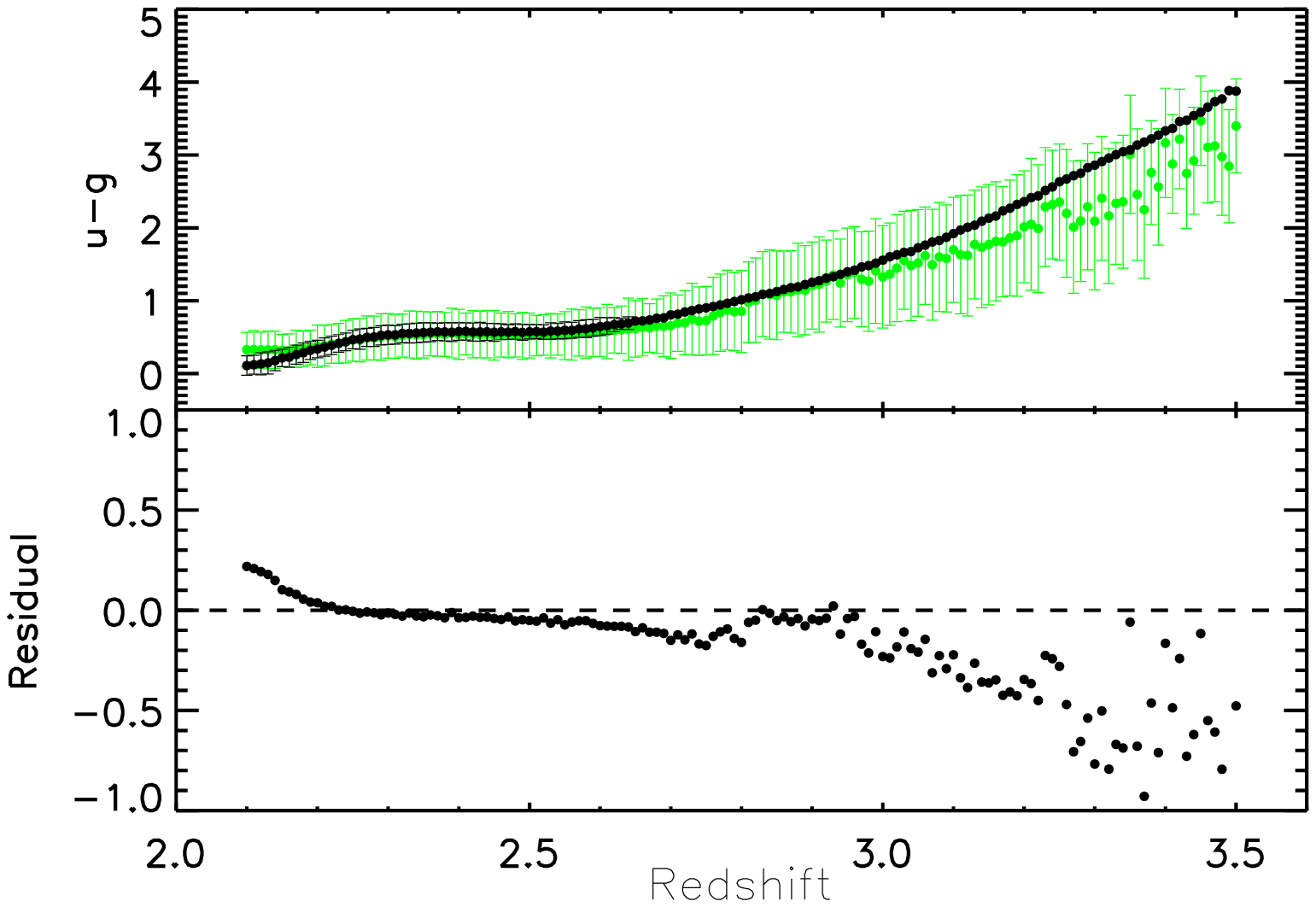}
\includegraphics[width=.5\textwidth]{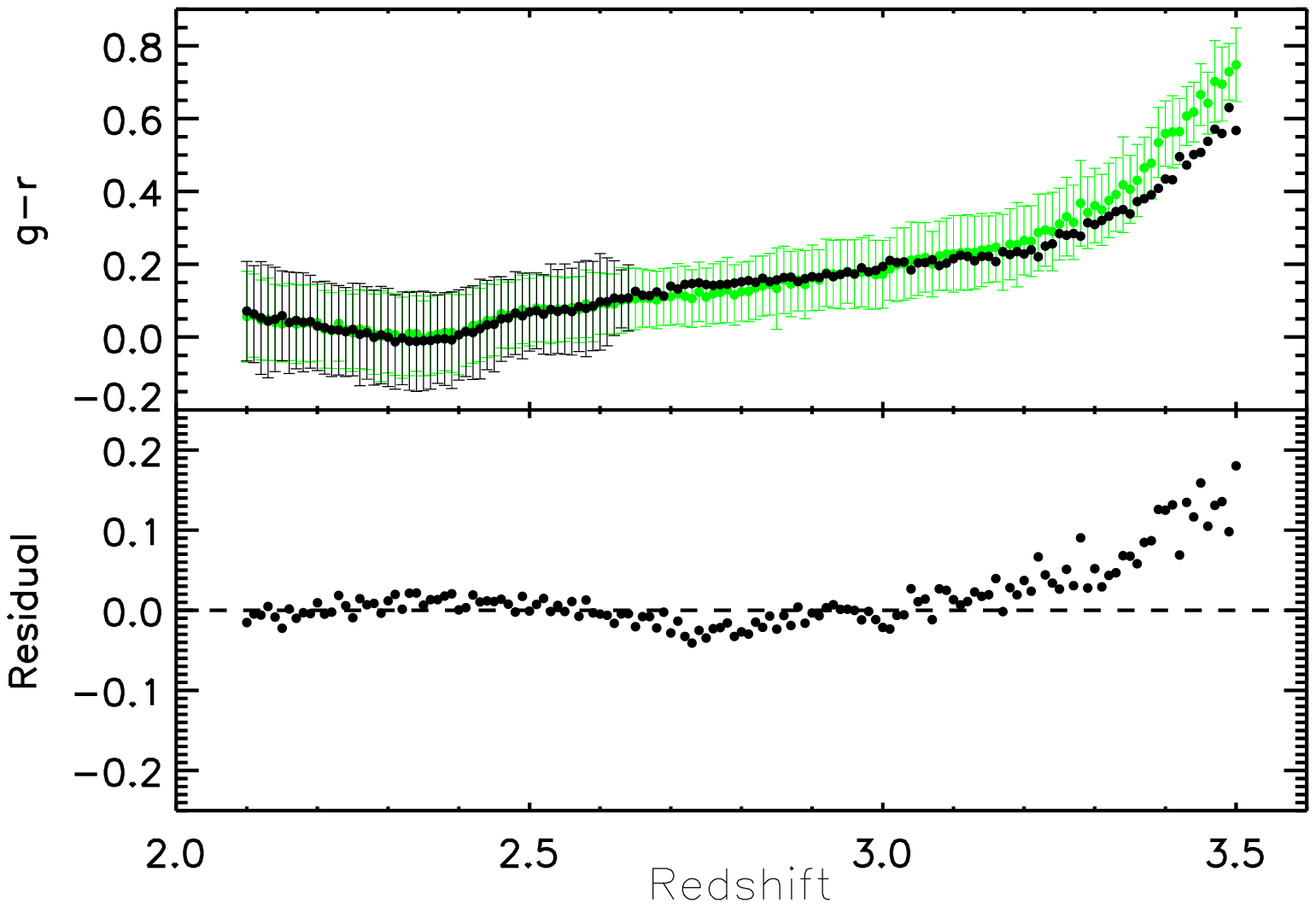}
\caption{The progression of the composite through color-redshift space.  The photometric color of the quasar sample is shown in green, along with the $+1\sigma$ and $-1\sigma$ of the distribution.  The synthetic colors of the spectral index matched composites are shown in black, along with the synthetic colors of the spectral index matched composites of the $+1\sigma$ and $-1\sigma$ of the distribution of the actual quasar sample.  {\bf First Panel:}  $u-g$.  {\bf Second Panel:} $g-r$.}
\label{fig:fig10}
\end{figure}

\begin{figure}[h]
\centering
\includegraphics[width=.5\textwidth]{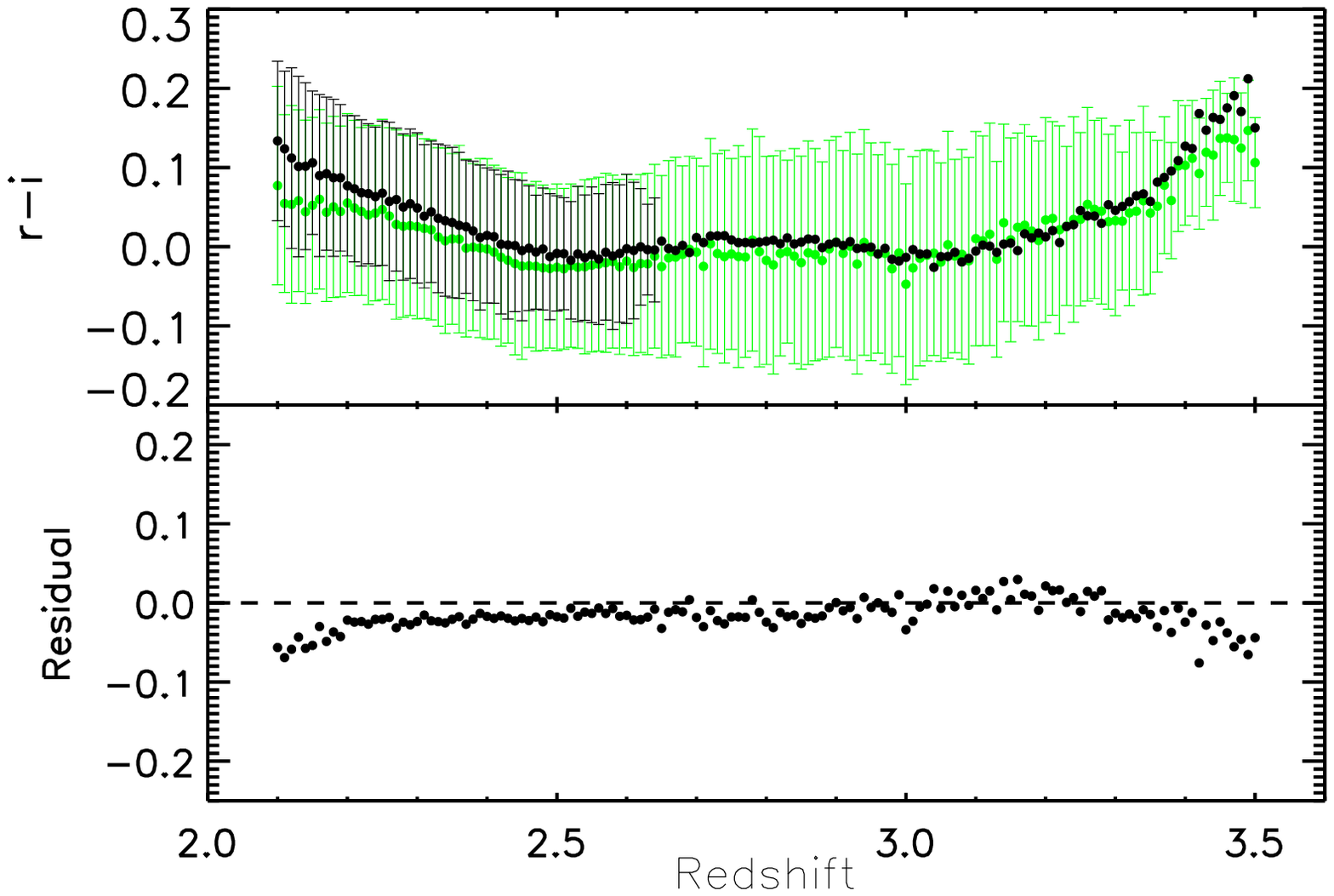}
\includegraphics[width=.5\textwidth]{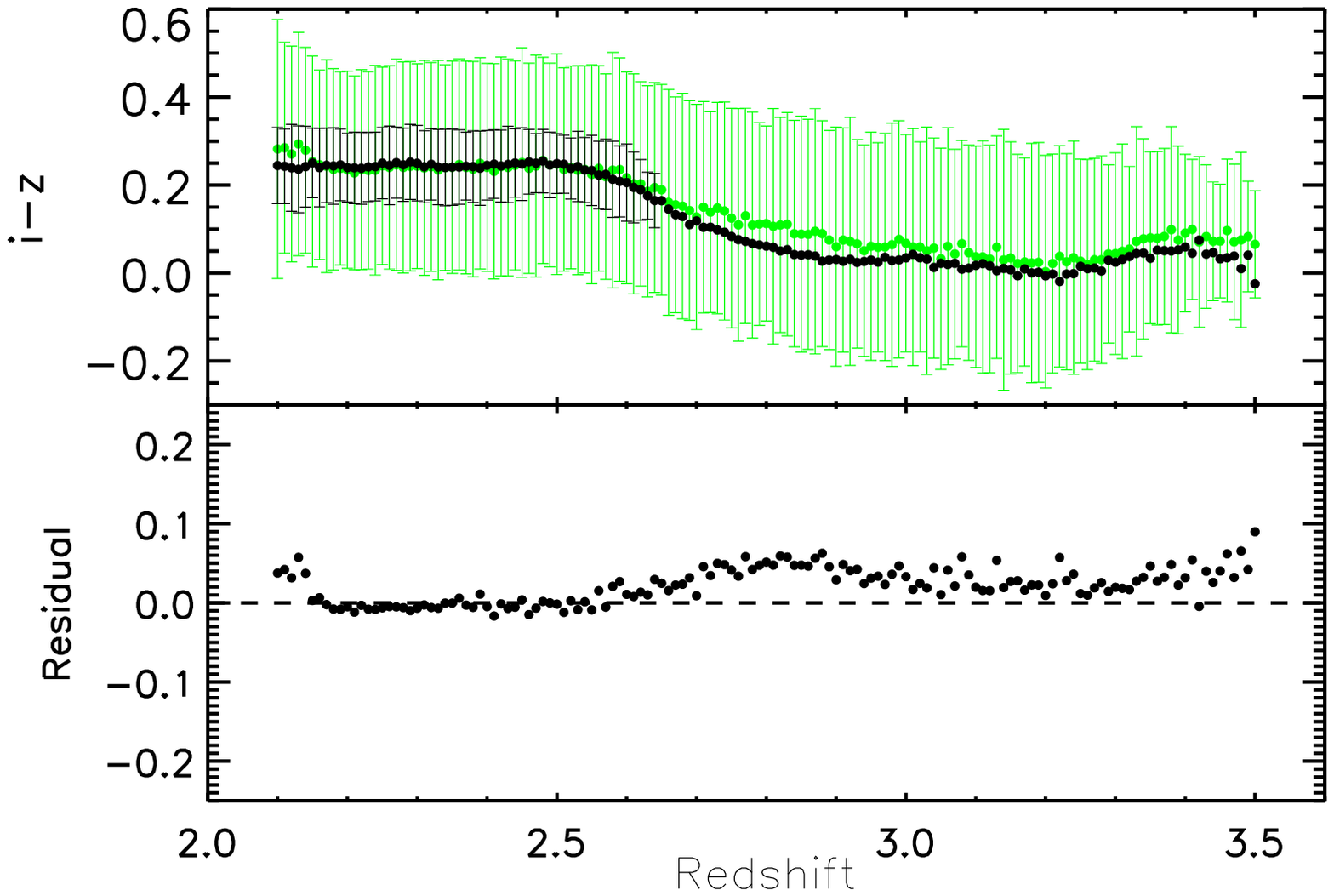}
\caption{The progression of the composite through color-redshift space.  The photometric color of the quasar sample is shown in green, along with the $+1\sigma$ and $-1\sigma$ of the distribution.  The synthetic colors of the spectral index matched composites are shown in black, along with the synthetic colors of the spectral index matched composites of the  $+1\sigma$ and $-1\sigma$ of the distribution of the actual quasar sample.  {\bf First Panel:}  $r-i$.  {\bf Second Panel:} $i-z$.}
\label{fig:fig11}
\end{figure}

The synthetic $u-g$, $g-r$, $r-i$, and $i-z$ colors of these modified composites for each redshift bin are presented in Figures~\ref{fig:fig10} and \ref{fig:fig11}.  Also displayed are the median photometric colors and the standard deviations of the imaging sample in each redshift bin.  

\begin{figure}[h]
\centering
\includegraphics[width=.5\textwidth]{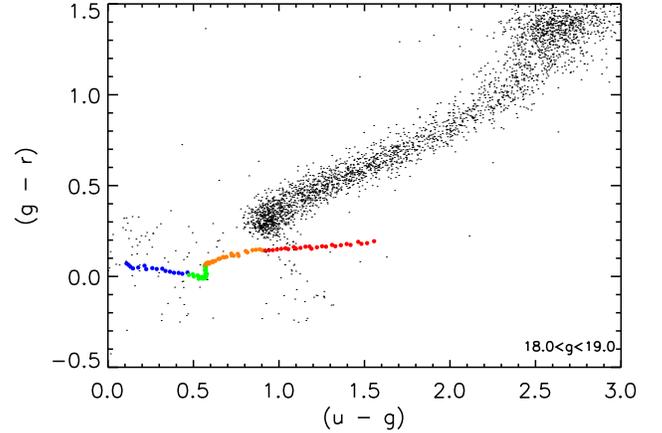}
\caption{The progression of the composite through ($g-r$) vs ($r-i$) color-color space.  The composite at $2.1 \leq z < 2.25$ is shown in blue, green indicates $2.25 \leq z <2.5$, orange indicates $2.5 \leq z <2.75$, and red indicates $2.75 \leq z \leq 3.0$ .  The black dots indicate stellar colors from SDSS photometry.}
\label{fig:fig12}
\end{figure}

Figure~\ref{fig:fig12} shows the trend in $u-g$ against $g-r$ of the spectral index and optical depth-corrected composite and displays the stellar locus for comparison.  The colors indicated here are found in the same manner as above.  Here, the data for the composite synthetic photometry are ignored at redshifts above 3.0 for the reasons stated in the following paragraph concerning the discrepancy in $u$ magnitude.

The colors of the composite spectra reproduce the photometric colors of the quasar sample across most of color and redshift space remarkably accurately.  In $g-r$ at a redshift of $3.2 \leq z \leq 3.5$, we see an offset of 0.15 magnitudes likely caused by incorrect modeling of mean optical depth of the Lyman-$\alpha$ forest.  In most colors, the difference between photometric and synthetic photometric data increases in the vicinity of $z = 2.7$ (in $u-g$ by $-0.2$, in $g-r$ by $-0.03$, in $i-z$ by +0.06, and in $r-i$ the difference is consistent with the scatter between bins).  At this redshift, the quasar locus approaches the stellar locus as shown in Figure~\ref{fig:fig12}.  As the quasar target selection schemes tend to avoid objects near the stellar locus, the quasars targeted at this redshift may not be as representative of the full quasar population around this redshift.  In $u-g$, the offset between the observed photometry and the synthetic photometry of the modified composite begins to grow around a redshift of 2.9 to approximately 0.5 magnitudes.  This behavior likely appears because the photometric bandpass extends to shorter wavelengths than the synthetic photometry and the two measurements have differing sensitivity to quasar flux beyond the Lyman limit at 912 \AA.  In addition, the composite is poorly constrained at $\lambda < 912$ \AA\  as shown in Figure~\ref{fig:fig5} by the non-physical break in continuum at short wavelengths in the composite spectrum that has been corrected for \tlya.  This result could potentially be improved with better models for absorption of ionizing photons over the redshifts covered by the BOSS quasar sample.

\section{Spectral Features}\label{sec:lines}

Having completed the characterization of the continuum to show that the broad band colors of the quasar spectrum are
consistent with quasar photometry, we now investigate quasar spectral lines. 
The high signal-to-noise quasar composite spectrum allows for detailed investigation of these lines, a comparison to previous work,
and the identification of new faint lines.

\subsection{Line Fitting}\label{sec:linefit}
Measurements of line features are subject to assumptions of continuum modelling
and complexity in line structures such as those in the Fe complex.  The work here represents one
specific approach to estimating continuum and parameterizing line profiles.  While the fits
performed here appear to describe the lines quite well, we encourage the
readers to make their own fits to the composite spectrum for lines of special interest.
In particular, we note that the fit in the C~\textsc{III}] region is complicated by
multiple blended features and a series of known Fe~\textsc{III} lines that we were unable to resolve.

We proceeded to measure the line profiles in a semi-manual fashion.  We measured each line and its underlying continuum independently, adding additional parameters to the fit when there was clear blending upon visual inspection.  In some cases, the additional parameters resolve blended lines with different central wavelengths.  In other cases, such as \lya, multiple components with effectively the same central wavelength but different widths are required to describe a single emission line.  Our procedure for fitting each line is as follows:
\begin{enumerate}
\item Two continuum fitting regions are identified in the vicinity of the line.  Each region is typically 5 \AA\ wide.
These regions are separated by the centroid of the line by at least one FWHM.  The continuum fitting regions are visually inspected
to ensure the absence of faint lines and to ensure an appropriate approximation of the local continuum.  A power-law is fit to the
composite spectrum over the wavelength regions identified to model the continuum.
\item The line region is fit with a single Gaussian.  A wavelength range (typically covering a wavelength region
spanning twice the FWHM) is chosen to perform the fit of the line.  A single Gaussian fit is first performed and visually inspected
to ensure that emission line wings are well described.  
\item If clear residuals are found by visual inspection, a second Gaussian is introduced to improve the fit.
Likewise, additional Gaussians are introduced until there are no obvious residuals.  In cases of slightly blended lines, these additional
parameters allow us to accurately describe the individual components of overlapping lines.  In the case of complex line profiles, these
additional parameters allow is to separate broadline regions with velocity offset and/or varying dynamics into discrete components.
\item We begin at the bluest region of the spectrum and continue until 3000 \AA, where the composite spectrum begins to show degradation.
\end{enumerate}

The results of the line fitting are reported in Table~\ref{tab:linetable}.  In that table, we list the properties of the detected lines
in the rest frame followed by the central wavelength, equivalent width, and the Gaussian RMS width associated with each component of the fit.
In total, 40 distinct emission features were identified.  Of those, nine features were blends that we were unable to resolve.
The six strongest features (Ly$\beta$ / O~\textsc{IV}, Ly$\alpha$, Si~\textsc{IV}/O~\textsc{IV},
C~\textsc{IV}, C~\textsc{III}], Mg~\textsc{II}) each required multiple components to accurately describe the line profile.

\begin{deluxetable*}{cr||rrrr|rrrr}
\centering
\tablewidth{0pt}
\tabletypesize{\footnotesize}
\tablecaption{\label{tab:linetable} Emission lines detected in the composite spectrum generated in this work compared to known wavelengths and the measurements
from V01.  Entries that required multiple fits to describe the emission feature are listed with an additional suffix (e.g. Ly$\alpha$-1 and Ly$\alpha$-2).
Blends are presented with each discrete component in the lab frame and by the multiple components in the observed data (e.g. Si~\textsc{IV} / O~\textsc{IV}).
}
\tablehead{
\multicolumn{2}{c}{Laboratory Frame}   & \multicolumn{4}{c}{This Work}    &     \multicolumn{4}{c}{Vanden Berk 2001} \\
\hline
\colhead{ID} &  \colhead{$\lambda$} &  \colhead{$\lambda$} &   \colhead{equivalent width} &  \colhead{$\sigma$}  &  \colhead{ID}  &  \colhead{$\lambda$}  &  \colhead{equivalent width} &  \colhead{$\sigma$} &   \colhead{ID} 
}
\startdata
Ly$\epsilon$ & 937.80 & 942.66 & 1.50 $\pm$ 0.07 &    5.33  & Ly$\epsilon$/Ly$\delta$  & 940.93 & 2.95 $\pm$ 0.54 & 4.73 & Ly$\epsilon$/Ly$\delta$           \\ 
Ly$\delta$ & 949.74 & & & & & & & & \\ \hline
 C~\textsc{III} & 977.02 & 977.74 & 1.53 $\pm$ 0.03 & 4.12 & C~\textsc{III} & 985.46 & 6.55 $\pm$ 0.58 & 8.95 & C~\textsc{III}/N~\textsc{III} \\ 
 N~\textsc{III} & 990.69 & 989.73 & 1.50 $\pm$ 0.03 & 4.14 & N~\textsc{III} & & & &                                                           \\ \hline
 Ly$\beta$ & 1025.72 & 1031.48 & 13.53 $\pm$ 0.10 & 13.15 & Ly$\beta$ / O~\textsc{VI}-1 & 1033.03 & 9.77 $\pm$ 0.49 & 7.76 & Ly$\beta$ / O~\textsc{VI}  \\ 
 O~\textsc{VI} & 1033.83 & 1034.07 & 4.26 $\pm$ 0.15 & 4.84 & Ly$\beta$ / O~\textsc{VI}-2 & & & &                                                     \\ \hline
 unknown & 1062.7 & 1064.01 & 2.90 $\pm$ 0.04 & 7.66 & unknown & 1065.10 & 0.80 $\pm$ 0.27 & 4.17 & Ar~\textsc{I}                       \\ \hline
 unknown & 1073.0 & 1073.53 & 0.71 $\pm$ 0.08 & 3.51 & unknown & & & &                                                                           \\ \hline
 N~\textsc{II}\tablenotemark{a} & 1083.99 & 1083.31 & 1.32 $\pm$ 0.04 & 5.32 & N~\textsc{II} & & & &                                                                    \\ \hline
 Fe~\textsc{III} & UV 1 & 1117.85 & 0.76 $\pm$ 0.07 & 5.26 & Fe~\textsc{III} & 1117.26 & 3.66 $\pm$ 0.34 & 8.49 & Fe~\textsc{III}                    \\ \hline
 Fe~\textsc{III}\tablenotemark{b} & UV 1 & 1127.55 & 0.46 $\pm$ 0.06 & 4.11 & Fe~\textsc{III} & & & &                                                                   \\ \hline
 C~\textsc{III} & 1175.7 & 1174.91 & 2.49 $\pm$ 0.01 & 7.68 & C~\textsc{III} & 1175.35 & 0.83 $\pm$ 0.14 & 2.28 & C~\textsc{III}                      \\ \hline
 Ly$\alpha$ & 1215.67 & 1215.86 & 19.08 $\pm$ 1.03 & 4.85 & Ly$\alpha$-1 & 1216.25 & 92.91 $\pm$ 0.72 & 19.46 & Ly$\alpha$    \\
 & & 1216.94 & 66.63 $\pm$ 4.01 & 16.56 & Ly$\alpha$-2 & & & &                                                                  \\ \hline
 N~\textsc{V} & 1240.14 & 1241.73 & 9.22 $\pm$ 0.02 & 6.89 & N~\textsc{V} & 1239.85 & 1.11 $\pm$ 0.09 & 2.71 & N~\textsc{V}                             \\ \hline
 Si~\textsc{II} & 1262.59 & 1260.81 & 4.47 $\pm$ 0.008 & 7.91 & Si~\textsc{II} & 1265.22 & 0.21 $\pm$ 0.06 & 2.74 & Si~\textsc{II}                        \\ \hline
 O~\textsc{I} & 1304.85 & 1305.31 & 1.73 $\pm$ 0.02 & 5.74 & Si~\textsc{II}/O~\textsc{I} & 1305.42 & 1.66 $\pm$ 0.06 & 5.42 & Si~\textsc{II}/O~\textsc{I}                           \\ 
 Si~\textsc{II} & 1306.82 &  &  &  & & & & &                                                              \\ \hline
 C~\textsc{II} & 1335.3 & 1335.43 & 0.77 $\pm$ 0.01 & 4.73 & C~\textsc{II} & 1336.60 & 0.59 $\pm$ 0.05 & 3.86 & C~\textsc{II}                                  \\ \hline
 Si~\textsc{IV} & 1396.76 & 1398.16 & 5.07 $\pm$ 0.12 & 19.54 & Si~\textsc{IV}/O~\textsc{IV}-1 & 1398.33 & 8.13 $\pm$ 0.09 & 12.50 & Si~\textsc{IV}/O~\textsc{IV}        \\ 
 O~\textsc{IV} & 1402.06 & 1398.22 & 5.01 $\pm$ 0.17 & 8.81 & Si~\textsc{IV}/O~\textsc{IV}-2 & & & &                      \\ \hline
 N~\textsc{IV}]\tablenotemark{a}  & 1486.5 & 1485.50 & 0.15 $\pm$ 0.01 & 3.25 & N~\textsc{IV}] & & & &                                                                           \\ \hline
 C~\textsc{IV}-1 & 1548.19 & 1545.57 & 28.51 $\pm$ 0.03 & 17.00 & C~\textsc{IV}-1 & 1546.15 & 23.78 $\pm$ 0.10 & 14.33 & C~\textsc{IV}                      \\
 C~\textsc{IV}-2 & 1550.77 & 1548.28 & 8.92 $\pm$ 0.04 & 6.11 & C~\textsc{IV}-2 & & & &                                                            \\ \hline
 He~\textsc{II} & 1640.42 & 1639.46 & 1.53 $\pm$ 0.01 & 4.35 & He~\textsc{II} & 1637.84 & 0.51 $\pm$ 0.03 & 4.43 & He~\textsc{II}                       \\ \hline
 O~\textsc{III}] & 1663.48 & 1663.21 & 1.94 $\pm$ 0.01 & 8.22 & O~\textsc{III}]/Al~\textsc{II} & 1664.74 & 0.50 $\pm$ 0.03 & 5.50 & O~\textsc{III}]/Al~\textsc{II}          \\ 
 Al~\textsc{II} & 1670.79 & & & & & & & &                                                                                           \\ \hline
 N~\textsc{IV} & 1718.55 & 1717.47 & 0.13 $\pm$ 0.01 & 5.20 & N~\textsc{IV}/Fe~\textsc{II} & 1716.88 & 0.30 $\pm$ 0.03 & 7.36 & N~\textsc{IV}/Fe~\textsc{II}                \\
 Fe~\textsc{II} & UV 37 & & & & & & & &                                                                                             \\ \hline
 N~\textsc{III}] & 1750.26 & 1748.94 & 0.81 $\pm$ 0.01 & 6.37 & N~\textsc{III}] & 1748.31 & 0.44 $\pm$ 0.03 & 5.12 & N~\textsc{III}]                    \\ \hline
 Fe~\textsc{II} & UV 191 & 1787.41 & 0.30 $\pm$ 0.02 & 5.37 & Fe~\textsc{II} & 1788.73 & 0.28 $\pm$ 0.02 & 6.06 & Fe~\textsc{II}                        \\ \hline
 [Ne~\textsc{III}] & 1814.73 & 1816.34 & 0.54 $\pm$ 0.02 & 8.86 & Si~\textsc{II}/Ne~\textsc{III} & 1818.17 & 0.16 $\pm$ 0.02 & 5.72 & Si~\textsc{II}/Ne~\textsc{III}        \\
 Si~\textsc{II} & 1816.98 & & & & & & & &                                                                                           \\ \hline
 Al~\textsc{III} & 1857.4 & 1857.40 & 3.47 $\pm$ 0.08 & 11.01 & Al~\textsc{III} & 1856.76 & 0.40 $\pm$ 0.03 & 4.95 & Al~\textsc{III}                    \\ \hline
 Si~\textsc{III} & 1892.03 & 1893.24 & 0.55 $\pm$ 0.03 & 3.99 & Si~\textsc{III}/Fe~\textsc{IV} & 1892.64 & 0.16 $\pm$ 0.02 & 3.09 & Si~\textsc{III}/Fe~\textsc{IV}          \\
 Fe~\textsc{III} & UV 34 & & & & & & & &                                                                                             \\ \hline
 C~\textsc{III}] & 1908.73 & 1903.61 & 20.51 $\pm$ 0.09 & 20.81 & C~\textsc{III}]-1 & 1905.97 & 21.19 $\pm$ 0.05 & 23.58 & C~\textsc{III}]                \\
 & & 1907.00 & 2.26 $\pm$ 0.09 & 3.55 & C~\textsc{III}]-2 & & & &                                                                 \\ 
 & & 1908.36 & 3.10 $\pm$ 0.03 & 5.18 & C~\textsc{III}]-3 & & & &                                                                 \\ \hline
 Fe~\textsc{III}\tablenotemark{b} & UV 50 & 1988.81 & 0.13 $\pm$ 0.01 & 8.96 & Fe~\textsc{III} & 1991.83 & 0.20 $\pm$ 0.02 & 6.73 & Fe~\textsc{III}                   \\ \hline
 Fe~\textsc{III}\tablenotemark{b} & UV 48 & 2068.51 & 1.65 $\pm$ 2.13 & 13.47 & Fe~\textsc{III}  & & & &                                                               \\ \hline
 Fe~\textsc{II}\tablenotemark{b}  & UV 91 & 2094.27 & 2.39 $\pm$ 2.46 & 20.28 & Fe~\textsc{II}  & & & &                                                               \\ \hline
 Fe~\textsc{II}\tablenotemark{b}  & UV 212 & 2135.23 & 0.70 $\pm$ 0.02 & 12.46 & Fe~\textsc{II}  & & & &                                                    \\ \hline
 Fe~\textsc{II} & UV 79 & 2176.27 & 0.39 $\pm$ 0.01 & 9.84 & Fe~\textsc{II} & 2175.62 & 0.25 $\pm$ 0.02 & 5.85 & Fe~\textsc{II}                         \\ \hline
 Fe~\textsc{II}\tablenotemark{b}  & UV 118 & 2219.70 & 0.34 $\pm$ 0.03 & 9.41 & Fe~\textsc{II} & & & &                                                                      \\ \hline
 Fe~\textsc{II}\tablenotemark{b} & UV 183 & 2314.91 & 5.11 $\pm$ 0.05 & 32.64 & Fe~\textsc{II} & & & &                                                                    \\ \hline
 C~\textsc{II}] & 2326.44 & 2326.96 & 0.46 $\pm$ 0.02 & 5.49 & C~\textsc{II}] & 2327.34 & 0.31 $\pm$ 0.02 & 4.95 & C~\textsc{II}]                     \\ \hline
 [Ne~\textsc{IV}] & 2423.83 & 2423.93 & 0.94 $\pm$ 0.01 & 9.52 & [Ne~\textsc{IV}] & 2423.46 & 0.77 $\pm$ 0.02 & 8.42 & [Ne~\textsc{IV}]                 \\ \hline
 [O~\textsc{II}] & 2471.03 & 2470.99 & 0.29 $\pm$ 0.01 & 8.79 & [O~\textsc{II}] & 2467.98 & 0.16 $\pm$ 0.02 & 4.54 & [O~\textsc{II}]                    \\ \hline
 Fe~\textsc{II} & UV 1 & 2624.05 & 1.12 $\pm$ 0.03 & 12.55 & Fe~\textsc{II} & 2626.92 & 0.81 $\pm$ 0.03 & 9.93 & Fe~\textsc{II}                         \\ \hline
 Al~\textsc{II}] & 2669.95 & 2671.64 & 0.52 $\pm$ 0.12 & 7.20 & AlII]/O~\textsc{III} & 2671.89 & 0.14 $\pm$ 0.02 & 5.10 & AlII]/O~\textsc{III}           \\
 O~\textsc{III} & 2672.04 & & & & & & & &                                                                                          \\ \hline
 Mg~\textsc{II}-1 & 2796 & 2798.90 & 9.55 $\pm$ 0.13 & 10.26 & Mg~\textsc{II}-1 & 2800.26 & 32.28 $\pm$ 0.07 & 34.95 & Mg~\textsc{II}                  \\
 Mg~\textsc{II}-2 & 2803 & 2800.14 & 36.67 $\pm$ 0.13 & 54.40 & Mg~\textsc{II}-2 & & & &                                                                  \\
 & & 2804.55 & 6.49 $\pm$ 0.31 & 18.48 & Mg~\textsc{II}-3 & & & &                                                                   \\ \hline
 Fe~\textsc{II} & UV 78 & 2958.36 & 8.44 $\pm$ 0.31 & 33.16 & Fe~\textsc{II} & 2964.28 & 4.93 $\pm$ 0.04 & 22.92 & Fe~\textsc{II}                       
\enddata
\tablenotetext{a}{Line identification and wavelength from \citet{shull12a}.}
\tablenotetext{b}{Line identification and wavelength from \citet{vestergaard01a}.}
\end{deluxetable*}

\subsection{Comparisons to V01}\label{sec:lineV01}

For the sake of comparison, we include the line measurements from V01 in the final four columns of Table~\ref{tab:linetable}.
A number of lines exhibited differences between the V01 composite spectrum and this work.
First, the higher redshift range and higher signal-to-noise of this composite spectrum allowed us to resolve
the C~\textsc{III} $\lambda\lambda$977 and N~\textsc{III} $\lambda\lambda$990 lines that were blended in the V01 analysis.
Second, we decompose the complex line profiles associated with the six strongest emission features
into multiple Gaussian components, a measurement not possible in the V01 approach.

A more significant difference between V01 and this work is this work is the equivalent width
and velocity dispersion measurements for several of the line species.
Lines which were found to have at least 50\%
larger equivalent width and a 50\% larger $\sigma$ in this work compared to V01 include O~\textsc{IV} $\lambda\lambda$1034,
an unknown line at $\lambda\lambda$1067, C~\textsc{III} $\lambda\lambda$1176, N~\textsc{V} $\lambda\lambda$1240,
Si~\textsc{II} $\lambda\lambda$1262, Si~\textsc{II}/Ne~\textsc{III}, Al~\textsc{III} $\lambda\lambda$1857, and [O~\textsc{II}] $\lambda\lambda$2471.
These lines are found over a wide range of wavelength and ionization energy.
The difference in these line strengths is likely due to the different target selection and the lower luminosity of quasars in this
work relative to V01.
There are several instances where the V01 lines are systematically more pronounced then our measurements.  Those
occur in the far blue where V01 has fewer quasars to sample and in the Fe regions, where there are likely
different assumptions about continuum fitting and blending of lines.

The most interesting aspect of the line measurements introduced here is the identification of faint lines
not recognized in V01.  We find three lines in the \lya\ forest region that were not identified in
V01: an unknown line at $\lambda\lambda$1063, N~\textsc{II} $\lambda\lambda$1084, and Fe~\textsc{III} $\lambda\lambda$UV 1.
Understanding of these lines becomes increasingly important as cosmology with the \lya\ forest in
quasar spectra advances beyond measurements of BAO.
An additional unknown line at $\lambda\lambda$1073 is also found.  
The N~\textsc{IV}] $\lambda\lambda$1486 line becomes apparent in this composite spectrum, likely because
of the lower luminosity of the parent quasar sample.
Finally, we resolve seven distinct lines over the wavelength range $1980<\lambda<2320$ which we attribute to
either Fe~\textsc{II} or Fe~\textsc{III}.  V01 isolated two lines over this wavelength range with
different line profiles.
In total, we characterize nine lines that were not included in that previous work.

\section{Conclusion}\label{sec:conclusion}

This work presents a very high signal-to-noise quasar composite spectrum using 102,150 BOSS quasar spectra.   We have made the final composite spectrum available via download at a publicly accessible website\footnote{http://data.sdss3.org/sas/dr12/boss/qso/composite/}.  This method of calculating a composite is usable for any large sample of quasar spectra in BOSS.  As long as every two consecutive sets of 500 spectra each covers a sufficiently large wavelength range to allow the above calculations, this method can be used for any redshift range.  The spectra used to create this composite spectrum are a uniform sample with $2.1 \leq z \leq 3.5$, excluding BALs and DLAs, and an observed airmass less than 1.2.  The large number of quasar spectra used in this work and the uniformity of the sample represent a significant improvement over previous works, such as V01, and samples a different wavelength range than other recent works, such as S14.

The composite spectrum is shifted to match the median spectral index of the Lyman-$\alpha$ forest sample after performing spectrophotometric corrections.  We have made the final spectrophotometric correction available via download at a publicly accessible website\footnote{http://data.sdss3.org/sas/dr12/boss/qso/composite/}.  The spectrophotometric correction is parameterized on only airmass and reduces the color bias to less than 0.02 magnitudes.  This correction can be used to allow direct comparison between SDSS and BOSS spectra, which is useful in variability studies.  The measured spectral index of $-1.4613 \pm 0.0017$ differs from some other studies \citep{wilhite05a,ruan14}, most likely due to sampling a different region of redshift-luminosity space.  We examine the properties of this composite and the initial sample in color-color space to show that the synthetic photometry of this quasar composite spectrum closely recreates the photometry of the quasar sample.

As pointed out earlier in the text, selection of quasars over the high
redshift range conducive to \lya\ forest
studies is not trivial.  Comparison of the number density of BOSS \lya\
forest quasars to the full population
expected over this redshift range
\citep[e.g.][]{palanque-delabrouille15a} reveals significant gaps in
the completeness of the BOSS quasar sample.  The composite spectrum
presented in this work is therefore
representative not of the entire population of $2.1<z<3.5$ quasars, but
instead, representative of the quasar spectra used in the BOSS \lya\ forest cosmology studies
\citep{delubac15a}.

Current programs such as the extended Baryon Oscillation Spectroscopic
Survey \citep[eBOSS;][]{dawson15a} and future programs such the Dark
Energy Spectroscopic Instrument \citep[DESI;][]{levi13a}
will use new imaging data and selection techniques
\citep[e.g.][]{myers15a} to improve the completeness
of high redshift quasars.  Future work with the BOSS and eBOSS samples
will quantitavily address selection effects
by assessing quasar spectral properties after tuning the samples to have
well-controlled populations
binned by observable quantities.  This work will include the examination
of luminosity-redshift binned subsamples,
and examination of trends in quasar properties in luminosity and
redshift, and thus trends in quasar evolution.

Finally, the techniques introduced here will allow more detailed
characterizations of quasars, such
as assessment of the differences between different populations and tests
of accretion models.
For example, normal and BAL quasars can be examined by contrasting
composite spectra of subsamples binned on balnicity.
The composite we created was used to study relative line strengths,
widths, and velocity
offsets in quasars at $z = 2.457$, opening up new avenues to the study
of quasar structure at high redshift.
This composite spectrum and table of linewidths can be used to assess
accretion models, theoretical line
templates, and contribution from the stellar populations of the quasar
host galaxy.
More generally, the large samples of quasars produced in BOSS and future
surveys can be
divided varying populations, thus enabling even more thorough
investigations of quasar
physics via composite spectra.

Funding for SDSS-III has been provided by the Alfred P. Sloan Foundation, the Participating Institutions,
the National Science Foundation, and the U.S. Department of Energy Office of Science.
The SDSS-III web site is http://www.sdss3.org/.

SDSS-III is managed by the Astrophysical Research Consortium for the Participating Institutions of the
SDSS-III Collaboration including the University of Arizona, the Brazilian Participation Group,
Brookhaven National Laboratory, University of Cambridge, Carnegie Mellon University, University of Florida,
the French Participation Group, the German Participation Group, Harvard University, the Instituto de Astrofisica de Canarias,
the Michigan State/Notre Dame/JINA Participation Group, Johns Hopkins University,
Lawrence Berkeley National Laboratory, Max Planck Institute for Astrophysics,
Max Planck Institute for Extraterrestrial Physics, New Mexico State University,
New York University, Ohio State University, Pennsylvania State University, University of Portsmouth,
Princeton University, the Spanish Participation Group, University of Tokyo, University of Utah,
Vanderbilt University, University of Virginia, University of Washington, and Yale University.

The work of DH and KD was supported in part
by the U.S. Department of Energy under Grant
DE-SC0009959. The support and resources from the
Center for High Performance Computing at the University of Utah is gratefully acknowledged.
Please contact the author(s) to request access to research
materials discussed in this paper.

\appendix

\section{{\bf A.} Flux Calibration}\label{sec:fluxcal}
In this appendix, we describe the process by which the spectrophotometric errors in BOSS quasar spectra are quantified and corrected.  We compare BOSS spectroscopic data to SDSS photometric and spectroscopic data to quantify the errors and test the derived correction.  The model for the correction is determined using only BOSS spectroscopy.

\subsection{{\bf A.1} Spectrophotometric Errors}\label{subsec:specphot_err}

Two independent methods are used to quantify the spectrophotometric errors introduced by the quasar fiber hole offsets.  In the first method, we calculate the average ratio between BOSS stellar spectra observed in the blue focal plane and the corresponding SDSS spectra for the same objects.  In the second method, we compare the imaging data from SDSS to synthetic magnitudes calculated from BOSS spectra observed in the blue focal plane.  These techniques are also used to test the spectrophotometric corrections in Appendix~\ref{subsubsec:initial_test} and Appendix~\ref{subsubsec:finalcorrection}.

\subsubsection{{\bf A.1.1} Spectral Comparison}\label{subsubsec:spectral_comparison}

Approximately 2000 stars targeted as quasars in the blue focal plane of BOSS (which we call ``stellar contaminants'') have previous spectra from SDSS.  These matches are found by locating pairs of objects within 0.36$''$ of each other from the DR12 catalog \citep[for BOSS;][]{alam15a} and the DR8 catalog \citep[for SDSS;][]{agol11a}.  This sample is trimmed to include only A, F, and G type stars with no redshift pipeline problems, and a $g$-band PSF magnitude brighter than 22.  The objects satisfying these criteria are visually inspected in both the SDSS and BOSS spectra.  Objects with radically different spectra (possibly from bad matching between spectra) and objects with significant and unreasonable flux differences are also rejected.  Flux differences were considered unreasonable and rejected if one or more of the following conditions were met:
\begin{itemize}
\item A flux ratio of greater than three or less than one third in the region 6000 \AA\ to 6500 \AA
\item A ratio which changed by more than a factor of three from less than 4500 \AA\ to greater than 8000 \AA
\item Having a calculated standard deviation of more than 0.5 in the flux ratio over the region 6000 \AA\ to 6500 \AA.  
\end{itemize}
After these cuts, 543 objects remained in the catalog.  

\begin{figure}[h]
\centering
\includegraphics[width=0.5\textwidth]{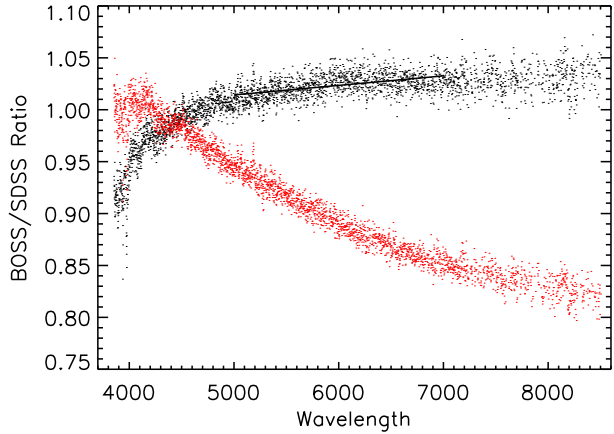}
\caption{The ratio between BOSS stellar contaminant spectra and their counterparts in SDSS as a function of wavelength.  The median ratio between uncorrected BOSS and SDSS spectra is shown in red, while the median ratio between BOSS spectra after the initial correction and SDSS spectra is shown in black (see Appendix~\ref{subsec:correct_specphot_error}).  A fit to the corrected data ratio is shown in black, with a slope of 0.0092 per thousand Angstroms.}
\label{fig:fig13}
\end{figure}

The ratio of each BOSS spectrum to its corresponding SDSS spectrum provides an independent non-variable sample by which one can judge changes in flux calibration between SDSS and BOSS.   Since SDSS spectroscopy used a shorter wavelength range than BOSS, both sets of spectra are clipped to include only the wavelength range of SDSS spectra (3850 \AA\ to 8500 \AA).  The ratio between the BOSS and SDSS spectra is calculated for each star as a function of wavelength.  The median value of all BOSS/SDSS star ratios is taken at each wavelength bin.  This median ratio is shown in Figure~\ref{fig:fig13} in red.  The BOSS/SDSS ratio is lowest at long wavelengths, corresponding to a deficit in flux in BOSS of roughly 20\% relative to SDSS.  At shorter wavelengths, the ratio is of order unity, having a peak of about unity at about 4000 \AA\ and declining to either side.  
In this figure, pixels flagged by the Lyman alpha skymask are masked \citep{delubac15a}.  Stellar variability could cause shifts between
the SDSS epoch and the BOSS epoch, but due to the large number of stars in this test, the average shift is expected to be small.
For example, even variability as high as 25\% between SDSS and BOSS would result in only a 1\% disrepancy in the average.

\subsubsection{{\bf A.1.2} Quantification by Photometric Comparison}\label{subsubsec:photometric_comparison}
A second test of the spectrophotometric errors involves the difference between the photometry from the SDSS imaging data (using PSF magnitudes) and the synthetic photometry from BOSS spectroscopy.

As in Appendix~\ref{subsubsec:spectral_comparison}, we use stellar contaminants for the comparison but no longer restrict ourselves to objects with spectra in SDSS.  The corresponding plate for each stellar contaminant is identified, and the median photometric difference is found for each plate in $g$, $r$, and $i$.  The median photometric difference is plotted against the airmass at which the BOSS spectra were observed in the first row of panels in Figure~\ref{fig:fig3}.
There is a clear trend in the difference in magnitudes of stellar contaminants.
In $g$, $r$, and $i$, BOSS synthetic photometry indicates that BOSS views these stellar contaminants as being dimmer than in SDSS photometry with a difference that increases with airmass.
Only data for filters $g$, $r$, and $i$ are shown because these have the most consistent coverage between SDSS filter bandpasses and BOSS spectral coverage.  
See Table~\ref{tab:correction_table} for the linear fits to the data.  
In general we expect a color bias in $g-r$ of 0.059 magnitudes and a color bias of 0.058 magnitudes in $r-i$, corresponding to the offsets observed at the median airmass of BOSS spectra.

\begin{deluxetable}{lcccccccccc}
\centering
\tablewidth{0pt}
\tabletypesize{\footnotesize}
\tablecaption{\label{tab:correction_table} Data showing the fitted slope, intercept, and color bias at the average airmass of 1.095 for (SDSS PSF magnitudes - BOSS synthetic magnitudes) in the $g$, $r$, and $i$ filters.  The second column is the characteristic error in magnitude for each filter.  The first set of columns is for the uncorrected raw data, the second set of columns shows the values after performing the first round of corrections, and the third set of columns shows the values after the final round of corrections.}

\tablehead{
\multicolumn{2}{c}{} & \multicolumn{3}{c}{Uncorrected} & \multicolumn{3}{c}{Corrected} & \multicolumn{3}{c}{Final Correction} \\
\hline \\
\colhead{Filter} & \colhead{Error} & \colhead{Slope} & \colhead{Intercept} & \colhead{At Mean} & \colhead{Slope} & \colhead{Intercept} & \colhead{At Mean} & \colhead{Slope} & \colhead{Intercept} & \colhead{At Mean}
}
\startdata
   $g$ & $\pm$0.02 & $-$0.15 & $-$0.02 & $-$0.03 & 0.09 & $-$0.02 & $-$0.01 & 0.05 & $-$0.02 & $-$0.01 \\
   $r$ & $\pm$0.03 & $-$0.80 & $-$0.02 & $-$0.09 & 0.28 & 0.00 & 0.02 & 0.12 & 0.00 & 0.00 \\
   $i$ & $\pm$0.04 & $-$1.16 & $-$0.04 & $-$0.15 & 0.28 & $-$0.01 & 0.01 & 0.08 & $-$0.02 & $-$0.01 \\
\enddata
\end{deluxetable}

\subsection{{\bf A.2} Correcting the Spectrophotometric Errors}\label{subsec:correct_specphot_error}

Having quantified the spectrophotometric errors, we now present an independent, empirical correction to the spectral response of the fibers in the blue focal plane.  First, we generate a correction using BOSS spectroscopic information from a series of specially-designed plates.  Second, we test this correction by comparing synthetic photometry from corrected BOSS spectra to SDSS photometry.  Third, we apply a scaling factor to the amplitude of the correction to minimize the color bias of the synthetic magnitudes of the corrected BOSS spectra.

To better understand the fluxing problems in the blue focal plane, a few plates (listed in Table~\ref{tab:platetable}) were drilled with standard stars in the blue focal plane.  The data reduction pipeline was rerun using these standard stars instead of the normal (``red'') standard stars.  We call this the ``blue reduction'', whereas the usual data processing is referred to here as the ``red reduction''.  The blue reduction includes 52 plates, each with more than 10 standard stars in the blue focal plane.  These standard stars have the same spectral throughput as the quasars to provide direct flux calibration for the blue focal plane.  The difference in the standard star spectra between the blue reductions and the red reductions captures the spectral distortion introduced by offsetting the fibers in the telescope focal plane.  
$F_{BLUE}$ represents the flux of the custom standard stars in the blue reduction, while $F_{RED}$ represents the flux in the red reduction.

\subsubsection{{\bf A.2.1} Parameterizing the Correction}\label{subsubsec:param_corr}

\begin{figure}[h]
\centering
\includegraphics[width=0.45\textwidth]{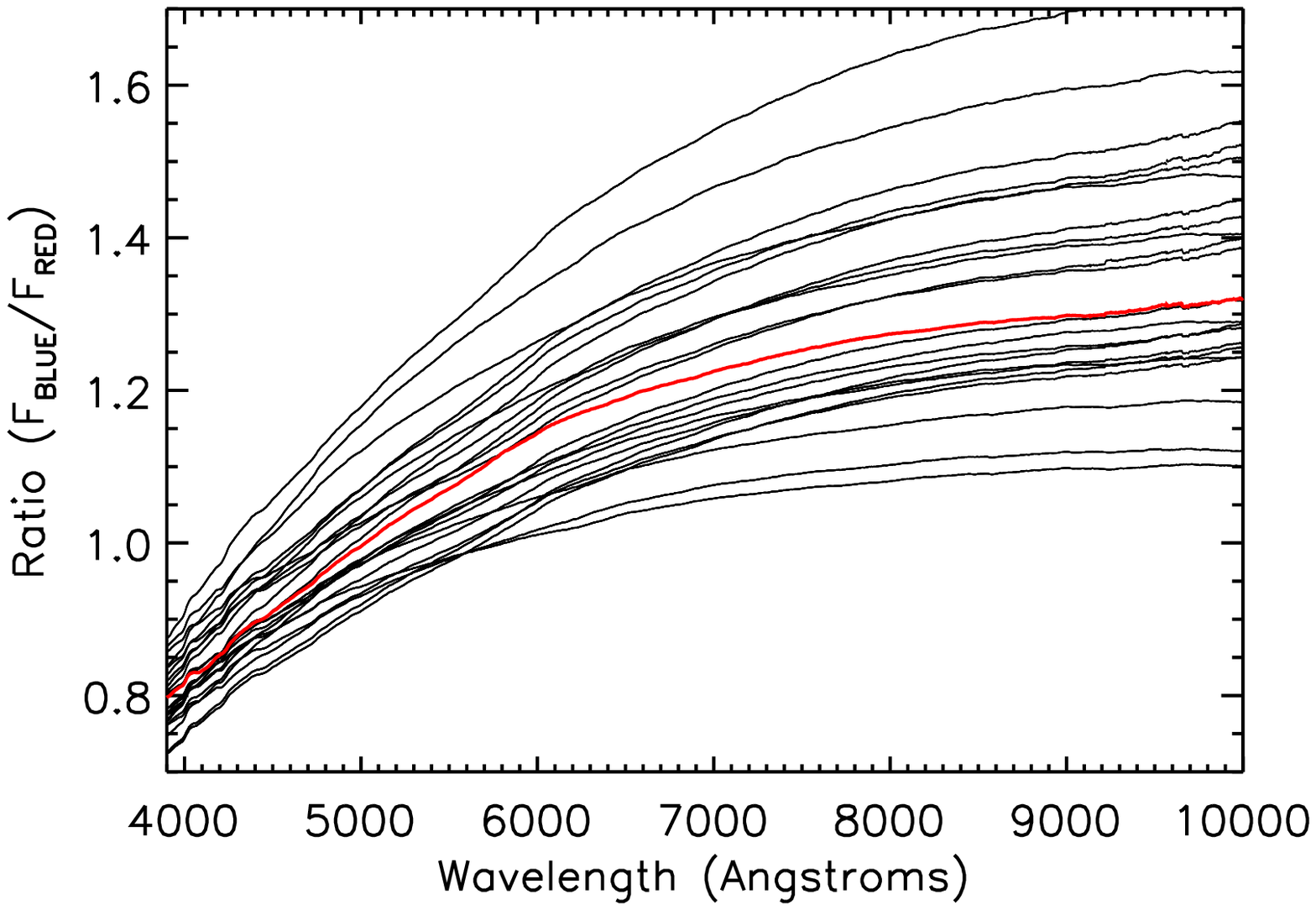}
\includegraphics[width=0.45\textwidth]{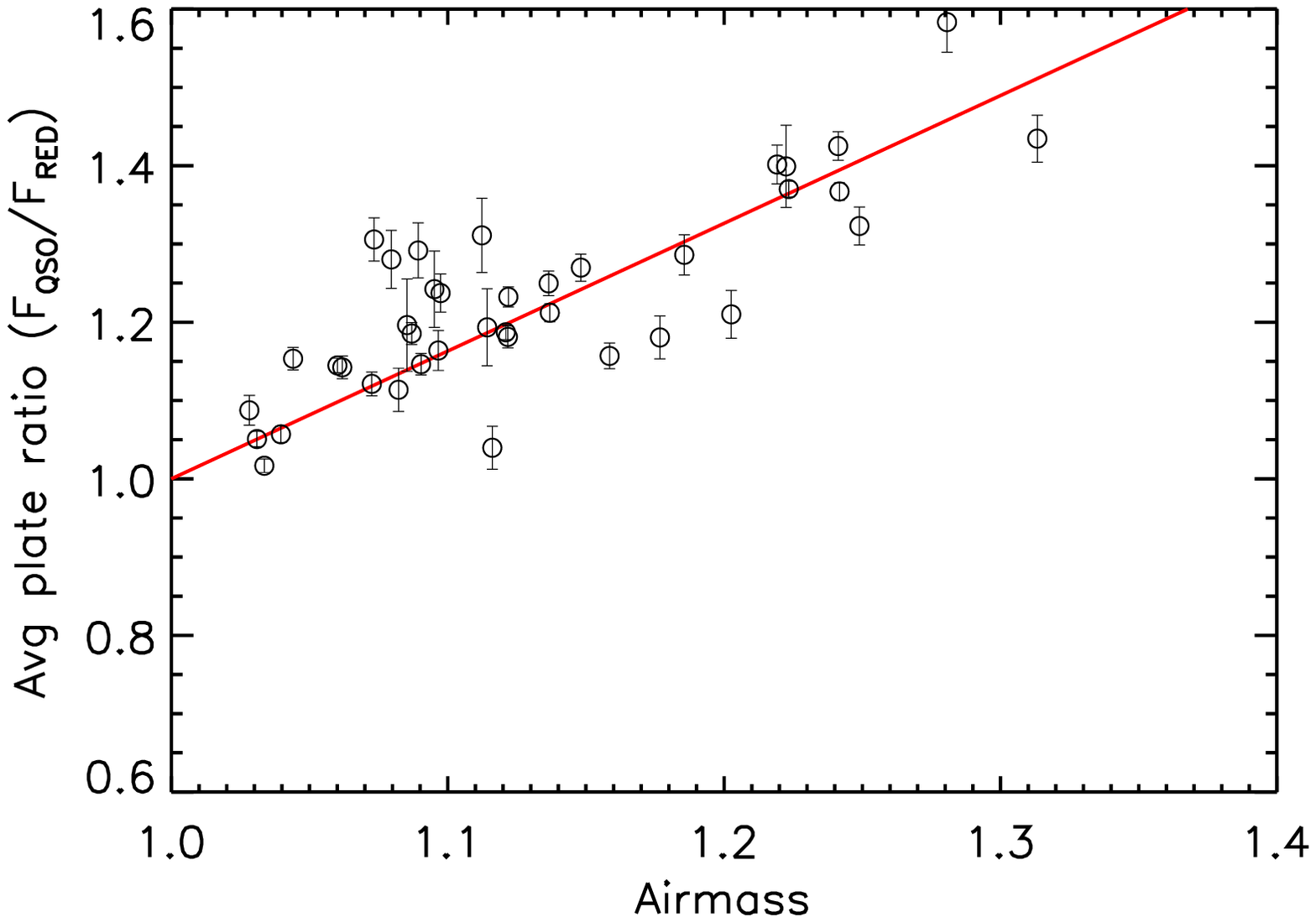}
\caption[plate]{{\bf First Panel:}  the ratio between BOSS and SDSS standard stars on plate 6149 (black) and the median ratio (red) for that plate.  {\bf Second Panel:}  A plot of median plate ratios vs. airmass at a wavelength of 7133 \AA.  Circles represent the median plate ratios at this wavelength, and the error bars represent the RMS scatter of that plate at that wavelength.}
\label{fig:fig14}
\end{figure}

Before comparing the blue reduction to the red reduction, each spectrum is smoothed with a 201 pixel weighted mean smoothing routine.  We then calculate the ratio between the spectrum of each blue focal plane standard star in the blue reduction to the same spectrum in the red reduction.  If no correction was necessary, the ratio would reveal a flat continuum valued at one.  
Figure~\ref{fig:fig14} shows the ratio from all standard stars on plate 6149 and the median ratio for that plate.
In this case, which is representative of the population of plates, uncorrected flux errors reach 20\% at long wavelengths (broadband 8500 \AA), which is similar to the trend in Figure~\ref{fig:fig13}.
This large difference would cause an incorrect interpretation of features such as quasar spectral indices.
This procedure is performed for each plate listed in Table~\ref{tab:platetable}.
We refer to the median spectral ratio for each plate as the ``plate ratio''.  Note also the range of flux errors for this single plate; it is this effect that motivated the decision to model $\sigma_{sys}$ in Section~\ref{subsec:mediansamplespectralindex}.

At each wavelength interval measured in BOSS spectra, we record the relationship between the plate ratio and the airmass.  This relationship is fit assuming a simple linear model that is constrained to intersect a ratio of one at an airmass of one, thus providing no correction where there is no ADR offset between the red and blue focal planes.  The slope of the fit with respect to airmass is recorded for each wavelength.  An example of one of these fits is presented in the second panel of Figure~\ref{fig:fig14}.  Some plates consistently provided fits that were outliers from the rest of the plates when considered at all wavelengths.  Plates that fall more than 3$\sigma$ from the rest of the sample over more than 40\% of wavelengths are rejected and this process is run once again.  A total of 12 plates are excluded by this criterion.  These accepted data are then used to correct BOSS quasar spectra in the following manner: 
\begin{equation}
F_c(\lambda) = F(\lambda)[S(\lambda)X+I(\lambda)]
\end{equation}
where $F_c$ is the corrected flux density value at a given wavelength, $F$ is the original flux density value at the given wavelength, $S$ is the slope of the fit fixed to one as discussed above, $X$ is the airmass of the plate on which the given quasar was observed, and $I$ is the intercept of the fit discussed above.  The best fit value of $S(\lambda)$ is determined at each wavelength using the sample of specially designed plates.

Table \ref{tab:platetable} lists the list of plates used for this analysis.  It also reports the dispersion of the plate which is calculated by taking the rms dispersion of all the spectra on a plate at each wavelength from observer frame 5000 \AA\ to 6000 \AA, then taking the median of those wavelength-by-wavelength dispersions.  Finally, we report the median slope of the ratio spectrum per thousand angstroms.

\begin{deluxetable}{ccrrrrr}
\centering
\tablewidth{0pt}
\tabletypesize{\footnotesize}
\tablecaption{\label{tab:platetable} Plates used in making the spectrophotometric correction to the data using the blue reduction.}
\tablehead{
\colhead{Plate Number} & \colhead{Blue Standards} & \colhead{Airmass} & \colhead{RA ($\deg$)} & \colhead{DEC ($\deg$)} & \colhead{Dispersion} & \colhead{Median Slope x $10^{3}$}
}
\startdata
    6116 & 13 & 1.121 & 354.5 & 21.41 & 0.0231 & 0.062 \\
    6118 & 11 & 1.265 & 337.3 & 21.36 & 0.0611 & 0.227 \\
    6119 & 22 & 1.281 & 339.6 & 21.16 & 0.0574 & 0.220 \\
    6120 & 23 & 1.313 & 342.2 & 20.94 & 0.0479 & 0.121 \\
    6121 & 17 & 1.249 & 344.8 & 20.71 & 0.0341 & 0.149 \\
    6125 & 15 & 1.223 & 354.1 & 19.75 & 0.0199 & 0.125 \\
    6126 & 23 & 1.040 & 356.5 & 19.54 & 0.0115 & 0.025 \\
    6128 & 12 & 1.034 & 345.7 & 18.72 & 0.0184 & 0.011 \\
    6130 & 21 & 1.121 & 350.1 & 18.30 & 0.0186 & 0.062 \\
    6131 & 22 & 1.136 & 352.3 & 18.07 & 0.0259 & 0.096 \\
    6133 & 15 & 1.085 & 357.6 & 17.73 & 0.0987 & 0.064 \\
    6136 & 21 & 1.177 & 351.1 & 16.38 & 0.0400 & 0.047 \\
    6137 & 24 & 1.044 & 353.5 & 16.19 & 0.0233 & 0.032 \\
    6139 & 21 & 1.080 & 358.7 & 15.96 & 0.0444 & 0.107 \\
    6140 & 11 & 1.219 & 345.4 & 15.02 & 0.0404 & 0.117 \\
    6141 & 20 & 1.073 & 347.2 & 14.85 & 0.0267 & 0.064 \\
    6142 & 22 & 1.137 & 349.4 & 14.67 & 0.0184 & 0.078 \\
    6146 & 15 & 1.186 & 346.6 & 13.10 & 0.0375 & 0.106 \\
    6148 & 11 & 1.097 & 351.1 & 12.79 & 0.0284 & 0.107 \\
    6149 & 23 & 1.122 & 353.6 & 12.65 & 0.0177 & 0.120 \\
    6157 & 22 & 1.097 & 352.9 & 10.94 & 0.0454 & 0.101 \\
    6161 & 15 & 1.112 & 355.2 & 7.32 & 0.0634 & 0.083 \\
    6162 & 17 & 1.089 & 346.9 & 9.43 & 0.0639 & 0.080 \\
    6163 & 16 & 1.090 & 349.4 & 9.32 & 0.0187 & 0.075 \\
    6165 & 22 & 1.095 & 353.8 & 9.11 & 0.0746 & 0.069 \\
    6168 & 11 & 1.116 & 347.9 & 7.55 & 0.0407 & 0.027 \\
    6169 & 14 & 1.114 & 350.2 & 7.46 & 0.0745 & 0.11 \\
    6176 & 15 & 1.073 & 1.6 & 12.49 & 0.0396 & 0.065 \\
    6185 & 14 & 1.062 & 5.6 & 16.19 & 0.0174 & 0.0863 \\
    6189 & 11 & 1.082 & 5.4 & 10.83 & 0.0406 & 0.0531 \\
    6198 & 19 & 1.203 & 10.3 & 16.58 & 0.0385 & 0.0986 \\
    6290 & 27 & 1.060 & 341.9 & 24.81 & 0.0166 & 0.0436 \\
    6291 & 20 & 1.222 & 338.5 & 23.21 & 0.0699 & 0.130 \\
    6296 & 23 & 1.159 & 340.4 & 26.92 & 0.0293 & 0.0608 \\
    6297 & 24 & 1.148 & 338.0 & 27.16 & 0.0244 & 0.0799 \\
    6298 & 27 & 1.241 & 339.7 & 25.06 & 0.0258 & 0.119 \\
    6308 & 22 & 1.242 & 343.3 & 22.74 & 0.0172 & 0.0867 \\
    6735 & 18 & 1.087 & 232.7 & 46.46 & 0.0182 & 0.0594 \\
    6737 & 16 & 1.028 & 223.7 & 46.19 & 0.0275 & -0.0141 \\
    6752 & 21 & 1.031 & 217.2 & 46.82 & 0.0197 & 0.0206 \\
\enddata
\end{deluxetable}

\subsubsection{{\bf A.2.2} Tests to the Initial Correction}\label{subsubsec:initial_test}

To test the quality of the initial correction developed in Appendix~\ref{subsec:correct_specphot_error}, we rerun the analysis presented in Appendix~\ref{subsec:specphot_err} on the data after the initial correction is applied.  The results of these tests are presented in Figure~\ref{fig:fig13} and in the middle row of panels in Figure~\ref{fig:fig3}.

Figure~\ref{fig:fig13} demonstrates that after the correction, the ratio of BOSS to SDSS spectra is close to one for stellar contaminants observed in both programs over most of the spectrum, but falls to 0.9 at short wavelengths.  The average slope of this curve after the correction is 0.0092 per thousand angstroms in the 5000 \AA\ to 7000~\AA\ range.

The second row of panels in Figure~\ref{fig:fig3} display the magnitude difference between SDSS imaging photometry and BOSS synthetic photometry after applying the correction.  The results of this linear fit are reported in Table~\ref{tab:correction_table}.
Evaluating the results at the median airmass of $X = 1.095$, we expect a bias in color in $g-r$ of $-0.033$ and in $r-i$ of $-0.012$ magnitudes after the correction.  This is a significant improvement over the differences before the correction.

One may expect the fit in each of the $g$, $r$, and $i$ bands to have the same value at an airmass of one regardless of the spectrophotometric correction, as there is no correction at an airmass of one.  However, this is only true for objects with a flat Spectral Energy Distribution (SED).  In this case, the complex structure of the stellar SED, as well as variations in SED between stars, adds a nonlinearity to the correction at higher airmass which changes the slope and intercept of the fit by a small amount.  This is the origin of the 0.02 magnitude shifts in the intercept of the fit in the $r$ band and 0.03 magnitudes in the $i$ band.

\subsubsection{{\bf A.2.3} Refining the Initial Correction to Minimize Color Bias}\label{subsubsec:color_bias}

Due to the small number of plates available from the Blue Reduction, we expect statistical uncertainty that is larger than desired for the much larger sample of plates covering the quasar spectra used in the composite analysis.
The residual errors are demonstrated in the second row of Figure~\ref{fig:fig3} as a population with non-zero slope.
The spectrophotometric correction appears to overcorrect the bias introduced by fiber offsets.

We assume that the wavelength dependence of the correction is correct, but the amplitude is too high.
We next seek to minimize the broad-band errors by keeping the shape of the correction but reducing its magnitude.
The new modified correction is analyzed in the same manner as in Appendix~\ref{subsec:specphot_err}.  In particular, the offset between the PSF magnitudes from SDSS and the corrected BOSS synthetic magnitudes at the mean airmass of 1.095 is recorded at each reduced correction.  Since one of the main results of this work is a measurement of the spectral index of the quasar sample, we choose to optimize the correction such that the colors measured in BOSS (using synthetic photometry) most closely match the colors measured in SDSS at the median airmass.  We record the difference in $g-r$, $r-i$, or $g-i$ after scaling the slope ($S(\lambda)$) across all wavelengths by the same constant factor and identify the scale factor that minimizes the largest color bias (in $g-r$, $r-i$, or $g-i$).
The best fit is a scale factor of  0.83, at which the color bias in $r-i$ is equal to 0.0186 magnitudes.  The results after applying this consistent scale factor  are shown in Figure~\ref{fig:fig3} in the third row of panels.
These results include the correction outlined in Appendix~\ref{subsec:correct_specphot_error}, and the correction amplitude adjustment of 0.83.  Refer to Table~\ref{tab:correction_table} for a summary of the color bias with no correction, with the initial correction, and with the final correction.

\subsubsection{Performance of the Final Correction}\label{subsubsec:finalcorrection}

\begin{figure}[h]
\centering
\includegraphics[width=0.5\textwidth]{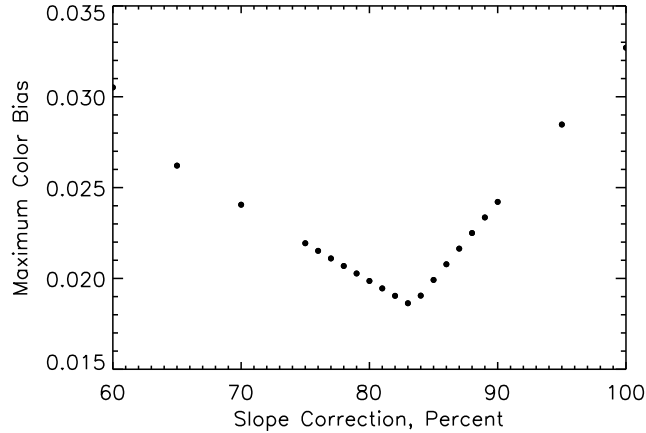}
\caption{The maximum color bias is displayed as a function of the arbitrary scaling factor applied to the slope of the correction.  A slope adjustment factor of 0.83 minimizes the maximum offset between PSF - BOSS data for the various filters.}
\label{fig:fig15}
\end{figure}

\begin{figure}[h]
\centering
\includegraphics[width=0.45\textwidth]{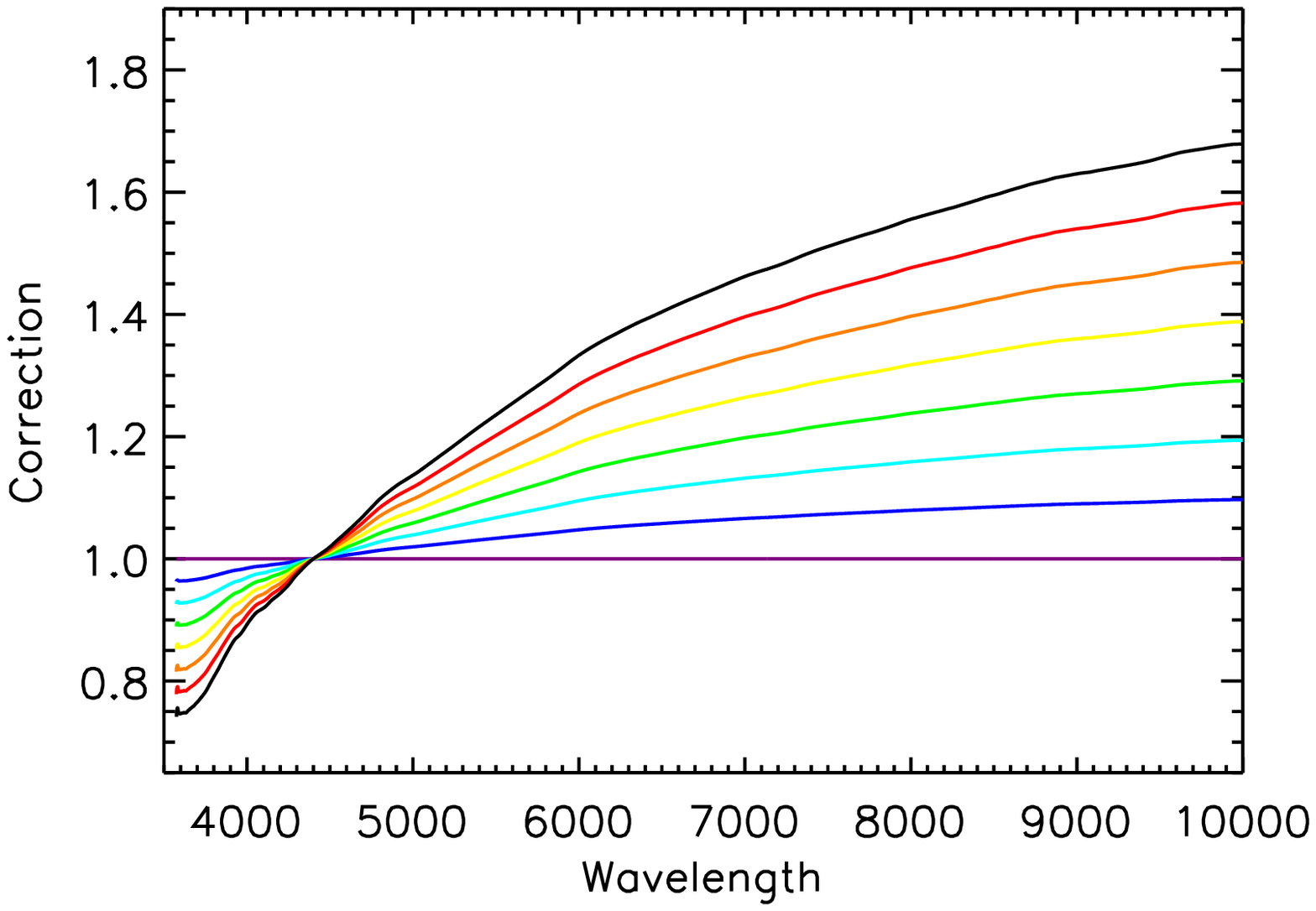}
\includegraphics[width=0.45\textwidth]{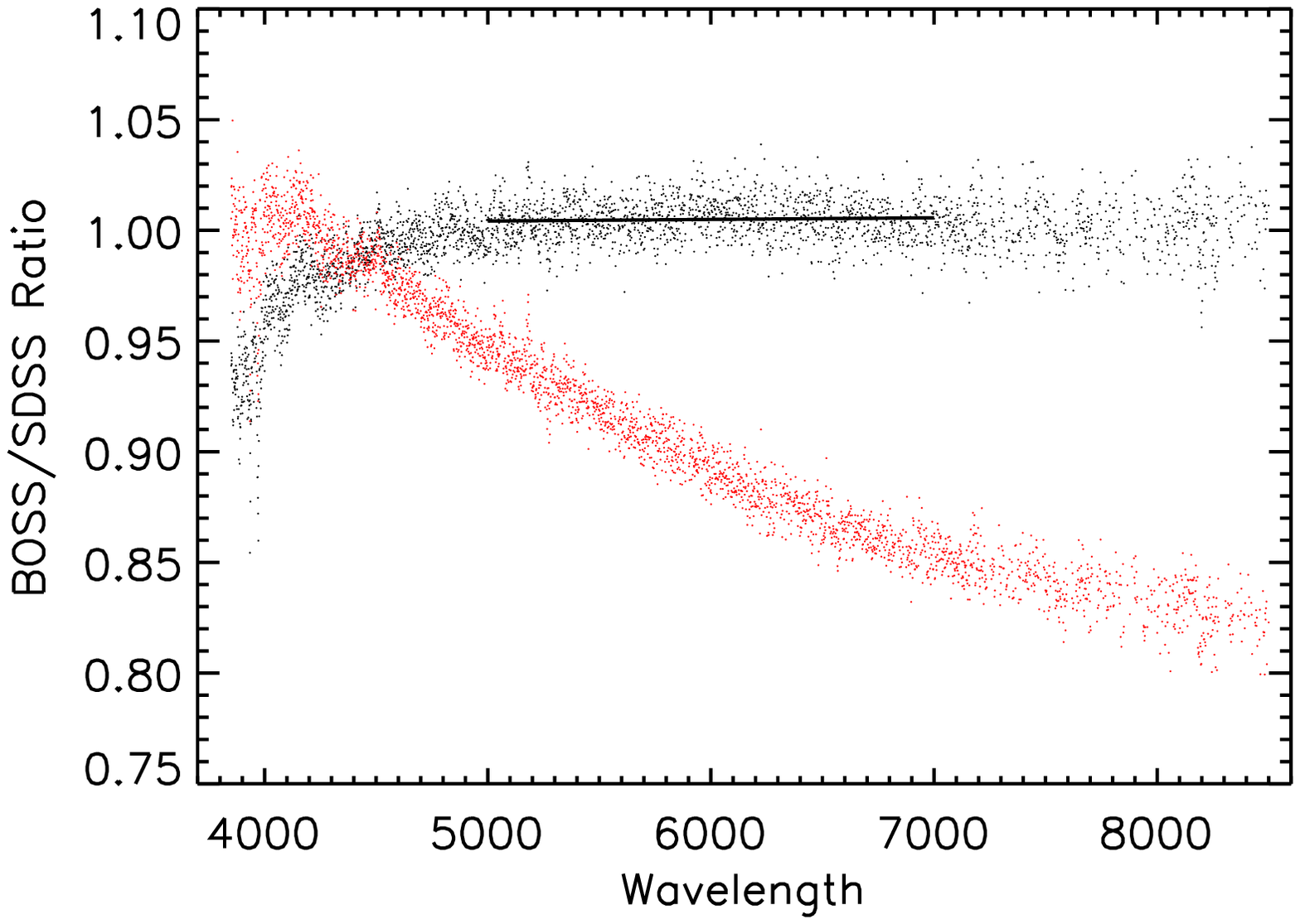}
\caption[fanplot2]{{\bf First Panel:} Example corrections for spectra at different airmasses using the final spectrophotometric correction laid out in Appendix~\ref{subsubsec:finalcorrection}.  Airmasses of bottom (purple) to top (black) lines: 1.00, 1.05, 1.10, 1.15, 1.20, 1.25, 1.30, 1.35.  {\bf Second Panel:} The horizontal axis is the wavelength in angstroms; the vertical axis shows the ratio between BOSS standard star spectra and their counterparts in SDSS.  The median ratio between uncorrected BOSS and SDSS spectra is shown in red, while the median ratio between final corrected BOSS and SDSS spectra is shown in black (see Appendix~\ref{subsubsec:finalcorrection}).  The slope of the fit to the corrected ratio, shown as the black line, is 0.00073 per thousand angstroms between 5000 \AA\ and 7000 \AA, and the fit at 6000 \AA\ is 1.005.}
\label{fig:fig16}
\end{figure}

The correction model evaluated at eight different airmasses is presented in Figure~\ref{fig:fig16}.  At an airmass of one, there is no correction, by design; as the airmass increases, the correction becomes more significant.  The correction suppresses flux at wavelengths $\lambda < 4500$ \AA\  and increases flux at longer wavelengths.  At the average airmass of the quasar sample of $X = 1.095$, the correction at 4000 \AA\ is 0.96 and the correction at 8500 \AA\ is 1.21.

After the correction is applied, the ratio between BOSS and SDSS standard stars is close to one, as shown in Figure~\ref{fig:fig16}.  The slope of the ratio between 5000 \AA\ and 7000 \AA\ is 0.00073 per thousand angstroms.  Over this wavelength range, the ratio between BOSS and SDSS is one to the sub-percent level.
In general we expect a bias in color in $g-r$ of $-0.018$ magnitudes and in $r-i$ of +0.019 magnitudes after this final correction, a significant improvement over the previous results.

\section{{\bf B.} BOSS/SDSS Ratio}\label{sec:boss_sdss_ratio}

\begin{figure}[h]
\centering
\includegraphics[width=0.45\textwidth]{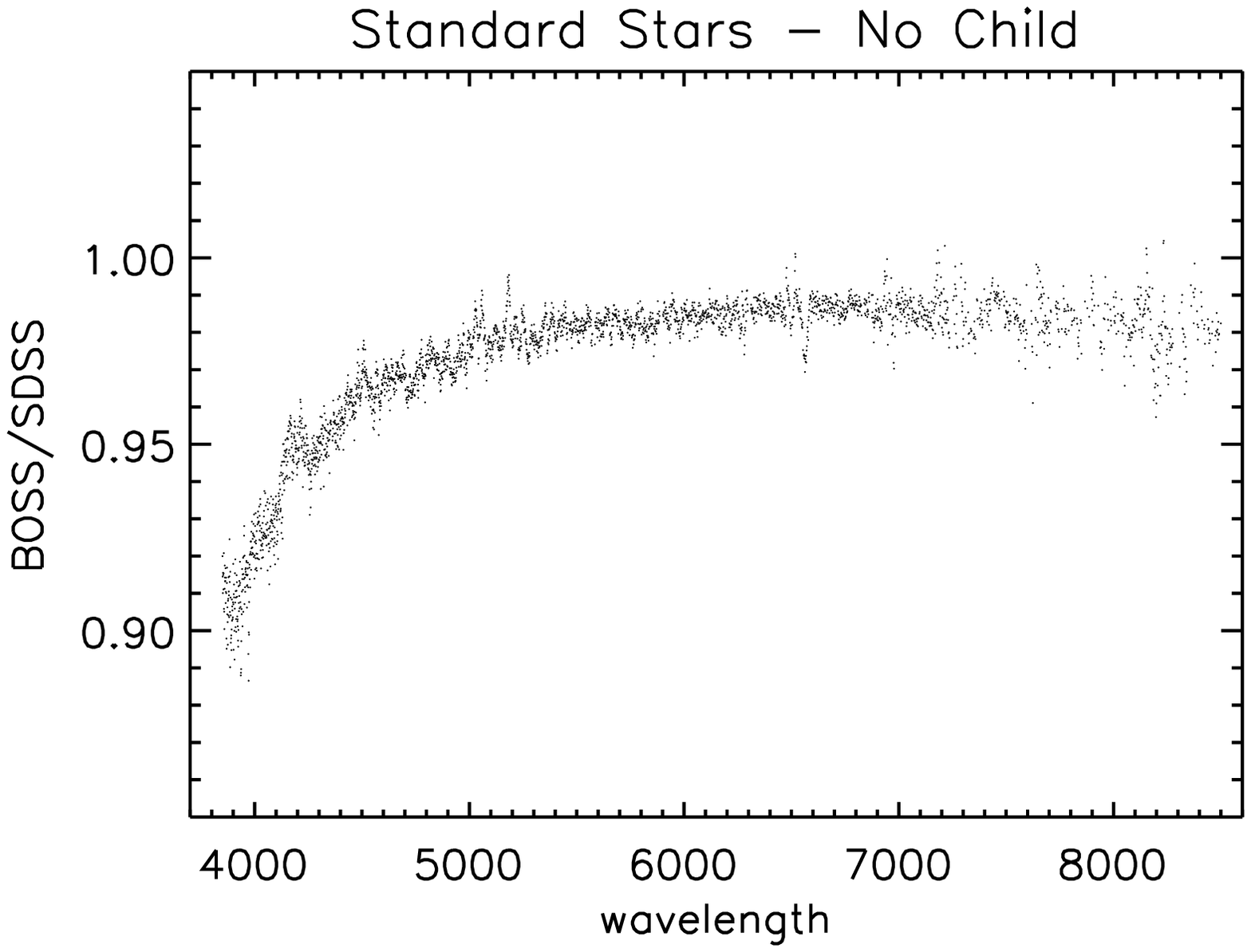}
\includegraphics[width=0.45\textwidth]{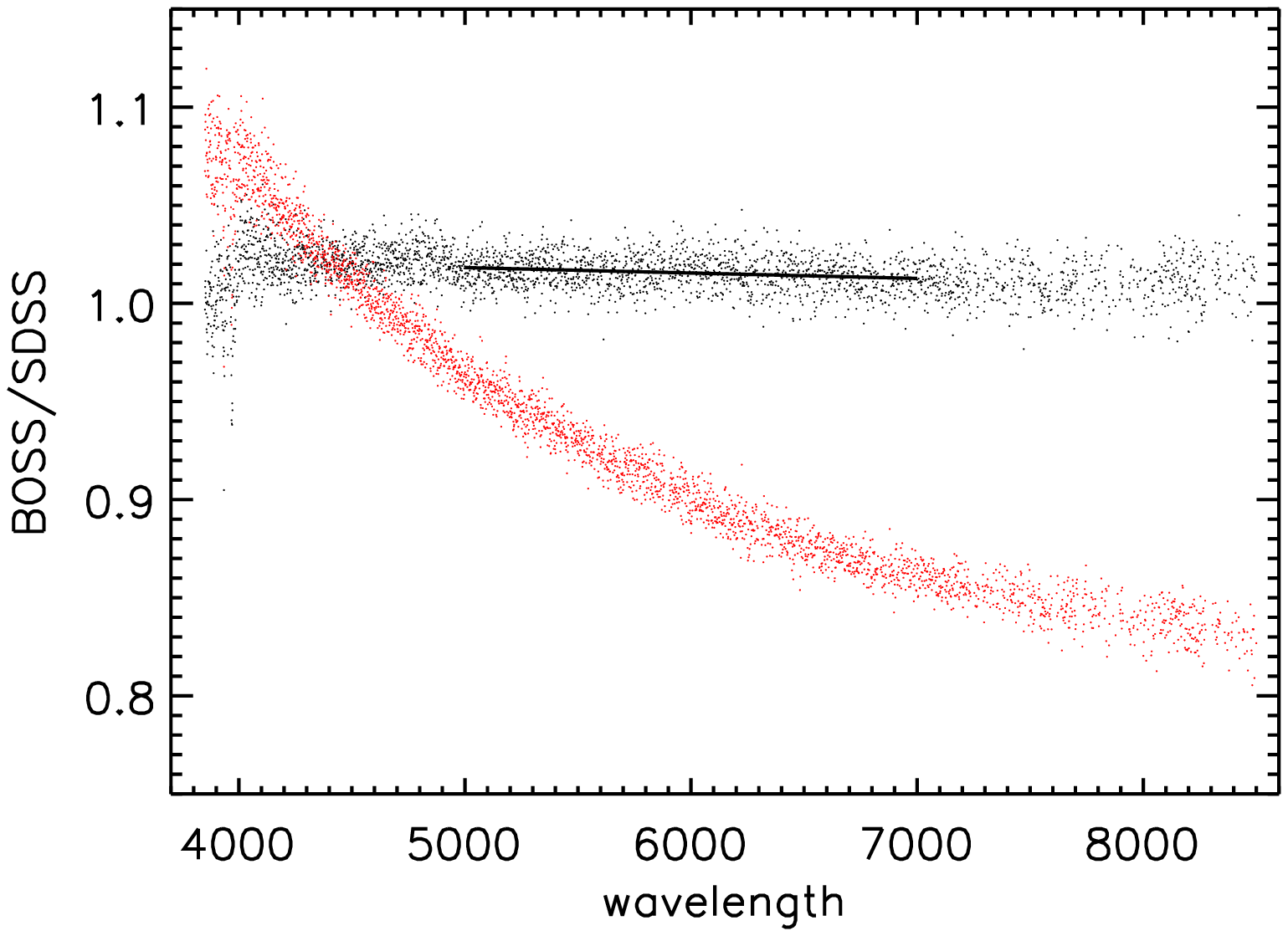}
\caption[ic]{{\bf First Panel:} The ratio between BOSS and SDSS standard star spectra for standard stars at the red focal plane holes.  {\bf Second Panel:} The BOSS/SDSS ratio after the ratio spectrum on the right has been divided out for two cases: before and after the spectrophotometric correction has been applied, with red showing before and black showing after.}
\label{fig:fig17}
\end{figure}

In Figures~\ref{fig:fig13} and \ref{fig:fig16}, a residual offset is visible between the final BOSS corrections and the SDSS spectra at wavelengths below 4500 \AA; the offset exceeds 5\% at 3850 \AA.  To assess the origin of this residual, we derive a median spectrum of the ratio between standard stars observed in BOSS and SDSS.  We exclude objects near other photometrically-detected objects (objects with the CHILD flag set) as SDSS fibers are larger than BOSS fibers and may have included extraneous light from other objects.  Even though these stars provide the reference calibration for both BOSS and SDSS, Figure~\ref{fig:fig17} shows the same trend in the BOSS/SDSS spectral ratio below 4500 \AA\ as presented in Figures~\ref{fig:fig13} and \ref{fig:fig16}. This result indicates the observed residual is not caused by the correction; instead, it is the result of an intrinsic difference between SDSS and BOSS flux calibrations.  

\begin{figure}[h]
\centering
\includegraphics[width=0.6\textwidth]{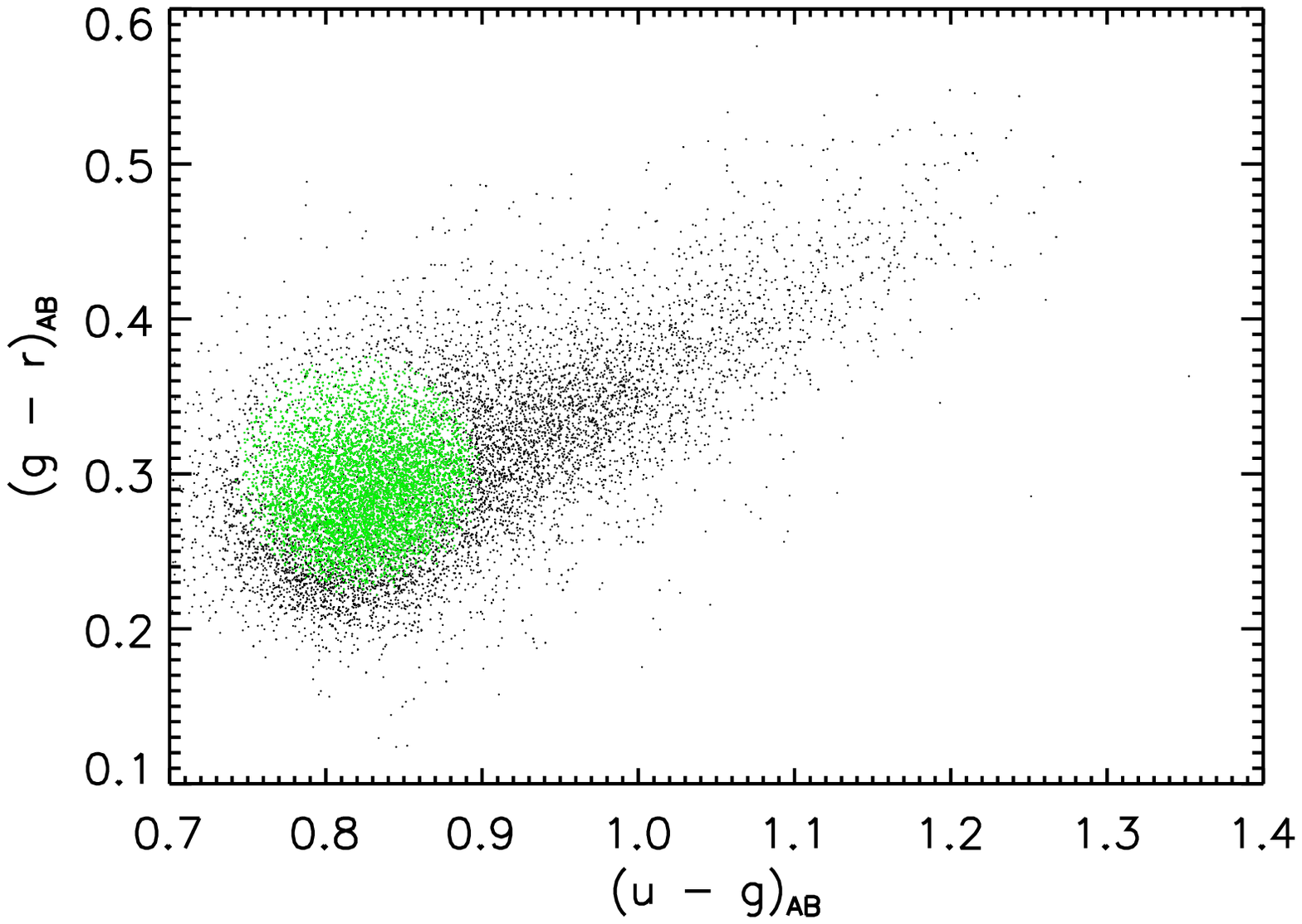}
\includegraphics[width=0.6\textwidth]{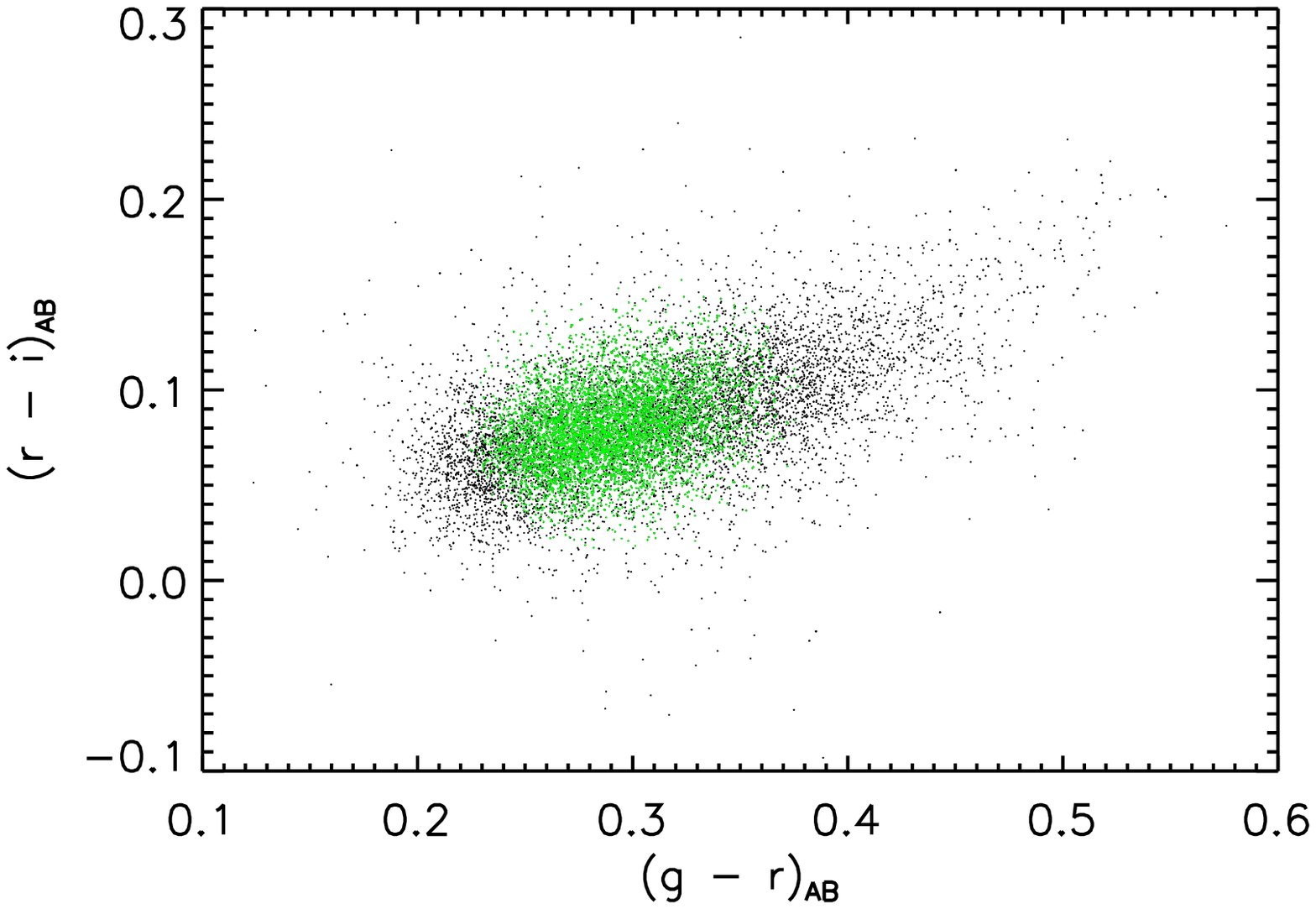}
\caption[ic]{Color-color diagrams of the standard stars in SDSS.  SDSS stars are shown in black.  SDSS stars which also make the color-space cut to be considered as BOSS standards are shown in green.}
\label{fig:fig18}
\end{figure}

Stellar spectral models, such as the Kurucz models, are sensitive to $T_{\rm eff}$ and [Fe/H], especially in the near UV continuum \citep{prieto00a}.  Good standard star target selection must account for this effect by using a set of standard stars with similar properties.  We examine the color-color distribution of SDSS and BOSS standard stars and find that the BOSS standard stars are targeted in a more discriminating manner, restricting their range in color-color space compared to SDSS standard stars, as shown in Figure~\ref{fig:fig18}.  We hypothesize this residual offset between BOSS and SDSS flux calibration in the near UV is caused by the sensitivity of Kurucz models to differences in $T_{eff}$ and [Fe/H] of the standard stars coupled to the differing populations of stars selected between SDSS and BOSS.  Improvements to the stellar models for calibration stars is beyond the scope of this work and left to a future effort.

\begin{figure}[h]
\centering
\includegraphics[width=0.6\textwidth]{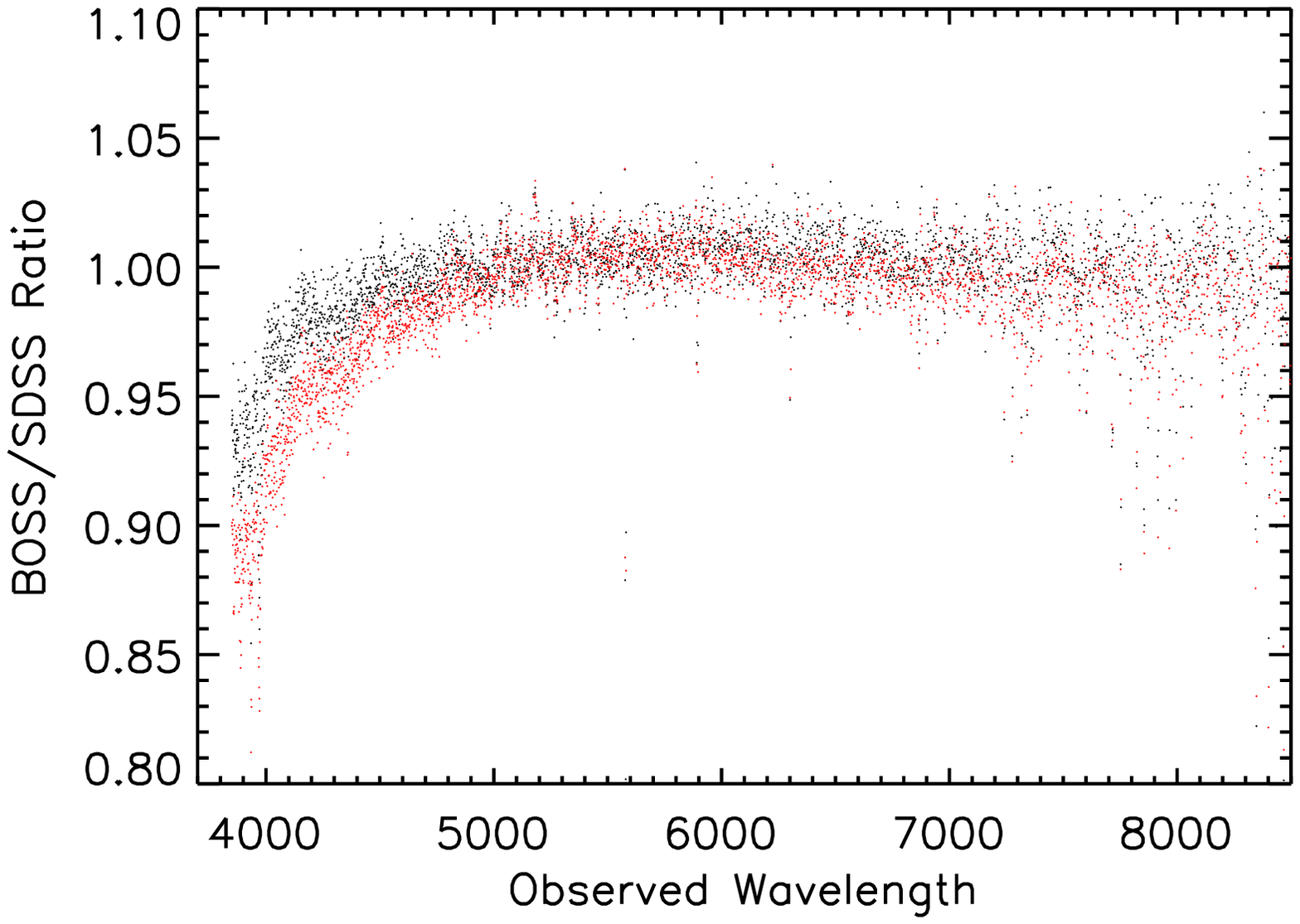}
\caption[ic]{The BOSS flux of stellar contaminants relative to the SDSS flux of the same objects as a function of wavelength.  The median ratio between BOSS and SDSS spectra corrected using the method in this work is shown in black, while the median ratio between corrected BOSS and SDSS spectra using the correction in \citet{margala15a} is indicated in red.}
\label{fig:fig19}
\end{figure}

\citet{margala15a} takes an independent approach to correcting spectrophotometric calibration of BOSS quasars.  In that analysis, the effects of ADR across the focal plane is predicted from first principles given optimal guiding, seeing estimates, and plate design.  We apply that correction to the same set of stellar contaminant spectra used in Appendix~\ref{subsubsec:spectral_comparison}, and create a BOSS/SDSS median ratio spectrum similar to the one shown in Figure~\ref{fig:fig16}.  We compare our correction and the \citet{margala15a} correction in Figure~\ref{fig:fig19}.  The two deviate most significantly at 3850 \AA; the BOSS flux from our analysis is suppressed by 7\% relative to SDSS while \citet{margala15a} is suppressed by 11\%.  Over the wavelength range $4800 $\AA\ $ \leq \lambda \leq 8200 $\AA, the two corrections agree to better than 1\%.

\bibliographystyle{mnras}
\bibliography{archive}

\begin{thebibliography}{60}
\expandafter\ifx\csname natexlab\endcsname\relax\def\natexlab#1{#1}\fi

\bibitem[{Aihara} et~al.(2011{\natexlab{a}}){Aihara}, {Allende Prieto}, {An}
  et~al.]{aihara11a}
{Aihara} H., {Allende Prieto} C., {An} D., et~al., 2011{\natexlab{a}}, \apjs,
  193, 29

\bibitem[{Aihara} et~al.(2011{\natexlab{b}}){Aihara}, {Allende Prieto}, {An}
  et~al.]{agol11a}
{Aihara} H., {Allende Prieto} C., {An} D., et~al., 2011{\natexlab{b}}, {The
  Eighth Data Release of the Sloan Digital Sky Survey: First Data from
  SDSS-III}

\bibitem[{Alam} et~al.(2015){Alam}, {Albareti}, {Allende Prieto}
  et~al.]{alam15a}
{Alam} S., {Albareti} F.~D., {Allende Prieto} C., et~al., 2015, \apjs, 219, 12

\bibitem[{Allende Prieto} \& {Lambert}(2000)]{prieto00a}
{Allende Prieto} C., {Lambert} D.~L., 2000, \aj, 119, 2445

\bibitem[{Anderson} et~al.(2014){Anderson}, {Aubourg}, {Bailey}
  et~al.]{anderson14a}
{Anderson} L., {Aubourg} {\'E}., {Bailey} S., et~al., 2014, \mnras, 441, 24

\bibitem[{Baldwin}(1977)]{baldwin77a}
{Baldwin} J.~A., 1977, \apj, 214, 679

\bibitem[{Becker} et~al.(2013){Becker}, {Hewett}, {Worseck} \&
  {Prochaska}]{becker13a}
{Becker} G.~D., {Hewett} P.~C., {Worseck} G., {Prochaska} J.~X., 2013, \mnras,
  430, 2067

\bibitem[{Bolton} et~al.(2012){Bolton}, {Schlegel}, {Aubourg}
  et~al.]{bolton12a}
{Bolton} A.~S., {Schlegel} D.~J., {Aubourg} {\'E}., et~al., 2012, \aj, 144, 144

\bibitem[{Bovy} et~al.(2011){Bovy}, {Hennawi}, {Hogg} et~al.]{bovy11a}
{Bovy} J., {Hennawi} J.~F., {Hogg} D.~W., et~al., 2011, \apj, 729, 141

\bibitem[{Brotherton} et~al.(2001){Brotherton}, {Tran}, {Becker}, {Gregg},
  {Laurent-Muehleisen} \& {White}]{brotherton01a}
{Brotherton} M.~S., {Tran} H.~D., {Becker} R.~H., {Gregg} M.~D.,
  {Laurent-Muehleisen} S.~A., {White} R.~L., 2001, \apj, 546, 775

\bibitem[{Bruhweiler} \& {Verner}(2008)]{bruhweiler08a}
{Bruhweiler} F., {Verner} E., 2008, \apj, 675, 83

\bibitem[{Dawson} et~al.(2015){Dawson}, {Kneib}, {Percival} et~al.]{dawson15a}
{Dawson} K.~S., {Kneib} J.-P., {Percival} W.~J., et~al., 2015, arXiv:1508.04473

\bibitem[{Dawson} et~al.(2013){Dawson}, {Schlegel}, {Ahn} et~al.]{dawson13a}
{Dawson} K.~S., {Schlegel} D.~J., {Ahn} C.~P., et~al., 2013, \aj, 145, 10

\bibitem[{Delubac} et~al.(2015){Delubac}, {Bautista}, {Busca}
  et~al.]{delubac15a}
{Delubac} T., {Bautista} J.~E., {Busca} N.~G., et~al., 2015, \aap, 574, A59

\bibitem[{Eisenstein} et~al.(2011){Eisenstein}, {Weinberg}, {Agol}
  et~al.]{eisenstein11a}
{Eisenstein} D.~J., {Weinberg} D.~H., {Agol} E., et~al., 2011, \aj, 142, 72

\bibitem[{Fan}(1999)]{fan99b}
{Fan} X., 1999, \aj, 117, 2528

\bibitem[{Faucher-Gigu{\`e}re} et~al.(2008){Faucher-Gigu{\`e}re}, {Prochaska},
  {Lidz}, {Hernquist} \& {Zaldarriaga}]{faucher08a}
{Faucher-Gigu{\`e}re} C.-A., {Prochaska} J.~X., {Lidz} A., {Hernquist} L.,
  {Zaldarriaga} M., 2008, \apj, 681, 831

\bibitem[{Fitzpatrick}(1999)]{fitzpatrick99}
{Fitzpatrick} E.~L., 1999, \pasp, 111, 63

\bibitem[{Foltz} et~al.(1989){Foltz}, {Chaffee}, {Hewett}, {Weymann},
  {Anderson} \& {MacAlpine}]{foltz89}
{Foltz} C.~B., {Chaffee} F.~H., {Hewett} P.~C., {Weymann} R.~J., {Anderson}
  S.~F., {MacAlpine} G.~M., 1989, \aj, 98, 1959

\bibitem[{Francis} et~al.(1991){Francis}, {Hewett}, {Foltz}, {Chaffee},
  {Weymann} \& {Morris}]{francis91a}
{Francis} P.~J., {Hewett} P.~C., {Foltz} C.~B., {Chaffee} F.~H., {Weymann}
  R.~J., {Morris} S.~L., 1991, \apj, 373, 465

\bibitem[{Fukugita} et~al.(1996){Fukugita}, {Ichikawa}, {Gunn}, {Doi},
  {Shimasaku} \& {Schneider}]{fukugita96a}
{Fukugita} M., {Ichikawa} T., {Gunn} J.~E., {Doi} M., {Shimasaku} K.,
  {Schneider} D.~P., 1996, \aj, 111, 1748

\bibitem[{Green} et~al.(2012){Green}, {Froning}, {Osterman} et~al.]{green12a}
{Green} J.~C., {Froning} C.~S., {Osterman} S., et~al., 2012, \apj, 744, 60

\bibitem[{Grier} et~al.(2015){Grier}, {Hall}, {Brandt} et~al.]{grier15a}
{Grier} C.~J., {Hall} P.~B., {Brandt} W.~N., et~al., 2015, \apj, 806, 111

\bibitem[{Gunn} et~al.(2006){Gunn}, {Siegmund}, {Mannery} et~al.]{gunn06a}
{Gunn} J.~E., {Siegmund} W.~A., {Mannery} E.~J., et~al., 2006, \aj, 131, 2332

\bibitem[{Hinshaw} et~al.(2013){Hinshaw}, {Larson}, {Komatsu}
  et~al.]{hinshaw13a}
{Hinshaw} G., {Larson} D., {Komatsu} E., et~al., 2013, \apjs, 208, 19

\bibitem[{Hogg}(2001)]{hogg01a}
{Hogg} D.~W., 2001, in { The Extragalactic Infrared Background and its
  Cosmological Implications\/}, edited by M.~{Harwit}, M.~G. {Hauser}, vol. 204
  of { IAU Symposium\/},  209

\bibitem[{Ivezi{\'c}} et~al.(2004){Ivezi{\'c}}, {Lupton}, {Schlegel}
  et~al.]{ivezic04a}
{Ivezi{\'c}} {\v Z}., {Lupton} R.~H., {Schlegel} D., et~al., 2004,
  Astronomische Nachrichten, 325, 583

\bibitem[{Kirkpatrick} et~al.(2011){Kirkpatrick}, {Schlegel}, {Ross}
  et~al.]{kirkpatrick11a}
{Kirkpatrick} J.~A., {Schlegel} D.~J., {Ross} N.~P., et~al., 2011, \apj, 743,
  125

\bibitem[{Levi} et~al.(2013){Levi}, {Bebek}, {Beers} et~al.]{levi13a}
{Levi} M., {Bebek} C., {Beers} T., et~al., 2013, arXiv:1308.0847

\bibitem[{Lupton} et~al.(2001){Lupton}, {Gunn}, {Ivezi{\'c}}, {Knapp}, {Kent}
  \& {Yasuda}]{lupton01a}
{Lupton} R., {Gunn} J.~E., {Ivezi{\'c}} Z., {Knapp} G.~R., {Kent} S., {Yasuda}
  N., 2001, in { Astronomical Data Analysis Software and Systems X\/}, edited
  by {F.~R.~Harnden Jr., F.~A.~Primini, \& H.~E.~Payne}, vol. 238 of {
  Astronomical Society of the Pacific Conference Series\/},  269

\bibitem[{Margala} et~al.(2015){Margala}, {Kirkby}, {Dawson}, {Bailey},
  {Blanton} \& {Schneider}]{margala15a}
{Margala} D., {Kirkby} D., {Dawson} K., {Bailey} S., {Blanton} M., {Schneider}
  D.~P., 2015, arXiv:1506.04790

\bibitem[{Myers} et~al.(2015){Myers}, {Palanque-Delabrouille}, {Prakash}
  et~al.]{myers15a}
{Myers} A.~D., {Palanque-Delabrouille} N., {Prakash} A., et~al., 2015, \apjs,
  221, 27

\bibitem[{Nagao} et~al.(2006){Nagao}, {Marconi} \& {Maiolino}]{nagao06}
{Nagao} T., {Marconi} A., {Maiolino} R., 2006, \aap, 447, 157

\bibitem[{Noterdaeme} et~al.(2012){Noterdaeme}, {Petitjean}, {Carithers}
  et~al.]{noterdaeme12a}
{Noterdaeme} P., {Petitjean} P., {Carithers} W.~C., et~al., 2012, \aap, 547, L1

\bibitem[{Oke} \& {Gunn}(1983)]{oke83a}
{Oke} J.~B., {Gunn} J.~E., 1983, \apj, 266, 713

\bibitem[{Padmanabhan} et~al.(2008){Padmanabhan}, {Schlegel}, {Finkbeiner}
  et~al.]{padmanabhan08a}
{Padmanabhan} N., {Schlegel} D.~J., {Finkbeiner} D.~P., et~al., 2008, \apj,
  674, 1217

\bibitem[{Palanque-Delabrouille} et~al.(2015){Palanque-Delabrouille},
  {Magneville}, {Y{\`e}che} et~al.]{palanque-delabrouille15a}
{Palanque-Delabrouille} N., {Magneville} C., {Y{\`e}che} C., et~al., 2015,
  arXiv:1509.05607

\bibitem[{P{\^a}ris} et~al.(2012){P{\^a}ris}, {Petitjean}, {Aubourg}
  et~al.]{paris12a}
{P{\^a}ris} I., {Petitjean} P., {Aubourg} {\'E}., et~al., 2012, \aap, 548, A66

\bibitem[{P{\^a}ris} et~al.(2014){P{\^a}ris}, {Petitjean}, {Aubourg}
  et~al.]{paris14a}
{P{\^a}ris} I., {Petitjean} P., {Aubourg} {\'E}., et~al., 2014, \aap, 563, A54

\bibitem[{Pereyra} et~al.(2006){Pereyra}, {Vanden Berk}, {Turnshek}
  et~al.]{pereyra06}
{Pereyra} N.~A., {Vanden Berk} D.~E., {Turnshek} D.~A., et~al., 2006, \apj,
  642, 87

\bibitem[{Pier} et~al.(2003){Pier}, {Munn}, {Hindsley} et~al.]{pier03a}
{Pier} J.~R., {Munn} J.~A., {Hindsley} R.~B., et~al., 2003, \aj, 125, 1559

\bibitem[{Planck Collaboration} et~al.(2015){Planck Collaboration}, {Ade},
  {Aghanim} et~al.]{ade15a}
{Planck Collaboration}, {Ade} P.~A.~R., {Aghanim} N., et~al., 2015,
  arXiv:1502.01589

\bibitem[{Richards} et~al.(2002){Richards}, {Fan}, {Newberg}
  et~al.]{richards02a}
{Richards} G.~T., {Fan} X., {Newberg} H.~J., et~al., 2002, \aj, 123, 2945

\bibitem[{Ross} et~al.(2013){Ross}, {McGreer}, {White} et~al.]{ross13a}
{Ross} N.~P., {McGreer} I.~D., {White} M., et~al., 2013, \apj, 773, 14

\bibitem[{Ross} et~al.(2012){Ross}, {Myers}, {Sheldon} et~al.]{ross12a}
{Ross} N.~P., {Myers} A.~D., {Sheldon} E.~S., et~al., 2012, \apjs, 199, 3

\bibitem[{Ruan} et~al.(2014){Ruan}, {Anderson}, {Dexter} \& {Agol}]{ruan14}
{Ruan} J.~J., {Anderson} S.~F., {Dexter} J., {Agol} E., 2014, \apj, 783, 105

\bibitem[{Sakata} et~al.(2011){Sakata}, {Morokuma}, {Minezaki}
  et~al.]{sakata11}
{Sakata} Y., {Morokuma} T., {Minezaki} T., et~al., 2011, \apj, 731, 50

\bibitem[{Schlegel} et~al.(1998){Schlegel}, {Finkbeiner} \&
  {Davis}]{schlegel98a}
{Schlegel} D.~J., {Finkbeiner} D.~P., {Davis} M., 1998, \apj, 500, 525

\bibitem[{Shull} et~al.(2012){Shull}, {Stevans} \& {Danforth}]{shull12a}
{Shull} J.~M., {Stevans} M., {Danforth} C.~W., 2012, \apj, 752, 162

\bibitem[{Smee} et~al.(2013){Smee}, {Gunn}, {Uomoto} et~al.]{smee13a}
{Smee} S.~A., {Gunn} J.~E., {Uomoto} A., et~al., 2013, \aj, 146, 32

\bibitem[{Smith} et~al.(2002){Smith}, {Tucker}, {Kent} et~al.]{smith02a}
{Smith} J.~A., {Tucker} D.~L., {Kent} S., et~al., 2002, \aj, 123, 2121

\bibitem[{Stevans} et~al.(2014){Stevans}, {Shull}, {Danforth} \&
  {Tilton}]{stevans14a}
{Stevans} M.~L., {Shull} J.~M., {Danforth} C.~W., {Tilton} E.~M., 2014, \apj,
  794, 75

\bibitem[{Stoughton} et~al.(2002){Stoughton}, {Lupton}, {Bernardi}
  et~al.]{stoughton02a}
{Stoughton} C., {Lupton} R.~H., {Bernardi} M., et~al., 2002, \aj, 123, 485

\bibitem[{Tucker} et~al.(2006){Tucker}, {Kent}, {Richmond} et~al.]{tucker06a}
{Tucker} D.~L., {Kent} S., {Richmond} M.~W., et~al., 2006, Astronomische
  Nachrichten, 327, 821

\bibitem[{Vanden Berk} et~al.(2001){Vanden Berk}, {Richards}, {Bauer}
  et~al.]{vandenberk01a}
{Vanden Berk} D.~E., {Richards} G.~T., {Bauer} A., et~al., 2001, \aj, 122, 549

\bibitem[{Vanden Berk} et~al.(2000){Vanden Berk}, {Richards} \& {SDSS
  Collaboration}]{vandenberk01b}
{Vanden Berk} D.~E., {Richards} G.~T., {SDSS Collaboration}, 2000, in {
  American Astronomical Society Meeting Abstracts \#196\/}, vol.~32 of {
  Bulletin of the American Astronomical Society\/},  751

\bibitem[{Vestergaard} \& {Wilkes}(2001)]{vestergaard01a}
{Vestergaard} M., {Wilkes} B.~J., 2001, \apjs, 134, 1

\bibitem[{Wilhite} et~al.(2005){Wilhite}, {Vanden Berk}, {Kron}
  et~al.]{wilhite05a}
{Wilhite} B.~C., {Vanden Berk} D.~E., {Kron} R.~G., et~al., 2005, \apj, 633,
  638

\bibitem[{Wills} et~al.(1985){Wills}, {Netzer} \& {Wills}]{wills85a}
{Wills} B.~J., {Netzer} H., {Wills} D., 1985, \apj, 288, 94

\bibitem[{York} et~al.(2000){York}, {Adelman}, {Anderson} et~al.]{york00a}
{York} D.~G., {Adelman} J., {Anderson} J.~E., et~al., 2000, \aj, 120, 1579

\end{thebibliography}

\end{document}